%% file: main.tex
\documentclass{article}
\usepackage{dilab_arxiv}

\usepackage{amsfonts}
\usepackage{algorithm}
\usepackage{algpseudocode}
\usepackage{tabularx}
\usepackage{array}
\usepackage{float}
\usepackage{placeins}
\usepackage{pifont}
\newcommand{\yes}{\textcolor{green!50!black}{\ding{51}}}
\newcommand{\no}{\textcolor{red!70!black}{\ding{55}}}
\usepackage{aliascnt}
\theoremstyle{plain}
\newtheorem{theorem}{Theorem}
\newaliascnt{lemma}{theorem}\newtheorem{lemma}[lemma]{Lemma}\aliascntresetthe{lemma}
\newaliascnt{corollary}{theorem}\newtheorem{corollary}[corollary]{Corollary}\aliascntresetthe{corollary}
\newaliascnt{proposition}{theorem}\newtheorem{proposition}[proposition]{Proposition}\aliascntresetthe{proposition}
\theoremstyle{definition}
\newaliascnt{definition}{theorem}\aliascntresetthe{definition}
\newaliascnt{assumption}{theorem}\newtheorem{assumption}[assumption]{Assumption}\aliascntresetthe{assumption}
\theoremstyle{remark}
\newaliascnt{remark}{theorem}\aliascntresetthe{remark}

\newcommand{\R}{\mathbb{R}}
\newcommand{\norm}[1]{\lVert #1 \rVert}
\newcommand{\abs}[1]{\lvert #1 \rvert}
\newcommand{\ip}[2]{\langle #1, #2 \rangle}
\newcommand{\pos}[1]{[#1]_{+}}
\newcommand{\one}{\mathbf{1}}
\newcommand{\N}{\mathcal{N}}
\newcommand{\Gap}{\mathrm{Gap}}
\newcommand{\At}{\tilde{A}}
\newcommand{\Umat}{U}
\newcommand{\eps}{\varepsilon}
\newcommand{\rhoiro}{\rho_{\mathrm{IRO}}}
\newcommand{\IRO}{\textsc{IRO-PRM$^+$}}
\newcommand{\IREG}{\textsc{IREG-PRM$^+$}}
\newcommand{\RMp}{\textsc{RM$^+$}}
\newcommand{\PRMp}{\textsc{PRM$^+$}}
\DeclareMathOperator{\supp}{supp}
\DeclareMathOperator{\diag}{diag}
\DeclareMathOperator{\spec}{spec}

\DeclareMathOperator*{\argmax}{arg\,max}
\newcommand{\Description}[1]{}   % ACM accessibility macro: no-op here

\usepackage[capitalise]{cleveref}
\crefname{theorem}{Theorem}{Theorems}
\crefname{lemma}{Lemma}{Lemmas}
\crefname{corollary}{Corollary}{Corollaries}
\crefname{proposition}{Proposition}{Propositions}
\crefname{definition}{Definition}{Definitions}
\crefname{remark}{Remark}{Remarks}
\crefname{assumption}{Assumption}{Assumptions}
\crefname{algorithm}{Algorithm}{Algorithms}

\makeatletter
\AtBeginDocument{%
  \begingroup\usefont{T1}{XCharter-TLF}{m}{n}\endgroup
  \@ifundefined{T1+XCharter-TLF}{}{%
    \DeclareFontShape{T1}{XCharter-TLF}{m}{scit}{<->ssub*XCharter-TLF/m/sc}{}}%
}
\makeatother

\input{tables/numbers}

\title{Linear Last-Iterate Convergence of Single-Call Scale-Invariant Regret Matching}
\runningtitle{Last-Iterate Convergence of Single-Call Scale-Invariant Regret Matching}
\date{\today}
\paperlogo{\includegraphics[height=1.5cm]{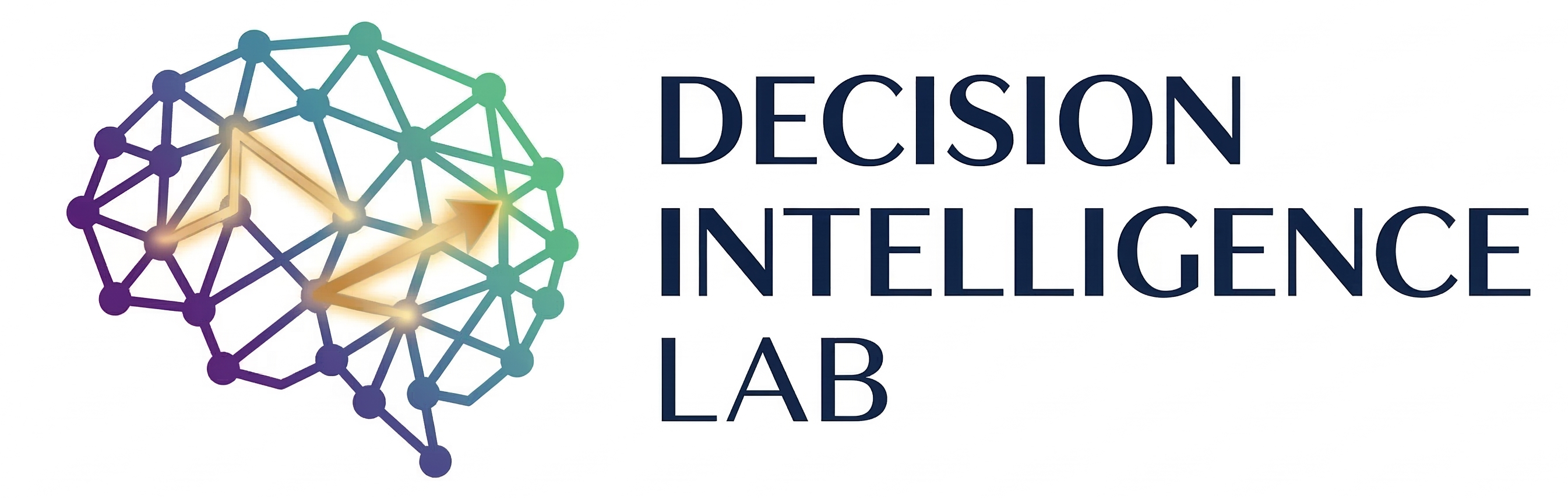}}
\author{
  Boning Li$^{1}$ and Longbo Huang$^{1\,\text{\faEnvelope}}$
  \\[0.3em]\normalfont
  $^1$Institute for Interdisciplinary Information Sciences, Tsinghua University
  \\
  \text{\faEnvelope}\ Correspondence: longbohuang@tsinghua.edu.cn
}

\begin{document}
\maketitle
\thispagestyle{fancy}

\begin{abstract}
\input{sections/00_abstract}
\end{abstract}

\input{sections/01_intro}
\input{sections/02_related}
\input{sections/03_setup}
\input{sections/04_saturation}

\input{sections/05_face}
\input{sections/06_threshold}
\input{sections/07_experiments}
\FloatBarrier
\FloatBarrier
\input{sections/08_conclusion}

\FloatBarrier
\bibliography{references}
\bibliographystyle{dilab_ref}

\appendix
\makeappendixtoc

\input{sections/16_imported}
\input{sections/07b_beyond}
\input{sections/15_extra}
\input{sections/14_proofs}

\input{sections/11_sharpproof}
\input{sections/13_protocol}
\input{sections/09_ledger}
\input{sections/10_instances}
\input{sections/12_lyapproof}

\end{document}

%% file: tables/numbers.tex
\newcommand{\nInst}{240}
\newcommand{\nConvIRO}{236}
\newcommand{\nConvIREG}{240}

\newcommand{\boundViolIRO}{0}

\newcommand{\ratioBoundMedIRO}{$3.7\times10^{-5}$}
\newcommand{\ratioBoundMaxIRO}{$1.6\times10^{-3}$}

\newcommand{\nCtwo}{83}
\newcommand{\nCtwoOK}{83}

\newcommand{\nDisc}{35}
\newcommand{\nDiscIRO}{35}
\newcommand{\nDiscEG}{0}

\newcommand{\growthPostMax}{$5.7\times10^{-11}$}

\newcommand{\nAboveEsc}{860}

\newcommand{\nDegen}{64}
\newcommand{\nDegenPure}{57}

\newcommand{\NlevelRatioMed}{1.70}
\newcommand{\cfiveDimthree}{2.08}

\newcommand{\cfiveDimfifty}{0.68}

\newcommand{\figInstEta}{0.394}

\newcommand{\fdcfN}{83}
\newcommand{\fdcfMedian}{$8.7\times10^{-10}$}
\newcommand{\fgcN}{6}

\newcommand{\tTwoSamp}{2000}

\newcommand{\exXstar}{\left(\tfrac{2}{5},\tfrac{3}{5}\right)}
\newcommand{\exValue}{\tfrac{1}{5}}
\newcommand{\exQone}{\left(\tfrac{1}{4},0\right)}
\newcommand{\exQm}{\left(\tfrac{3}{4},0\right)}
\newcommand{\exGamma}{\tfrac{1}{2}}
\newcommand{\exP}{\left(\tfrac{1}{4},0\right)}
\newcommand{\exQnorm}{\tfrac{1}{4}}
\newcommand{\exZ}{\left(1,0\right)}
\newcommand{\nAltConv}{238}
\newcommand{\nAltMixed}{161}

\newcommand{\nAltAboveSim}{16}

\newcommand{\nAltBoth}{132}

\newcommand{\altSpeedQpfifty}{1.65}

\newcommand{\nIREGcf}{136}
\newcommand{\iregCfWithinFive}{122}

\newcommand{\landGapCrossMed}{1.01}
\newcommand{\landGrowthAfterMed}{27\%}

\newcommand{\oracleFracDimthree}{0.28}

\newcommand{\oracleFracDimfifty}{0.63}

\newcommand{\pureTfixMed}{2}
\newcommand{\nOgdaCmp}{132}
\newcommand{\ogdaGrid}{32}
\newcommand{\ogdaFracMed}{0.72}

\newcommand{\ogdaExtra}{3}

\newcommand{\nUniqueRank}{186}
\newcommand{\nUniqueMixed}{163}

\newcommand{\nLedgerRows}{720}
\newcommand{\nLedgerCols}{46}

\newcommand{\nUniqueRankDisagree}{0}
\newcommand{\rankTolMax}{$1.0\times10^{-8}$}
\newcommand{\sigmaRetMin}{0.048}
\newcommand{\sigmaDropMax}{$7.4\times10^{-13}$}
\newcommand{\timeN}{50}

\newcommand{\timeRatio}{6.0}
\newcommand{\altFigN}{142}

\newcommand{\altFigSimN}{135}

\newcommand{\landFigAfter}{52\% / 43\% / 25\%}

\newcommand{\altSpecFullN}{161}
\newcommand{\altSpecFullOK}{161}

\newcommand{\sgFarinaIRORhoCf}{0.9567}

\newcommand{\sgFarinaIRORateRatio}{1.25}

\newcommand{\sgBrpsIRORateRatio}{0.99}

\newcommand{\sgWeiIRORateRatio}{1.00}

\newcommand{\sgBestDev}{$2.8\times10^{-16}$}
\newcommand{\efgKuhnRMNcT}{$1.6\times10^{-1}$}

\newcommand{\efgKuhnIROTconv}{114}

\newcommand{\efgLeducRMNcT}{$1.6\times10^{-1}$}

\newcommand{\efgT}{20000}
\newcommand{\efgLeducCfrPlus}{$1.8\times10^{-3}$}
\newcommand{\efgLeducPublished}{-0.0856}
\newcommand{\efgLeducNinf}{144}
\newcommand{\efgKuhnBrDev}{$2.8\times10^{-16}$}
\newcommand{\efgKuhnNpure}{64}
\newcommand{\efgLeducValueDev}{$2.4\times10^{-5}$}
\newcommand{\efgLeducValue}{-0.08562}

\newcommand{\sgFarinaKappa}{1.99}
\newcommand{\sgBrpsKappa}{1.00}

\newcommand{\sgFarinaWinLate}{0.98}

\newcommand{\sgFarinaRhoSecond}{0.9416}

\newcommand{\tttReachIROsix}{218}
\newcommand{\tttReachIROaltsix}{228}
\newcommand{\tttReachIREGsix}{236}

\newcommand{\tttReachPRMsix}{167}

\newcommand{\fitLateN}{131}

\newcommand{\altFitLateN}{143}

\newcommand{\sgFarinaModeRatioStart}{43}
\newcommand{\sgFarinaModeRatioFourth}{0.13}

\newcommand{\sharpLemRounds}{$6.7\times10^{6}$}
\newcommand{\sharpLemRoundsAlt}{$4.8\times10^{6}$}
\newcommand{\sharpLemWorst}{0.49}
\newcommand{\sharpBelowHalf}{123}
\newcommand{\sharpTrig}{119}
\newcommand{\sharpNoTrig}{4}
\newcommand{\sharpTrigGapMed}{$7.7\times10^{-8}$}
\newcommand{\sharpShrinkMed}{$2.6\times10^{-13}$}
\newcommand{\sharpShrinkMax}{$2.9\times10^{-7}$}

\newcommand{\sharpLoDecades}{4.4}

\newcommand{\gstarRestartRelMed}{8}
\newcommand{\nuSimStrict}{35}
\newcommand{\nuSimNotStrict}{19}
\newcommand{\nuSimDmax}{17}
\newcommand{\nuSimUnequal}{20}
\newcommand{\nuSimModes}{13}
\newcommand{\nuSimUzero}{22}
\newcommand{\nuSimRhoDev}{$5.9\times10^{-11}$}
\newcommand{\nuSimFit}{13}
\newcommand{\nuSimFitOK}{12}
\newcommand{\nuAltStrict}{39}

\newcommand{\nuAltModes}{13}

\newcommand{\nuAltRhoDev}{$5.1\times10^{-11}$}
\newcommand{\nuAltFit}{13}
\newcommand{\nuAltFitOK}{13}
\newcommand{\nuNonUnique}{54}
\newcommand{\weiMargin}{0.135}
\newcommand{\weiKone}{5}
\newcommand{\weiKtwo}{3}
\newcommand{\weiRank}{2}
\newcommand{\weiDone}{2}
\newcommand{\weiDtwo}{0}
\newcommand{\weiUnit}{4}
\newcommand{\rbStarts}{five}

\newcommand{\rbSimConv}{795}

\newcommand{\rbSimFit}{679}
\newcommand{\rbSimFitOK}{611}
\newcommand{\rbSimSpreadMed}{16\%}

\newcommand{\rbAltConv}{803}

\newcommand{\rbAltFit}{721}
\newcommand{\rbAltFitOK}{663}
\newcommand{\rbAltSpreadMed}{32\%}

\newcommand{\rbColConv}{162}

\newcommand{\tttReachSOGRMsix}{219}

\newcommand{\smoothGridN}{seven}

\newcommand{\efgKuhnAltTconv}{90}

\newcommand{\efgLeducAltMed}{$2.2\times10^{-5}$}
\newcommand{\efgLeducIROMed}{$1.2\times10^{-2}$}

\newcommand{\efgLongT}{$10^{5}$}
\newcommand{\efgLongAltNc}{$3.2\times10^{-13}$}
\newcommand{\efgLongAltStay}{52978}
\newcommand{\efgLongIregNc}{$9.0\times10^{-14}$}

\newcommand{\efgLongIroNc}{$1.3\times10^{-6}$}
\newcommand{\efgLongRmNc}{$1.5\times10^{-1}$}
\newcommand{\absN}{68}

\newcommand{\absClean}{9}
\newcommand{\absFirstViol}{90}
\newcommand{\lyTrigSim}{157}
\newcommand{\lyTrigAlt}{158}
\newcommand{\lyNoTrigSim}{6}
\newcommand{\lyNoTrigAlt}{5}
\newcommand{\lyHiSim}{40}
\newcommand{\lyHiTrigSim}{39}
\newcommand{\lyHiAlt}{34}
\newcommand{\lyHiTrigAlt}{33}
\newcommand{\lyEpsMaxSim}{0.5770}
\newcommand{\lyEpsMaxAlt}{0.6646}
\newcommand{\lyGapMedSim}{$2.8\times10^{-8}$}
\newcommand{\lyGapMedAlt}{$3.1\times10^{-8}$}
\newcommand{\lyDecayRounds}{$3.6\times10^{6}$}

\newcommand{\lyEnergyRatio}{2.1\%}
\newcommand{\lyPredTwo}{$9.4\times10^{-3}$}
\newcommand{\lyPredFour}{$1.1\times10^{-4}$}
\newcommand{\lyPredSeven}{$1.3\times10^{-7}$}

\newcommand{\frReconMed}{0.3\%}

\newcommand{\frFitDevTwo}{0.007}

\newcommand{\frTstart}{182}

%% file: sections/00_abstract.tex
Regret matching needs no step-size tuning and drives the counterfactual regret minimisation solvers behind superhuman poker agents by placing a learner at each decision point. Decision-time solvers need the last iterate that agents play in practice to converge, whereas classical guarantees cover only the time-averaged strategy, whose use requires storing a second strategy copy while the last iterate can cycle. In self-play on a two-player zero-sum matrix game, a scale-invariant modification restores an optimal parameter-free average-iterate convergence rate, yet its effective step size is the inverse regret-vector length. Because this trajectory-dependent quantity might not converge, fixed-step contraction arguments fail, while clipping makes updates nonsmooth and fixed points form a cone. Prior work proved a conditional last-iterate rate requiring two payoff evaluations per round, leaving open both how large the step size grows for a single-call learner under simultaneous or alternating schedules and a finite-round test for entering the convergence regime. Here, we analyse scale-invariant predictive regret matching plus (\IRO{}) with one payoff evaluation per round under both schedules, proving an explicit upper bound showing its regret norm saturates at a finite limit. We derive the exact local recursion on the equilibrium face and its closed-form convergence rate, finding sharp stability thresholds on the limiting step size of $1/\sqrt3$ and $2/3$ with a fastest contraction rate of $1/\sqrt2$. We also provide a single-round certificate applying whenever the algorithm lands strictly below its schedule's threshold from any initial regrets. Across \nInst{} random games, our closed-form rates match empirical slopes, all \nUniqueMixed{} games with a rank-unique mixed equilibrium are certified below threshold, and alternating updates approach a factor-of-two speedup on larger games. These results give explicit parameter-free last-iterate rates at one evaluation per round.

%% file: sections/01_intro.tex
\section{Introduction}
\label{sec:intro}

\looseness=-1
Large imperfect-information games, including heads-up and multiplayer poker, are solved by self-play using counterfactual regret minimisation \citep{zinkevich2007regret, bowling2015heads, brown2018superhuman, brown2019superhuman, li2024rl, li2025efficient, li2026effective, li2026abstraction}. These solvers place an independent regret-minimising learner at every decision point, typically using regret matching \citep{hart2000simple} or regret matching plus (\RMp{}) \citep{tammelin2014solving}. Because payoff scales vary widely across a game tree and across domains, a single step size does not transfer, and per-game tuning is impractical. Furthermore, classical theory guarantees only the time-averaged strategy, forcing implementations to maintain a second copy of the strategy table. Decision-time solvers recompute a strategy for the situation at hand \citep{moravvcik2017deepstack,li2026safe}. This paper asks whether the last iterate of a parameter-free learner converges on a two-player zero-sum matrix game, and how fast.

Classical regret matching offers no help here, as its last iterate can cycle \citep{farina2023regret, cai2025lastiterate}. Adding scale-invariant prediction \citep{zhang2025scale} restores an optimal average-iterate rate without tuning, while conditional last-iterate rates exist only for a two-evaluation variant (\IREG{}) per round \citep{li2026last}, and concurrent works establish qualitative convergence without rates \citep{sullins2026irprmpreviousutilitypredictionconstant, nguyen2026convergencealternatingregretmatching}. Obtaining explicit last-iterate rates for a single-call, parameter-free learner presents four open difficulties. First, the effective step size, the inverse of the accumulated regret vector's length, is a state variable that the trajectory itself moves, so fixed-step contraction arguments fail and the length itself might fail to converge. Second, clipping regrets at zero makes the update map nonsmooth, while the fixed points form a two-parameter cone of rays, one scaling ray per player, allowing contraction only toward a set. Third, nothing fixes how large the effective step size can grow before the iteration loses contractivity, leaving the stability boundary unknown. Fourth, the rate is asymptotic, and no finite-round test links a trajectory's current position to the regime in which the rate holds.

We resolve these four challenges for single-call scale-invariant predictive regret matching plus, denoted \IRO{}. Following optimistic online learning with predictable sequences \citep{rakhlin2013online} and predictive regret matching plus (\PRMp{}) \citep{farina2021faster}, \IRO{} predicts the upcoming utility using the previous utility and requires one payoff evaluation per round. First, adapting the saturation argument of \citet{li2026last} for the two-evaluation variant, we prove an unconditional saturation theorem giving an explicit upper bound on the accumulated regret vector's length under both simultaneous and alternating schedules. Second, we perform an exact linearisation on the equilibrium support where clipping switches off, deriving a companion-matrix recursion and a closed-form local rate, while showing that strict pure equilibria are reached in a single round. Third, we identify sharp stability thresholds for the asymptotic step size at $1/\sqrt3$ for simultaneous play and $2/3$ for alternation, proving that the fastest contraction factor is $1/\sqrt2$ and that alternation doubles the leading-order log-rate, the slope of log regret against rounds. Fourth, we construct a single-round certificate that verifies linear convergence for any landing, that is, any limiting step size, strictly below these thresholds from arbitrary initial regrets, and we extend our local analysis to limits admitting multiple equilibria. A detailed comparison with prior guarantees appears in \cref{tab:contrib}.

\looseness=-1Across \nInst{} randomly generated zero-sum games, the closed-form spectral rates match the measured last-iterate slopes under both simultaneous and alternating updates. Each of the \nUniqueMixed{} games with a rank-unique mixed equilibrium is certified strictly below the stability threshold of its schedule. Moreover, \nAltAboveSim{} alternating runs settle inside the interval between $1/\sqrt3$ and $2/3$, a region where simultaneous play has no attracting fixed point. When restarted inside this gap, trajectories escape under simultaneous updates yet converge under alternation, consistent with the predicted thresholds. Finally, without any parameter tuning, the self-selected step size of \IRO{} achieves a median \ogdaFracMed{} of the asymptotic log-rate of hand-tuned optimistic gradient descent-ascent.

\begin{itemize}
\item For the single-call learner, we prove the first explicit bound on the regret norms across all games and both update schedules. This guarantees saturation without any assumption on the payoff matrix and yields a closed-form bound when both regret vectors become nonzero in round two.
\item We derive the exact local dynamics near a strictly complementary equilibrium and obtain a closed-form linear convergence rate (\cref{sec:face}). Clipping switches off on the equilibrium support, reducing the update to a smooth two-step recursion governed by a companion matrix, while strict pure equilibria are reached in one round.
\item We identify sharp stability thresholds for the limiting effective step size, the largest landings at which the limit remains attracting (\cref{sec:threshold}). Measured in units of the top singular value of the payoff submatrix over the equilibrium support, these thresholds are $1/\sqrt3$ for simultaneous play and $2/3$ for alternation, both with fastest contraction $1/\sqrt2$, and alternation doubles the asymptotic log-rate.
\item We provide a single-round certificate\looseness=-1 (\cref{thm:lyap}) that verifies linear convergence from arbitrary initial regrets for any landing below the threshold of its schedule, bounding the remaining energy and step-size drift. We also show that our local rate theory requires strict complementarity of the limit, so every action outside the equilibrium support is strictly worse, dropping equilibrium uniqueness so the active spectrum lives on the nonzero singular values of the face.
\end{itemize}

%% file: sections/02_related.tex
\section{Related Work}
\label{sec:related}

\begin{table}[t]
\caption{Guarantees and thresholds by schedule for regret matching$^+$ and shifted variants in two-player zero-sum self-play.}
\label{tab:contrib}
\centering\footnotesize
\setlength{\tabcolsep}{3.5pt}
\begin{tabular}{@{}>{\raggedright\arraybackslash}p{4.6cm}ccccccccc@{}}\toprule
 & no step & evals / & & & norm & explicit & linear & closed-form & \\
learner & size & round & sim. & alt. & saturation & bound & last iterate & rate & threshold\\\midrule
\PRMp{} \citep{farina2021faster} & \yes & 1 & \yes & \yes & \no & \no & \no & \no & \\
smoothed \RMp{} with restarts \citep{cai2025lastiterate} & \no & 1 or 2 & \yes & \no & \no & \no & \yes & \no & \\
perturbed-regularized \RMp{} \citep{meng2025lastiterate,meng2025efficient} & \no & 1 & \yes & \no & \no & \no & \no & \no & \\
\IREG{} \citep{zhang2025scale} & \yes & 2 & \yes & \no & \no & \no & \no & \no & \\
\IREG{} \citep{li2026last} & \yes & 2 & \yes & \no & \yes & \yes & \yes & \yes & $1$\\
\IRO{} \citep{sullins2026irprmpreviousutilitypredictionconstant} & \yes & 1 & \yes & \yes & \yes & \no & \no & \no & \\
\IRO{} \citep{nguyen2026convergencealternatingregretmatching} & \yes & 1 & \no & \yes & \no & \no & \no & \no & \\\midrule
\IRO{} (this paper) & \yes & 1 & \yes & \yes & \yes & \yes & \yes & \yes & $1/\sqrt3$, $2/3$\\
\bottomrule
\end{tabular}
\end{table}

In regret matching, each player tracks a vector with one entry per action that accumulates missed payoff gains, then plays actions in proportion to its positive entries. Regret matching$^+$ (\RMp{}), which clips negative entries to zero, and its predictive variant (\PRMp{}), which adds a utility forecast before playing, drive equilibrium computation in practice \citep{tammelin2014solving,bowling2015heads,farina2021faster}. They operate inside counterfactual regret minimization (the standard solver for extensive-form games, running one regret-matching learner per decision point) and its optimistic and discounted refinements \citep{zinkevich2007regret,farina2019stable,farina2019optimistic,brown2019solving,brown2019deep,brown2020combining,xu2024dynamic,xu2024minimizing,meng2026faster,li2026real,li2026gpu}. Still, their last iterates, the strategies actually played, can cycle or diverge \citep{farina2023regret,cai2025lastiterate}.

Existing remedies that target linear last-iterate rates reintroduce a tuning parameter, for example with forgetting \citep{cai2024fast}, smoothing with restarts \citep{cai2025lastiterate}, a perturbed-regularized objective \citep{meng2025lastiterate,meng2025efficient}. A related line re-anchors the regrets without regularization \citep{zhang2026teaching}. \citet{zhang2025scale} eliminate the step size completely via an equal-norm shift of the regret vectors, achieving an optimal average-iterate rate without tuning. Building on this shifted dynamic, \citet{li2026last} study the two-evaluation extra-gradient learner. They prove both regret-norm saturation and a linear last-iterate rate conditioned on the saturated norm. However, how large the effective step size may grow for the single-call learner remains open. Meanwhile, two concurrent works investigate the single-call learner that reuses previous utilities. Specifically, \citet{sullins2026irprmpreviousutilitypredictionconstant} and \citet{nguyen2026convergencealternatingregretmatching} establish qualitative last-iterate convergence of the single-call learner, for simultaneous updates in one case and alternating updates in the other, though neither work provides rates.

As summarized in \cref{tab:contrib}, we provide closed-form last-iterate rates for both schedules, an explicit saturation certificate, and the two stability thresholds (the largest settled inverse norm at which the limit stays attracting). Combining our local rates with their asymptotic convergence yields a global rate (\cref{cor:global}). Our local guarantees require strict complementarity of the limit (every unused action is strictly suboptimal against the equilibrium) and allow multiple equilibria (\cref{sec:nonunique}). Optimistic, extra-gradient and alternating methods, and the Blackwell and Hoffman connections, are compared in \cref{app:relwork}.

%% file: sections/03_setup.tex
\section{Preliminaries and the Single-Call Learner}
\label{sec:setup}

A two-player zero-sum matrix game is given by $A\in\R^{n_1\times n_2}$. The row
player chooses $x\in\Delta_{n_1}$ to maximise $x^\top A y$, and the column player
chooses $y\in\Delta_{n_2}$ to minimise it. The utility vectors are $u_1=Ay$ and
$u_2=-A^\top x$. The duality gap, which measures how much either player could
gain by deviating, is $\Gap(x,y)=\max_a (Ay)_a-\min_b (x^\top A)_b\ge 0$, and
$\Gap(x,y)=0$ holds exactly at the Nash equilibria. Throughout,
$\sigma:=\sigma_{\max}(A)$, $K_i:=1+\sqrt{n_i}$, $\Delta_n:=\{x\in\R^n_{\ge0}:\ \one^\top x=1\}$ is the probability simplex, $\ip{\cdot}{\cdot}$ is the Euclidean inner product, $\N(q):=q/\norm{q}_1$ for
$q\in\R^n_+\setminus\{0\}$, and $\pos{\cdot}$ is the coordinatewise positive
part.

We defer all proofs omitted from the main text to \cref{app:proofs}. We now fix
the notation for our analysis. A subscript $i\in\{1,2\}$ indexes the player, as
in $q_i^t$, $N_{i,t}$, and $z_i^t$ with $z_1=x$ and $z_2=y$, and we set
$N_{i,\infty}:=\lim_tN_{i,t}$. A superscript ${\rm p}$ denotes a quantity stored
from the previous round. Moreover, $e_a$ denotes the $a$-th standard basis
vector, and $F(z)$ denotes the pair of utilities $(Ay,-A^\top x)$ at a profile
$z=(x,y)$. Let $t_i^*$ be the first round in which player $i$ plays from a
nonzero vector. Under simultaneous updates, this is the first $t$ with
$q_i^t\neq0$. Under alternation, the row player plays $x^t$ from $q_1^t$ and
the column player plays $y^t$ from $q_2^{t+1}$, so $t_2^*$ is the first $t$
with $q_2^{t+1}\neq0$. Let $N_i^*$ be the norm of the vector from which player $i$ first plays. Specifically, $N_1^*=N_{1,t_1^*}$, while $N_2^*=N_{2,t_2^*}$ under
simultaneous updates and $N_2^*=N_{2,t_2^*+1}$ under alternating updates. Let
$T_0:=\max(t_1^*,t_2^*)$ be the first round in which both plays are active.
\paragraph{The common update.}
Each player keeps a pseudo-regret vector $q^t\in\R^n_+$, which tracks how much
better each action would have performed than the strategies played so far. It
starts at $q^1=0$, with Euclidean norm $N_t:=\norm{q^t}_2$ and carried strategy
$\tilde z^t:=\N(q^t)$ (uniform while $q^t=0$). Given a prediction $m^t\in\R^n$
of the coming utility, the learner plays and accumulates according to the
following rules.
\begin{align}
  v^t &= q^t+m^t-\gamma^t\one,\qquad
  \gamma^t:=\gamma(q^t,m^t)\ \text{solves}\ \norm{\pos{v^t}}_2=N_t, \label{eq:shift}\\
  p^t &= \pos{v^t},\qquad z^t=\N(p^t)\ \text{(played)}, \label{eq:play}\\
  g^t &= (u^t-m^t)-\ip{u^t-m^t}{z^t}\one,\qquad
  q^{t+1}=\pos{v^t+g^t}. \label{eq:accum}
\end{align}
When $q^t=0$, the learner ignores $m^t$, plays $\tilde z^t$, and sets
$q^{t+1}=\pos{u^t-\ip{u^t}{\tilde z^t}\one}$, resetting negative entries to zero
(clipping). The shift \eqref{eq:shift} is the equal-norm shift of
\citet{zhang2025scale}. Because $\gamma\mapsto\norm{\pos{q+m-\gamma\one}}_2$ is
strictly decreasing where positive, $\gamma^t$ is unique whenever $q^t\neq0$, and the played vector
$p^t$ has the same Euclidean norm as $q^t$. The accumulation \eqref{eq:accum}
uses the unclipped $v^t$, as in \citet{zhang2025scale}.

With $m^t\equiv0$ and $\gamma^t\equiv0$, the update reduces to \RMp{}
\citep{tammelin2014solving}. Predictive \RMp{} \citep{farina2021faster} keeps
the accumulation $q^{t+1}=\pos{q^t+u^t-\ip{u^t}{z^t}\one}$ of \RMp{} and lets
the prediction enter only the played vector via
$z^t=\N(\pos{q^t+m^t-\ip{m^t}{z^{t-1}}\one})$. Setting $\gamma^t\equiv0$ in
\eqref{eq:shift} and \eqref{eq:accum} does not recover predictive \RMp{}.

\begin{algorithm}[t]
\caption{One round of the common update for one player with a nonzero regret vector.}
\label{alg:round}
\begin{algorithmic}[1]
\State $v^t\gets q^t+m^t-\gamma^t\one$, \quad $\gamma^t$ from \eqref{eq:shift}
\State play $z^t\gets\N(\pos{v^t})$ \Comment{$\tilde z^t=\N(q^t)$ is carried}
\State observe $u^t$; \quad $g^t\gets(u^t-m^t)-\ip{u^t-m^t}{z^t}\one$
\State $q^{t+1}\gets\pos{v^t+g^t}$, \quad $\tilde z^{t+1}\gets\N(q^{t+1})$
\State $m^{t+1}\gets u^t$ (\IRO{}) \textbf{or} $F(\tilde z^{t+1})$ (\IREG{})
\end{algorithmic}
\end{algorithm}

\paragraph{Two prediction rules.}
The learner is scale invariant. Multiplying $A$ and the initial regrets by $c>0$ scales every $q^t$, $v^t$, and $g^t$ by $c$, while leaving every $z^t$ unchanged. The variants we study differ in the choice of $m^t$.
\begin{itemize}
\item \IREG{} (extra-gradient prediction) sets $m_1^t=A\tilde y^t$ and
  $m_2^t=-A^\top\tilde x^t$, the utility at the carried profile. This requires
  two evaluations of the payoff per round.
\item \IRO{} (single-call prediction) sets $m^t=u^{t-1}$, the previous observed
  utility \citep{rakhlin2013online,farina2021faster}. This requires one
  evaluation per round.
\end{itemize}
\citet{li2026last} analyse \IREG{}. They show that its regret norm always saturates at a finite limit, with an explicit bound, and prove a local linear rate for the last iterate (the strategy played in the current round) under a step-size condition on that limit. This paper analyses \IRO{}, the same learner with half the evaluations and a different stability threshold.

\paragraph{Two schedules.}
Under simultaneous updates, both players play and then both observe. Under the
row-first alternating schedule of solvers
\citep{tammelin2014solving,nguyen2026convergencealternatingregretmatching}, the
row player first plays $x^t$ from $(q_1^t,m_1^t=Ay^{t-1})$ in round $t$. Next,
the column player observes $u_2=-A^\top x^t$ for its standing strategy
$y^{t-1}$, accumulates, and plays $y^t$ using that same utility as its
prediction. Finally, the row player observes $u_1^t=Ay^t$.

Each player still runs \eqref{eq:shift} and \eqref{eq:accum} on its own utility
sequence using the previous observed utility as prediction, and each still
evaluates the payoff once per round. In the column player's own round count, it
plays $y^k$ and observes $u_2^k=-A^\top x^{k+1}$ for $k\ge0$, with $y^0$
uniform. The only change is the opponent strategy at which the observed utility
is evaluated.

The round-boundary state of the alternating dynamics is
$(q_1,u_1^{\rm p},r_2,m_2^{\rm p})$, where $r_2$ is the column player's
unclipped played vector and $m_2^{\rm p}$ is the prediction it used. The column regret $q_2$ is omitted from the state because the next round rebuilds it from
$r_2$, the column player's stored unclipped vector (\cref{alg:altround}, line 2).
The sign pattern of $r_2$ at an alternating fixed point is part of \cref{lem:altfreeze} in \cref{app:extra}.

\begin{algorithm}[t]
\caption{One round of the row-first alternating schedule.}
\label{alg:altround}
\begin{algorithmic}[1]
\State \textbf{row plays} $x^t$ from $(q_1^t,m_1^t=Ay^{t-1})$: \cref{alg:round}, lines 1 and 2
\State \textbf{column observes} $u_2^t=-A^\top x^t$ at its standing $y^{t-1}=\N(\pos{r_2^t})$: $q_2^{t+1}=\pos{r_2^t+g_2^t}$
\State \textbf{column plays} $y^t$ from $(q_2^{t+1},m_2^t=u_2^t)$: \cref{alg:round}, lines 1 and 2
\State \textbf{row observes} $u_1^t=Ay^t$: $q_1^{t+1}=\pos{v_1^t+g_1^t}$ and $m_1^{t+1}=u_1^t$
\end{algorithmic}
\end{algorithm}

We first illustrate the shift on a two-by-two example. Take
$A=\begin{pmatrix}2&-1\\-1&1\end{pmatrix}$, with $x^\ast=y^\ast=\exXstar$ and
value $\exValue$. Both players open at the uniform profile, and after one round,
the row player reaches $q_1^2=\exQone$. The shift then becomes active in round
two. With $m_1^2=u_1^1$, the vector $q_1^2+m_1^2=\exQm$ solves \eqref{eq:shift}
at $\gamma_1^2=\exGamma$. Hence, the played vector is $p_1^2=\exP$ with
$\norm{p_1^2}=\norm{q_1^2}=\exQnorm$, inducing the pure strategy $\exZ$. Note
that the shift absorbs whatever the prediction adds, so the played vector
preserves the length of the regret. Moreover, because we read off the strategy
after the shift, the update has no step size to tune.

\paragraph{Terminology.}\looseness=-1 The regret norm $N_t$ sets the scale of a round. Because a round moves the state by an amount proportional to $1/N_t$, this norm is an inverse step size. At a fixed point $(c_1 x^*, c_2 y^*)$, where $c_i = \norm{q_i}_1$ is the $\ell_1$ mass of player $i$'s regret vector, we define the \emph{effective step size} as $\eta := 1/\sqrt{c_1 c_2}$. The \emph{landing} $\eta_\infty$ is the limit of the effective step size along a run, and we say that the run \emph{lands} at $\eta_\infty$. The \emph{support} of an equilibrium is the set of actions it plays with positive probability, written $S_1=\supp x^*$ and $S_2=\supp y^*$. The \emph{support face} collects the states whose regret mass lies on this support. Finally, a \emph{certificate} is a condition evaluated on one round of a run, together with the game, that implies the linear rate from that round onward. Our certificates are one-sided.

\looseness=-1Our analysis relies on four elementary facts about the common update, which we state along with the other imported lemmas in \cref{app:imported}.

%% file: sections/04_saturation.tex
\section{Saturation Is Unconditional}
\label{sec:saturation}

In this section, we establish unconditional saturation of the regret norm $N_t$, defined as the Euclidean length of the regret vector. This quantity is the inverse step size of the learner. We show that $N_t$ saturates (meaning it stays bounded and converges) on every payoff matrix. Specifically, an explicit bound holds in terms of $A$ and the two activation norms whenever both players become active. \Cref{lem:activation} covers the remaining case. For \IREG{}, \citet{li2026last} establish saturation with an explicit bound by building on the bounded path length of \citet{zhang2025scale}. Our results establish the corresponding guarantee for \IRO{}. For \IRO{}, \citet{sullins2026irprmpreviousutilitypredictionconstant} concurrently prove that the norms stay bounded, with a constant defined through the run. We compute our bound directly from $A$ and the state at activation, defined as the first round starting with both regret vectors nonzero. Single-call prediction incurs a logarithmic term.

Recall that $N_i^*$ is the norm from which player $i$ first plays, and that both players are active from round $T_0$ onward. Under simultaneous updates, both players choose the uniform strategy in round one. In the generic case with $n_1,n_2\ge2$, both activate in round two. This gives $T_0=2$ and $N_i^*=\norm{\pos{u_i^1-\bar u_i^1\one}}$ with $u_1^1=A\one/n_2$ and $u_2^1=-A^\top\one/n_1$, where $\bar u$ denotes the average entry of $u$. Both constants then have closed forms in terms of $A$. \Cref{lem:activation} in \cref{app:extra} handles the remaining cases and the alternating schedule. Before round $T_0$, an active player faces a constant utility, so its norm remains fixed at $N_i^*$. If $T_0=\infty$, at most one player ever activates. In that case, $N_{1,\infty}+N_{2,\infty}\le\max(N_1^*,N_2^*)$, where $N_i^*:=0$ for a player that never activates, so the norms are bounded for every $A$. We therefore assume $T_0<\infty$ below.

For $t\ge T_0$, we define $D_i:=\sum_t\norm{z_i^t-\tilde z_i^t}^2$, $G_i:=\sum_t\norm{g_i^t}^2$, $G:=G_1+G_2$, $a_i:=2n_i/N_i^*$, with every sum running over the rounds from $T_0$ onward except where stated below. Under alternation, the column player's energy sum $G_2$ runs over accumulations from a nonzero vector at or after $T_0$. Specifically, it runs over $t\ge T_0':=\max(T_0,t_2^*+1)$, where $N_{2,T_0'}=N_2^*$. Meanwhile, $D_2$ runs over its plays from round $T_0$ onward, all of which arise from nonzero vectors, with $\tilde y^t:=\N(q_2^{t+1})$ the carried strategy of the play $y^t$.

\begin{theorem}[\IRO{} saturates]\label{thm:sat-iro}
Consider \IRO{} on any $A$ with $T_0<\infty$. Then, we have $G\le G^*(A)<\infty$
and $N_{i,\infty}^2\le N_i^{*2}+G^*$, where $G^*$ is the unique positive root of
$G=\Phi(G)$,
\begin{align*}
  \Phi(G)&:=c_0+a\sqrt{G}+b\bigl[\log(1+G/N_1^{*2})+\log(1+G/N_2^{*2})\bigr],\\
  c_0&:=2\sigma^2\bigl[(K_1^2a_2+K_2^2a_1)(N_1^*+N_2^*)+2(K_1^2+K_2^2)\bigr],\\
  a&:=2\sqrt2\,\sigma^2(K_1^2a_2+K_2^2a_1),\qquad b:=2\sigma^2K_1^2K_2^2 .
\end{align*}
\end{theorem}

The proof follows the approach of \citet{li2026last} for \IREG{}, with one key modification. Under extra-gradient prediction, the error is a displacement $y^t-\tilde y^t$, which the shift correction dissipates linearly. Under single-call prediction, the error is an increment $y^t-y^{t-1}$. A part of this increment carries over from the previous round. We can control this carried-over part only through the norm increment $N_t^2-N_{t-1}^2$, which introduces the logarithmic term. We need no condition on $\sigma$ because the coupling term grows like $\sqrt G$, whereas the dissipation grows like $G$.

The constants in these bounds are loose (\cref{sec:experiments}). Nevertheless, the theorem establishes finiteness with an explicit bound, which we need for the local analysis below. Since the norms remain in $[N_i^*,N_{i,\infty}]$, at every fixed point (\cref{sec:face}) the dynamics selects its own effective step size $\eta:=1/\sqrt{c_1c_2}$, where $c_i:=\norm{q_i}_1$ and there is no parameter to tune.

Because our proofs use $T_0$ only as a round in which both players are active, we can restart each certificate at any later round $t_0>T_0$ with $N_{i,t_0}$ in place of $N_i^*$ to bound the energy from $t_0$ onward. We can drop the earlier nonnegative corrections because (F2) of \cref{lem:facts} holds on the full prefix. We give a sharper local certificate in \cref{thm:sharp}. The same bound extends to alternating updates (\cref{thm:sat-alt} in \cref{app:support}), and the nonzero-initialization bookkeeping with the limit-point lemma is collected in \cref{app:support}.

\paragraph{Alternating updates.}\looseness=-1
Under the schedule of \cref{sec:setup}, each player runs the same update on its own utility sequence. Every per-player ingredient above remains unchanged, with $g_1^t$ driven by $y^t-y^{t-1}$ and $g_2^k$ driven by $x^{k+1}-x^k$. Only (F2), stated in \cref{app:imported}, couples the two players, and under alternation the self-play regret sum acquires an extra signed term, which \cref{lem:improve,lem:altreg} in \cref{app:extra} control.

%% file: sections/05_face.tex
\section{The Frozen Face and Its Companion Jacobian}
\label{sec:face}

In this section, we consider a unique strictly complementary equilibrium, where every action outside the support plays strictly worse against the equilibrium. We first show that near such a point, the clipping in \eqref{eq:play} to \eqref{eq:accum} switches off on the support and freezes the off-support coordinates at zero. Thus, what remains is a smooth map on the face, namely the set of strategies using only the support actions. We then establish that its linearisation is the companion matrix that advances a two-step linear recursion by one step. In particular, this second statement shows where the single-call prediction cancels out at first order.

\begin{assumption}\label{ass:ne}
The game has a unique Nash equilibrium $(x^*,y^*)$ with value $v$. We denote its supports by $S_1=\supp x^*$ and $S_2=\supp y^*$, of sizes $k_i:=\abs{S_i}$. Moreover, the strict complementarity margin satisfies
$\mu:=\min\bigl(\min_{a\notin S_1}(v-(Ay^*)_a),\ \min_{b\notin S_2}((A^\top x^*)_b-v)\bigr)>0$. Here, we take a minimum over an empty set to be $+\infty$. In the rates below, we also take a maximum over an empty set of modes to be $0$.
\end{assumption}
We write $v$ for the game value and $v^t$, with a round index, for the unclipped shifted vector of \eqref{eq:shift}.

By \cref{lem:rank} of \citet{li2026last}, stated in \cref{app:imported}, the supports have equal size. We write $k:=k_1=k_2$.

We write the state of the \IRO{} dynamics as $s=(q_1,q_2,u_1^{\rm p},u_2^{\rm p})$, collecting the two regret vectors and the two previous utilities. One round is the map $F(s)=(q_1',q_2',u_1,u_2)$, and we define $R:=\{(c_1x^*,c_2y^*,u_1^*,u_2^*):c_1,c_2>0\}$ with $u_1^*:=Ay^*$ and $u_2^*:=-A^\top x^*$.

\begin{theorem}[One-step absorption into the face]\label{thm:freeze}
Let \cref{ass:ne} hold, and fix $s^*\in R$. Define $c_{\min}:=\min(c_1,c_2)$,
$\kappa_i:=2+5\sqrt{n_i}$, $A_{\max}:=\max_{a,b}\abs{A_{ab}}$,
$\ell:=A_{\max}\max_i 2k_i\kappa_i/c_{\min}$, and
\[
  r_0:=\min\Bigl(\frac{\mu}{\max_i(\kappa_i+2(\ell+1))},\
  \min_i\min_{a\in S_i}\frac{q^*_{i,a}}{\kappa_i+2(\ell+1)}\Bigr).
\]
Suppose that $\norm{q_i-q_i^*}_\infty\le r$ and $\norm{u_i^{\rm p}-u_i^*}_\infty\le r$ for
some $r<r_0$. Then, every $S_i$-coordinate of the unclipped played vector $v_i$
and of the unclipped accumulated vector $v_i+g_i$ is positive, while every
off-support coordinate of both vectors is negative. In particular, the played strategies
and the updated regrets are supported in $S_1\times S_2$, with $q_i'=v_i+g_i$ on $S_i$.
Moreover, we have $\norm{u_i-u_i^*}_\infty\le\ell r$.

Finally, consider the face $\mathcal F:=\{s:\supp q_i\subseteq S_i\}$. On this face near $R$,
the map $F$ coincides with the smooth map $F_S$ obtained by dropping $\pos{\cdot}$ on
the $S_i$-coordinates and setting the off-support coordinates of $p_i$ and $q_i'$ to zero.
\end{theorem}

\looseness=-1\Cref{thm:freeze} is the single-call counterpart to the one-step absorption result that \citet{li2026last} prove for \IREG{}. Because the learner stores the past utility, we need an additional hypothesis on $u^{\rm p}$. We defer the geometry of the fixed-point surface, the pure-equilibrium case, and the first-order identities underlying \cref{thm:companion} to \cref{app:support}. In our proof, we rely on two properties of the update. First, note that the off-support margin $-\mu$ resides in the unclipped $v^t$, and the accumulation \eqref{eq:accum} adds $g^t$ directly to $v^t$. Hence, a small correction cannot reopen a closed coordinate, whereas adding $g^t$ to $p^t$ could. Second, $\gamma$ depends on $m$ only through the active coordinates, so the off-support components of $u^{\rm p}$ never enter. On the face, the state therefore reduces to $(q_1|_{S_1},q_2|_{S_2},m_1,m_2)$ with $m_i:=u_i^{\rm p}|_{S_i}$, which we call the reduced coordinates of the face. The off-support coordinates are never read, so they contribute zero columns to $DF$ and add only zero eigenvalues.

To linearize on the face, we write $\At:=A[S_1,S_2]$, $x^*=q_1^*/c_1$, and $y^*=q_2^*/c_2$. We define $T_x:=\{\delta\in\R^{k_1}:\one^\top\delta=0\}$, $T_y$ analogously, $P_1:=I-\one x^{*\top}$, $\Pi_1:=I-x^*\one^\top=P_1^\top$, and $P_2,\Pi_2$ using $y^*$. Note that indifference on the support yields $\At y^*=v\one$ and $\At^\top x^*=v\one$.

\begin{theorem}[Companion spectrum]\label{thm:companion}
Let $J:=DF_S(s^*)$ denote the Jacobian of $F_S$ at $s^*$ on the face state space $\R^{2(k_1+k_2)}$, and let
\[
  L:=\begin{bmatrix}0&\Pi_1\At/c_1\\-\Pi_2\At^\top/c_2&0\end{bmatrix}:\ T_x\oplus T_y\to T_x\oplus T_y.
\]
Then, the following properties hold.
\begin{enumerate}
\item[(a)] Along the linearized \IRO{} dynamics, let $\zeta^t:=(dx^t,dy^t)$. Then
  $\zeta^{t+1}=(I+2L)\zeta^t-L\zeta^{t-1}$.
\item[(b)] The characteristic polynomial is given by $\operatorname{char}_J(x)=x^2(x-1)^2\prod_{\lambda\in\spec L}\bigl(x^2-(1+2\lambda)x+\lambda\bigr)$, where the product runs over the $k_1+k_2-2$ eigenvalues of $L$ on $T_x\oplus T_y$.
\item[(c)] Moreover, $\spec L=\{\pm i\,\eta\,\sigma_j(\Umat)\}_{j\le k-1}$, where
  $\Umat:=\At-v\one\one^\top$, $\eta:=1/\sqrt{c_1c_2}$, and $\sigma_1\ge\dots\ge\sigma_{k-1}>0$
  denote its singular values. In particular, by \cref{lem:rank}, the operator $L$ has no zero eigenvalue on
  $T_x\oplus T_y$. Hence, the only eigenvalues of $J$ equal to $1$ are the two along $R$.
\end{enumerate}
\end{theorem}

\begin{corollary}[Local rate]\label{cor:rate}
Let $\eps_j:=\eta\sigma_j(\Umat)$, and define
$\rhoiro:=\max_j\max\{\abs{x}:x^2-(1+2i\eps_j)x+i\eps_j=0\}$. Suppose that $\rhoiro<1$.
Then, for every $\rho\in(\rhoiro,1)$, there exists a neighborhood of $s^*$ in the
face from which every orbit converges to a point of $R$ with
$\Gap(z^t)=O(\rho^t)$. Note that this neighborhood depends on $\rho$.
\end{corollary}
The neighborhood statement is proved as \cref{thm:dichotomy}(a), with $C_\rho>0$. We write $\rhoiro(s^*)$ when the dependence on the point is explicit. Since the duality gap is $2\sigma$-Lipschitz in $(x,y)$, the convergence rate of the strategies transfers directly to the gap.

The operator $L$ and its purely imaginary spectrum in part (c) match those found by \citet{li2026last} for \IREG{}. The same value-centring identity yields them here. In the next section, this reduces the threshold analysis, where we find the largest settled effective step size at which the limit remains attracting, to a one-parameter question. Specifically, that parameter is the singular value of the value-centred face matrix $\Umat$ divided by the geometric mean of the saturated $\ell_1$-norms. Next, note that the two unit eigenvalues correspond to the directions along $R$. Since the dynamics admits a two-dimensional manifold of fixed points, convergence amounts to approaching one of its points, where the limit $\eta$ is selected by the trajectory. Finally, for extra-gradient prediction, \citet{li2026last} obtain the one-step recursion $p^{t+1}=(I+L+L^2)p^t$ with eigenvalues $1+\lambda+\lambda^2$. Hence, the two recursions share $L$ and differ only in the companion polynomial.

%% file: sections/06_threshold.tex
\section{Two Thresholds: $1/\sqrt3$ Simultaneous, $2/3$ Alternating}
\label{sec:threshold}

\looseness=-1We first examine the non-unit modes of the linearised dynamics. By \cref{thm:companion}, every such nonzero mode is, up to complex conjugation, a root of $x^2-(1+2i\eps)x+i\eps=0$ for some $\eps=\eta\sigma_j(\Umat)>0$. This is the companion polynomial of the optimistic gradient recursion $p^{t+1}=p^t+2\lambda p^t-\lambda p^{t-1}$ on an imaginary eigenvalue \citep{daskalakis2018limit,pmlr-v108-mokhtari20a}. We now derive its stability region via a short computation.

\begin{proposition}[Scalar recursion]\label{prop:scalar}
Fix $\eps>0$, and let $r(\eps)$ denote the largest modulus of a root of
$x^2-(1+2i\eps)x+i\eps$. Then, the following properties hold:
(i) We have $r(\eps)<1$ if and only if $\eps<1/\sqrt3$.
(ii) In closed form, we have $r(\eps)^2=\tfrac12\bigl(1+\sqrt{1-4\eps^2}\bigr)$ for $\eps\le\tfrac12$, and
$r(\eps)^2=\tfrac14\bigl(1+(2\eps+\sqrt{4\eps^2-1})^2\bigr)$ for $\eps\ge\tfrac12$.
(iii) The function $r$ decreases on $(0,\tfrac12]$ and increases on $[\tfrac12,1/\sqrt3)$.
In particular, its minimum is $r(\tfrac12)=1/\sqrt2$, which is attained at the double root $(1+i)/2$.
(iv) Moreover, we have $-\log r(\eps)=\eps^2/2+O(\eps^4)$.
\end{proposition}

\looseness=-1The limit set of the dynamics is a manifold of fixed points, and the round map is non-invertible. The convergence statement we establish below requires neither a hyperbolic fixed point nor a diffeomorphism (\cref{lem:attract} in \cref{app:extra}).
\begin{theorem}[Stability dichotomy of the fixed surface]\label{thm:dichotomy}
Suppose that \cref{ass:ne} holds. Fix $s^*\in R$ and let $\eps_{\max}:=\eta\,\sigma_{\max}(\Umat)$.
\begin{enumerate}
\item[(a)] Suppose that $\eps_{\max}<1/\sqrt3$. Then, for every $\rho\in(\rhoiro,1)$, there exists a
  neighborhood $V_\rho$ of $s^*$ in the face from which every orbit converges
  to a point in $R$ with $\norm{z^t-(x^*,y^*)}\le C_\rho\rho^t$.
\item[(b)] Suppose that $\eps_{\max}>1/\sqrt3$. Then, there exist a neighborhood $V$ of $s^*$ in the
  face and a $C^1$ submanifold $W\subset V$ of positive codimension with the following property.
  Every orbit that remains in $V$ and converges to $s^*$ starts in $W$. In particular, the set
  of such initial conditions has empty interior and measure zero in the face.
\end{enumerate}
\end{theorem}

Part (b) follows \citet{li2026last}'s centre-stable manifold argument for \IREG{} above threshold one.

\paragraph{Alternating updates.}
We now consider the alternating schedule of \cref{sec:setup}, where the round-boundary state on the face is $s=(q_1,u_1^{\rm p},r_2,m_2^{\rm p})\in\R^{2(k_1+k_2)}$. Here, the four blocks represent the row regret and stored utility, followed by the column player's unclipped played vector and the prediction it used (\cref{alg:altround}). We denote one round by the map $F_{\rm alt}$, which plays the row, lets the column observe and play, and then lets the row observe. Applying \cref{lem:ray} in \cref{app:support} to each player in turn, we see that the surface $R$, read as $\{(c_1x^*,u_1^*,c_2y^*,u_2^*)\}$, consists of fixed points of $F_{\rm alt}$. Let $R_{\rm alt}:=\{(c_1x^*,u_1^*,r_2^*(c_2),u_2^*):c_1,c_2>0\}$ with $r_2^*(c_2):=c_2y^*+u_2^*+v\one$ denote this surface in the round coordinates, so that $r_2^*=c_2y^*$ on $S_2$. \Cref{lem:altfreeze} in \cref{app:extra} adapts \cref{thm:freeze} to the alternating schedule. A single round maps every state near $R_{\rm alt}$ into the face. While the orbit stays in this neighbourhood, $4k$ active coordinates determine it.

\begin{theorem}[Alternating companion spectrum]\label{thm:altcompanion}
Suppose that \cref{ass:ne} holds. Then, at $s^*\in R_{\rm alt}$, the Jacobian $J_{\rm alt}$ of the smooth alternating round map on the face, written in the $4k$ active coordinates of \cref{lem:altfreeze}, has characteristic polynomial
\begin{equation}\label{eq:altchar}
  \operatorname{char}_{J_{\rm alt}}(x)=x^{k+1}(x-1)^2\prod_{j=1}^{k-1}\chi_{\eps_j}(x),\qquad
  \chi_\eps(x):=x^3-(2-4\eps^2)\,x^2+(1-4\eps^2)\,x+\eps^2,
\end{equation}
where $\eps_j:=\eta\sigma_j(\Umat)$. In particular, this Jacobian has $k+1$ zero eigenvalues and the double eigenvalue $1$ along the projection of $R_{\rm alt}$ to the active coordinates. Moreover, its remaining eigenvalues are exactly the roots of one real cubic per nonzero singular value of $\Umat$.
\end{theorem}
Observe that the alternating recursion is the two-step recursion of \cref{thm:companion}(a) for each player, except that the column player reads the row's current play.

\begin{proposition}[Alternating scalar recursion]\label{prop:altscalar}
For $\eps>0$, let $r_{\rm alt}(\eps)$ denote the largest modulus of a root of $\chi_\eps$. Then, the following properties hold.
(i) $\chi_\eps$ has one real root and one conjugate pair for every $\eps>0$.
(ii) We have $r_{\rm alt}(\eps)<1$ if and only if $\eps<2/3$, and the crossing occurs through the real root $x=-1$.
(iii) Write the real root as $-a$ with $a>0$. Then, $\eps^2=a(1+a)^2/(1+2a)^2$, which defines a
strictly increasing bijection $a\mapsto\eps$. Moreover, the conjugate pair has
modulus $(1+a)/(1+2a)$, so that $r_{\rm alt}(\eps)=\max\{a,(1+a)/(1+2a)\}$. The two branches
cross at $a=1/\sqrt2$, that is, at $\eps^*=2^{-3/4}$, where $r_{\rm alt}=1/\sqrt2$. Furthermore, $r_{\rm alt}$
decreases strictly on $(0,\eps^*]$ and increases strictly on $[\eps^*,\infty)$, and its minimum is a kink with finite one-sided slopes.
(iv) Finally, $-\log r_{\rm alt}(\eps)=\eps^2+O(\eps^4)$. In particular, at the same $\eps$, the alternating
contraction per round equals, to leading order, the square of the simultaneous one.
\end{proposition}

\begin{theorem}[Alternating dichotomy]\label{thm:altdichotomy}
Suppose that \cref{ass:ne} holds with $s^*\in R_{\rm alt}$ and $\eps_{\max}=\eta\sigma_{\max}(\Umat)$.
(a) Suppose that $\eps_{\max}<2/3$, and fix any $\rho\in(\rho_{\rm alt},1)$, where $\rho_{\rm alt}:=\max_jr_{\rm alt}(\eps_j)$. Then, there exists a neighborhood $V_\rho$ of $s^*$ in the face from which every orbit of the alternating map converges to a point of $R_{\rm alt}$ at rate $\rho$.
(b) Suppose that $\eps_{\max}>2/3$. Then, there exist a neighborhood $V$ of $s^*$ and a $C^1$ submanifold $W\subset V$ of positive codimension that contains every orbit staying in $V$ and converging to $s^*$. In particular, the set of such initial conditions has empty interior and measure zero in the face.
\end{theorem}
We establish this result by following the proof of \cref{thm:dichotomy}, using \cref{lem:altfreeze} in place of \cref{thm:freeze} and $\chi_\eps$ in place of the quadratic.

\paragraph{From local to global.}
The asymptotic theory of the single-call learner is available for both schedules. Under simultaneous updates from $q^1=0$, \citet[Theorem 2 and Corollary 3]{sullins2026irprmpreviousutilitypredictionconstant} show that the duality gap vanishes and the profile converges to the equilibrium set. Under alternating updates, \citet[Theorem 7.5]{nguyen2026convergencealternatingregretmatching} establish that the profile converges to a single equilibrium on every game. Using \cref{lem:limit} and \cref{prop:init}, both in \cref{app:support}, we obtain convergence to the equilibrium set from the zero start and from two nonzero initial vectors with zero first prediction. Combining these results, we determine the fate of a run up to one case.

\begin{corollary}[Eventual linear rate at a strictly stable landing]\label{cor:global}
Under \cref{ass:ne}, let $(s^t)$ be the orbit of either schedule starting from
$q^1_1=q^1_2=0$ with $T_0<\infty$, or from initial regrets as in
\cref{prop:init}. Let $\eps_{\rm thr}$ be $1/\sqrt3$ for the simultaneous schedule
or $2/3$ for the alternating schedule. Then $s^t\to s^*$, which is the point of $R$
(simultaneous) or of $R_{\rm alt}$ (alternating) with
$x_1^*:=x^*$, $x_2^*:=y^*$, and $c_i=N_{i,\infty}/\norm{x_i^*}_2$. At that limit, if
$\eps_{\max}(s^*):=\eta_\infty\sigma_{\max}(\Umat)<\eps_{\rm thr}$, then
$\Gap(z^t)=O(\rho^t)$ for every $\rho$ above the closed-form rate of the schedule
at $s^*$. If $\eps_{\max}(s^*)>\eps_{\rm thr}$, then from some round on, the orbit lies
on the submanifold $W$ of \cref{thm:dichotomy,thm:altdichotomy}(b). The
case $\eps_{\max}(s^*)=\eps_{\rm thr}$ is not covered.
\end{corollary}

This corollary is conditional on the landing, and that condition is observable along the run. Moreover, in \cref{thm:dichotomy}(b) and \cref{thm:altdichotomy}(b), the exceptional set is a manifold of positive codimension through $s^*$, and the energy identities leave open how the deterministic orbit from $q^1=0$ meets it.

The two steps of the next corollary, global entry followed by local capture, follow the pattern of \citet{li2026last}.

\begin{theorem}[Runtime certificate]\label{thm:certificate}
Assume \cref{ass:ne} with $k\ge2$ and a start as in \cref{cor:global}. For the
simultaneous schedule, define
$\eta_2(t):=\bigl(\norm{x^*}_2\norm{y^*}_2/(N_{1,t}N_{2,t})\bigr)^{1/2}$ for
$t\ge T_0$, and set $\eps_2(t):=\eta_2(t)\sigma_{\max}(\Umat)$. Then $\eps_2$ is
nonincreasing and has limit $\eps_{\max}(s^*)$. Hence $\eps_{\max}(s^*)<1/\sqrt3$ if and only if
$\eps_2(t)<1/\sqrt3$ for some $t$. For any such $t$,
\[
  \rhoiro(s^*)\le B(t):=\max\{r(\eps_{\rm lo}/\kappa),\,r(\eps_2(t))\},\qquad
  \eps_{\rm lo}:=\frac{\sigma_{\max}(\Umat)\sqrt{\norm{x^*}_2\norm{y^*}_2}}
  {\bigl((N_1^{*2}+G^*)(N_2^{*2}+G^*)\bigr)^{1/4}},\quad
  \kappa:=\frac{\sigma_{\max}(\Umat)}{\sigma^+_{\min}(\Umat)},
\]
where $\sigma^+_{\min}$ is the smallest nonzero singular value (since $\Umat$ has rank
$k-1$ by \cref{lem:rank}),
and $\Gap(z^t)=O(\rho^t)$ for every $\rho\in(B(t),1)$. The same conclusion holds for the
alternating schedule if we use $r_{\rm alt}$, $2/3$, \cref{thm:sat-alt} in \cref{app:support}, and the
column norm $N_{2,t+1}$ (the norm from which $y^t$ is played) in place of $N_{2,t}$. This certificate reads the
run through $N_{1,t}N_{2,t}$ and the game through $(x^*,y^*,\Umat)$. It is
one-sided. That is, it certifies a strictly stable landing if and when $\eps_2(t)$
crosses the threshold, whereas a run whose landing is at or above the threshold
is never certified. It is a test on an observed run and not a pre-run guarantee.
The case $k=1$ is covered by \cref{cor:pure} in \cref{app:support}.
\end{theorem}
It remains open to prove that the orbit from $q^1=0$
always lands strictly below the threshold and to resolve the boundary case
$\eps_{\max}(s^*)=\eps_{\rm thr}$. The monotone quantity $\eps_2$ is the same certificate that \citet{li2026last} introduce for \IREG{}, evaluated here against the thresholds of the single-call learner. The rate bound $B(t)$ is specific to \IRO{}.

\paragraph{Certifying the landing.} The runtime certificate on the limiting step size $\eta_\infty$ is loose because it uses the worst-case energy $G^*$, and \cref{app:landing} sharpens it. On rounds absorbed under the hypothesis of \cref{thm:freeze} an exact energy identity holds with $O(\rho^{2t})$ settling (\cref{prop:landing}), an energy certificate bounds simultaneous runs with $\eps_2<1/2$ (\cref{thm:tail,thm:sharp}), and a Lyapunov certificate bounds $\eta_\infty$ below each schedule's threshold when its state-and-game trigger holds near $R$ (\cref{thm:lyap}). Moreover, the remaining drift has the order of the squared distance (\cref{prop:drift}) and \cref{prop:oracle} compares the landing with the best possible one. In \cref{sec:nonunique}, we relax uniqueness and extend it to limit points that satisfy strict complementarity.

\looseness=-1Read together, the theorems give a unified picture of a run converging to a
strictly complementary unique equilibrium. The regret vectors settle at $\ell_1$-masses $(c_1,c_2)$, and the limit lies on $R$ with
$\eta=1/\sqrt{c_1c_2}$. The local rate depends only on the singular values of
$\Umat$, and among orbits staying near a limit above the schedule's threshold,
only a null set reaches it. Along an individual run, a crossing in
\cref{thm:certificate} certifies the linear rate. On the testbed, every
rank-unique mixed game crosses under both schedules, and at a later round where its trigger holds, \cref{thm:lyap} gives a bracket for the landing. Alternation raises the threshold
from $1/\sqrt3$ to $2/3$ and doubles the small-$\eps$ log-rate at a common
$\eps$ below $1/\sqrt3$, while the experiments measure where the landing falls.

If $\Umat=0$, every row of $\At$ is constant and the spectrum of $\Umat$ is $\{0\}$,
and the threshold condition holds vacuously, recovering \cref{cor:pure} for a
strict pure equilibrium. \Cref{thm:dichotomy}(b) names a $C^1$ submanifold $W$ that contains every nearby orbit which converges to $s^*$. Because $R$
consists of fixed points and $W$ is locally invariant, a restart placed on $R$ stays
fixed, so an experiment testing instability must perturb off that manifold
(\cref{sec:experiments}).

%% file: sections/07_experiments.tex
\section{Experiments}
\label{sec:experiments}

\paragraph{Testbed and regret saturation.}
We test the learning algorithms on \nInst{} zero-sum games with sizes $n\in\{3,5,10,25,50\}$. These games use Gaussian, uniform, or sparse entries in both plain and antisymmetric forms. The full experimental protocol appears in \cref{sec:protocol}. For every game, linear programming gives a reference equilibrium. We then apply a rank test to find games with a unique, strictly complementary equilibrium, which we call rank-unique below. This test checks for a strictly complementary reference equilibrium where both players use the same number of strategies and $\operatorname{rank}\Umat=k-1$, which guarantees uniqueness by \cref{lem:rankgen}. The test identifies \nUniqueRank{} rank-unique games, agreeing with a simpler heuristic that checks support size and the margin of the reference solution on all except \nUniqueRankDisagree{} instances. All runs start from zero regrets $q^1=0$. A run converges within the time limit if its final gap falls below $10^{-8}$. This happens on \nConvIRO{} simultaneous \IRO{} runs, \nAltConv{} alternating \IRO{} runs, and \nConvIREG{} \IREG{} runs. Every run that fails to converge in time has size $n=50$.

We test the saturation bound of \cref{thm:sat-iro} and the summed energy inequality of \cref{lem:energy}, on every run. Regret norms never decrease and saturate at finite values on all \nInst{} runs, with \boundViolIRO{} violations of the bound. The ratio of the final norm sum to the bound has median \ratioBoundMedIRO{} and maximum \ratioBoundMaxIRO{}. The worst-case bounds overestimate the final norms by two to five orders of magnitude, and restarting the certificate of \cref{sec:saturation} at the final norms still leaves a median gap of \gstarRestartRelMed{} decades. Exact saturation is governed by the face identities of \cref{thm:tail}. Once the iterate enters the face, the converged \IRO{} runs grow their final norm sum by at most \growthPostMax{}, consistent with \cref{cor:global}.

\paragraph{Spectral rate prediction.}
\begin{figure}[t]
\centering
\begin{minipage}[t]{0.48\linewidth}\centering
\includegraphics[width=\linewidth]{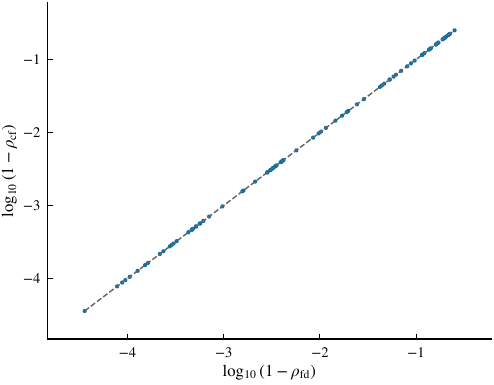}
\caption{Comparison of the closed-form rate $\rhoiro$ and the finite-difference Jacobian's spectral radius for the one-round map at the converged state.}
\label{fig:fdcf}
\end{minipage}\hfill
\begin{minipage}[t]{0.48\linewidth}\centering
\includegraphics[width=\linewidth]{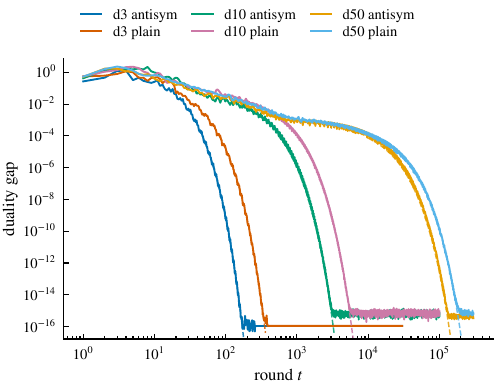}
\caption{Duality gap along \fgcN{} converged simultaneous runs at $n=3,10,50$, with a
dashed guide of slope $\rhoiro$ anchored at the first recorded point below $10^{-4}$.}
\label{fig:gap}
\end{minipage}
\end{figure}
\begin{figure}[t]
\centering
\begin{minipage}[t]{0.48\linewidth}\centering
\includegraphics[width=\linewidth]{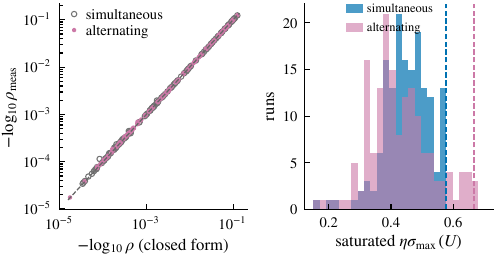}
\caption{Measured $\log\Gap$ slope on $[10^{-11},10^{-3}]$ ($R^2>0.98$) versus $\rhoiro$ (\altFigSimN{}) and $\rho_{\rm alt}$ from \cref{thm:altdichotomy} (\altFigN{}), alongside saturated $\eta\sigma_{\max}(\Umat)$.}
\label{fig:alt}
\end{minipage}\hfill
\begin{minipage}[t]{0.48\linewidth}\centering
\includegraphics[width=\linewidth]{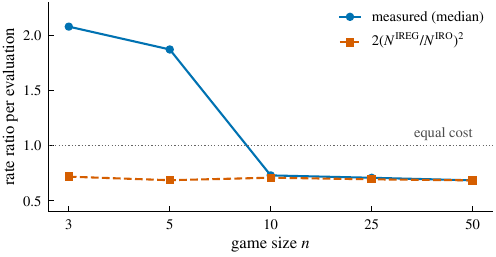}
\caption{Median per-evaluation rate ratio \IRO{}/\IREG{} by game size, with its large-game closed form.}
\label{fig:c5}
\end{minipage}
\end{figure}

\looseness=-1To verify \cref{thm:companion,thm:altcompanion}, we compare the companion matrix against finite-difference Jacobians at the converged states and against empirical decay rates from the gap curves (\cref{fig:gap}). Under simultaneous updates, the closed-form companion spectrum agrees with the numerical Jacobian within $10^{-3}$ on \nCtwoOK{} of \nCtwo{} instances (\cref{fig:fdcf}) and predicts the late-window empirical rate within 5\% on all \fitLateN{} fitted runs. Over the $n=\fdcfN$ instances with a mixed support and a consistent Jacobian, the median absolute $\log_{10}$ deviation is \fdcfMedian{}. Transient faster modes explain early-window deviations, matching the analysis in Appendix~\ref{sec:beyond}. Meanwhile, the extra-gradient polynomial $1+\lambda+\lambda^2$ predicts the rate on \nDiscEG{} of the \nDisc{} runs where the two predictions diverge by more than 5\%, against \nDiscIRO{} for the companion form. Under alternating updates, the characteristic cubic \eqref{eq:altchar} matches the empirical late-window rate within 5\% on all \altFitLateN{} fitted runs (\cref{fig:alt}, left). Optimal assignment matches every nontrivial Jacobian eigenvalue to the cubic roots or the two unit eigenvalues on \altSpecFullOK{} of \altSpecFullN{} states. For \IREG{}, the analogous closed form of \citet{li2026last} is within 5\% on \iregCfWithinFive{} of the \nIREGcf{} runs.

\paragraph{Landing thresholds and stability.}
\begin{figure}[t]
\centering
\begin{minipage}[t]{0.48\linewidth}\centering
\includegraphics[width=\linewidth]{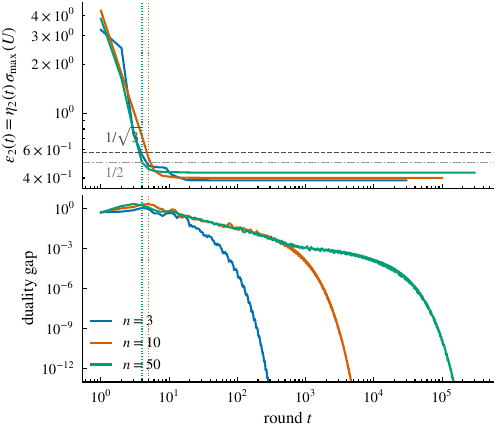}
\caption{Proxy $\eps_2(t)=\eta_2(t)\sigma_{\max}(\Umat)$ and gap across three simultaneous runs ($n=3,10,50$), with the threshold $1/\sqrt3$ dashed and post-crossing $N_1N_2$ growth share \landFigAfter{}.}
\label{fig:landing}
\end{minipage}\hfill
\begin{minipage}[t]{0.48\linewidth}\centering
\includegraphics[width=\linewidth]{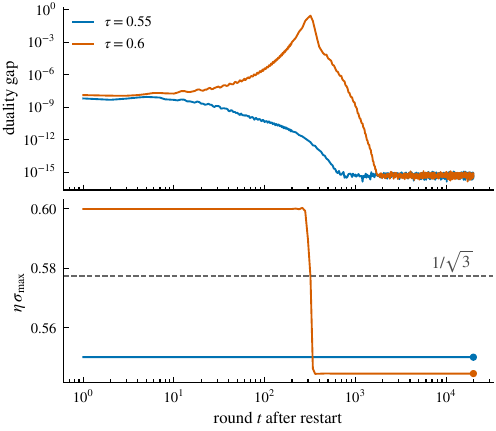}
\caption{Simultaneous restart of one $10\times10$ antisymmetric Gaussian game (saturated
$\eta\sigma_{\max}=\figInstEta$) at $\tau=0.55$ and $\tau=0.60$, showing the gap and
$\eta\sigma_{\max}$ along the run.}
\label{fig:instab}
\end{minipage}
\end{figure}

The inverse regret norm is an effective step size, settling at a limit called the landing. Each schedule has a stability threshold, the largest landing at which the equilibrium remains attracting (\cref{thm:dichotomy,thm:altdichotomy}). Every one of the \nUniqueMixed{} rank-unique mixed games crosses below its schedule's threshold, certifying linear convergence via \cref{thm:certificate} even for the few $n=50$ runs that require ten times the horizon to reach $10^{-8}$. Simultaneous runs always land below $1/\sqrt3$, whereas \nAltAboveSim{} of the \nAltMixed{} alternating runs land between $1/\sqrt3$ and $2/3$, operating at effective step sizes where simultaneous updates are unstable. Along the trajectories (\cref{fig:landing}), the effective step size decreases monotonically (\cref{prop:landing}). It crosses the threshold early, at a median gap of \landGapCrossMed{}. Afterward, it experiences a median \landGrowthAfterMed{} of its norm-product growth, placing the final landing inside the stable region. Moreover, a direct numerical grid confirms the root modulus curve of \cref{prop:scalar}. Games with $\Umat=0$ pose no threshold question. The number of pure equilibria equals the size of the support of the linear-programming reference solution (\nDegenPure{} of \nDegen{} runs). Strict pure equilibria reach the exact fixed point of \cref{cor:pure} in a median \pureTfixMed{} rounds.

\begin{figure}[t]
\centering
\begin{minipage}[t]{0.48\linewidth}\centering
\includegraphics[width=\linewidth]{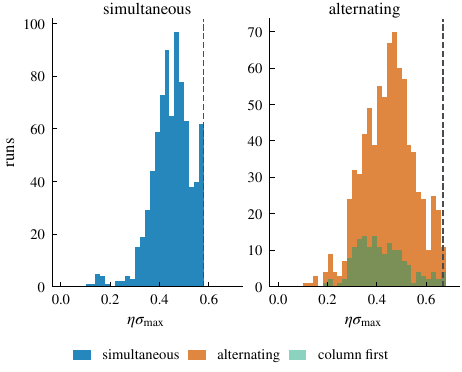}
\caption{Saturated $\eta\sigma_{\max}$ for converged simultaneous ($\rbSimConv{}$ runs), alternating ($\rbAltConv{}$), and column-first ($\rbColConv{}$) random restarts, with dashed lines at $1/\sqrt3$ and $2/3$.}
\label{fig:robust}
\end{minipage}\hfill
\begin{minipage}[t]{0.48\linewidth}\centering
\includegraphics[width=\linewidth]{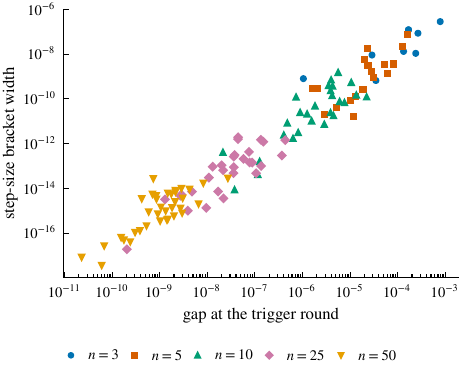}
\caption{Step-size bracket width $w_\eta$ of \cref{thm:sharp} versus trigger-round gap by size across triggered runs ($\sharpTrig$).}
\label{fig:sharp}
\end{minipage}
\end{figure}

To check dependence on initialization, we test \rbStarts{} nonzero random starts per schedule (\cref{prop:init} in \cref{app:support}) plus column-first alternation, in which the column player moves first, across the \nUniqueMixed{} rank-unique mixed games (\cref{fig:robust}). Every simultaneous, alternating, and column-first run crosses its respective threshold ($1/\sqrt3$ or $2/3$). Varying the initial state shifts the landing along the ray $R$ by a median \rbSimSpreadMed{} under simultaneous updates and \rbAltSpreadMed{} under alternation. Every selected point lies strictly inside the open stability region. Furthermore, the companion rate is within 5\% of the fitted one on \rbSimFitOK{} of \rbSimFit{} simultaneous and \rbAltFitOK{} of \rbAltFit{} alternating restarts.

\begin{table}[t]
\caption{Restart probe on converged simultaneous states after rescaling regret vectors by $c=\eta\sigma_{\max}/\tau$, with escaped and re-converged runs marked.}
\label{tab:instab-sim}
\centering\small
\input{tables/tab_instab}
\end{table}
\begin{table}[t]
\caption{Restart probe on alternating states under both schedules, with runs below $2/3$ and median landings.}
\label{tab:instab}
\centering\small
\input{tables/tab_instab_alt}
\end{table}

We test the instability of limits above the threshold by perturbing converged states that have rescaled effective step sizes $\tau$ (\cref{lem:ray} in \cref{app:support}). In \cref{tab:instab-sim}, rescaled simultaneous states escape on all \nAboveEsc{} trials above $1/\sqrt3$ and on none of the controls or sub-threshold trials. Escaped runs exceed $\max(10\,\mathrm{gap}_0,10^{-6})$ and re-converged runs have a final gap below $10^{-8}$. These escaping runs re-enter the stable region, as illustrated in \cref{fig:instab}. Restarting with a different schedule from alternating states shows the gap between the two boundaries in \cref{tab:instab}. For $\tau\in\{0.60,0.65\}$, every simultaneous continuation escapes while every alternating continuation stays stable. Above $2/3$, both schedules escape.

We also test the local absorption certificate of \cref{thm:freeze} by perturbing converged rank-unique states at $n\le10$ on a geometric grid of radii. Across all \absN{} tested games, paths stay on the equilibrium face up to \absClean{} times the theoretical radius. They first break support or positivity conditions at \absFirstViol{} times that radius. Consequently, the analytical radius is conservative by one to two orders of magnitude.

\paragraph{Alternating speedup and tuned baselines.}
\begin{table}[t]
\caption{Median rate ratio $\log\rho_{\rm alt}/\log\rho_{\rm sim}$ over runs with $R^2>0.98$, ratio to the closed form, and faster alternation counts.}
\label{tab:alt}
\centering\small
\input{tables/tab_alt}
\end{table}

Comparing the two schedules on the \nAltBoth{} games where both have fitted rates shows that alternation accelerates convergence per round by a median factor of \altSpeedQpfifty{} (\cref{tab:alt}). That speedup approaches the theoretical small-mode factor of two from \cref{prop:altscalar}(iv) as $n$ grows.

We compare parameter-free simultaneous \IRO{} against optimistic gradient descent-ascent (OGDA). OGDA is a standard gradient method with a step size hand-tuned for each game on a fine grid. We also compare against the theoretical step-size oracle of \cref{prop:oracle}. Self-tuned \IRO{} reaches a median \ogdaFracMed{} of the per-game tuned OGDA log-rate across \nOgdaCmp{} games. It reaches \oracleFracDimthree{} of its own family's oracle rate at $n=3$, rising to \oracleFracDimfifty{} at $n=50$ as the oracle moves toward the stability threshold. Computing the equal-norm shift requires a sort and a Newton refinement step in each round. This extra work makes an \IRO{} round \timeRatio{} times slower in wall-clock time than simplex projection at $n=\timeN{}$, while removing the need for step-size tuning.

\paragraph{Time to tolerance and per-evaluation efficiency.}
\begin{table}[t]
\caption{Median unfiltered time to tolerance by game size, with games that never reach tolerance counted in parentheses.}
\label{tab:ttt}
\centering\small
\input{tables/tab_ttt}
\end{table}

To measure robustness without any filter, we find the round where the iterate permanently enters a target gap tolerance during a run of $3\cdot10^4$ rounds, as shown in \cref{tab:ttt}. We compare our learners against OGDA, predictive RM+ (\PRMp{}), and the smooth step-size learners SPRM$^+$ \citep{farina2023regret} and SOGRM$^+$ \citep{meng2025lastiterate}. At a gap tolerance of $10^{-6}$, extra-gradient \IREG{} succeeds most often, reaching the target in \tttReachIREGsix{} of \nInst{} games. Alternating \IRO{} comes next with \tttReachIROaltsix{} games. Simultaneous \IRO{} succeeds in \tttReachIROsix{} games, matching tuned OGDA and best-fixed-parameter SOGRM$^+$ (\tttReachSOGRMsix{}) without needing a step-size parameter. It also beats \PRMp{} (\tttReachPRMsix{}) by a wide margin.

\IRO{} uses one payoff evaluation per round, while \IREG{} uses two. Accounting for this cost, \cref{fig:c5} shows that the single-call learner is faster per evaluation on small games and slower on large ones. The median per-evaluation rate ratio drops from \cfiveDimthree{} at $n=3$ to \cfiveDimfifty{} at $n=50$. Both characteristic polynomials share the same leading-order quadratic curvature, so the large-game ratio simplifies to $2(N^{\rm IREG}/N^{\rm IRO})^2$. Because \IRO{} saturates at a regret norm roughly \NlevelRatioMed{} times larger than \IREG{}, this value sits close to the $\sqrt3$ ratio of their stability thresholds. Consequently, the asymptotic per-evaluation ratio lands near $2/3$.

Under the relaxed conditions of \cref{sec:nonunique}, \nuNonUnique{} games in our testbed fail the rank test. Still, every run on these games converges under both simultaneous and alternating updates. We check the final state of each run and define the support using a relative threshold of $10^{-9}$. With simultaneous updates, \nuSimStrict{} of the limit points are strictly complementary. In these cases, $d_1+d_2$ reaches up to \nuSimDmax{}, and \nuSimUnequal{} limits have unequal supports. At every such limit, the finite-difference Jacobian has $2+d_1+d_2$ unit eigenvalues and none outside the unit disk. For the \nuSimModes{} limits with $r\ge1$, its largest remaining eigenvalue matches our closed-form formula within \nuSimRhoDev{}. On the \nuSimUzero{} limits where $\Umat=0$, this eigenvalue drops to zero.

\looseness=-1Under alternating updates, we obtain \nuAltStrict{} strictly complementary limits, all with $2+d_1+d_2$ unit eigenvalues, and agreement up to \nuAltRhoDev{} on the \nuAltModes{} limits with $r\ge1$. The measured rate lies within 5\% of the closed form on \nuSimFitOK{} of \nuSimFit{} simultaneous runs and \nuAltFitOK{} of \nuAltFit{} alternating runs with a fit. The remaining \nuSimNotStrict{} simultaneous limits have a zero margin. These occur on the sparse games with $n\le5$, where a best response outside the support ties.

Appendix~\ref{sec:beyond} reports results on the three published matrix games and the poker testbed.

%% file: tables/tab_instab.tex
%% generated by scripts/make_tables.py; do not edit
\begin{tabular}{lrrrrr}\toprule
$\tau$ & $n$ & esc. & reconv. & $\le\!1/\sqrt3$ at $T$ & med.\ $\eta\sigma_{\max}$\\\midrule
control & 172 & 0\% & 100\% & 100\% & 0.449 \\
0.50 & 130 & 0\% & 100\% & 100\% & 0.500 \\
0.55 & 157 & 0\% & 100\% & 100\% & 0.550 \\
0.60 & 172 & 100\% & 76\% & 98\% & 0.550 \\
0.65 & 172 & 100\% & 74\% & 98\% & 0.526 \\
0.75 & 172 & 100\% & 73\% & 98\% & 0.504 \\
0.90 & 172 & 100\% & 73\% & 99\% & 0.489 \\
1.20 & 172 & 100\% & 70\% & 98\% & 0.463 \\
\bottomrule\end{tabular}

%% file: tables/tab_instab_alt.tex
%% generated by scripts/make_tables.py; do not edit
\begin{tabular}{lrrrrr}\toprule
$\tau$ & $n$ & sim.\ esc. & alt.\ esc. & alt.\ $\le\!2/3$ at $T$ & alt.\ med.\ $\eta\sigma_{\max}$\\\midrule
control & 174 & 10\% & 0\% & 100\% & 0.409 \\
0.55 & 153 & 0\% & 0\% & 100\% & 0.550 \\
0.60 & 161 & 100\% & 0\% & 100\% & 0.600 \\
0.63 & 164 & 100\% & 0\% & 100\% & 0.630 \\
0.65 & 168 & 100\% & 0\% & 100\% & 0.650 \\
0.70 & 174 & 100\% & 100\% & 99\% & 0.634 \\
0.75 & 174 & 100\% & 100\% & 99\% & 0.595 \\
0.90 & 174 & 100\% & 100\% & 99\% & 0.532 \\
1.20 & 174 & 100\% & 100\% & 99\% & 0.484 \\
\bottomrule\end{tabular}

%% file: tables/tab_alt.tex
%% generated by scripts/make_tables.py; do not edit
\begin{tabular}{lrrrr}\toprule
$n$ & runs & speedup (meas.) & meas./c.f. & faster\\\midrule
3 & 17 & 1.46 & 0.88 & 13 \\
5 & 22 & 1.28 & 0.76 & 14 \\
10 & 27 & 1.62 & 0.82 & 24 \\
25 & 36 & 1.92 & 0.96 & 36 \\
50 & 30 & 1.70 & 0.85 & 30 \\
\bottomrule\end{tabular}

%% file: tables/tab_ttt.tex
%% generated by scripts/make_tables.py; do not edit
\begin{tabular}{lrrrrrr}\toprule
$n$ & 3 & 5 & 10 & 25 & 50 & not reached\\
\midrule
\multicolumn{7}{l}{\emph{gap $\le 10^{-6}$}}\\
simultaneous \IRO{} & 4 & 73 & 1008 & 5074 (1) & 14441 (21) & 22 \\
alternating \IRO{} & 5 & 80 & 346 & 3631 (1) & 11174 (11) & 12 \\
\IREG{} & 4 & 102 & 380 & 2402 & 8161 (4) & 4 \\
\PRMp{} & 2 (14) & 13 (10) & 664 (11) & 5761 (16) & 12309 (22) & 73 \\
OGDA & 5 (1) & 52 (2) & 660 (2) & 4040 (2) & 14273 (17) & 24 \\
SPRM$^+$ & 34 & 762 (1) & 2033 (2) & 5588 (11) & 11912 (26) & 40 \\
SOGRM$^+$ & 6 (1) & 82 & 994 & 5019 (3) & 12130 (17) & 21 \\
\midrule
\multicolumn{7}{l}{\emph{gap $\le 10^{-10}$}}\\
simultaneous \IRO{} & 4 & 120 & 1948 (1) & 6131 (12) & 23006 (39) & 52 \\
alternating \IRO{} & 5 & 134 & 643 & 5877 (6) & 17772 (30) & 36 \\
\IREG{} & 4 & 170 & 687 & 4216 (3) & 13994 (22) & 25 \\
\PRMp{} & 2 (14) & 13 (10) & 1228 (11) & 9417 (20) & 19214 (38) & 93 \\
OGDA & 5 (1) & 84 (2) & 1236 (2) & 6693 (7) & 19128 (35) & 47 \\
SPRM$^+$ & 34 (1) & 1338 (1) & 3834 (2) & 9388 (19) & 20617 (38) & 61 \\
SOGRM$^+$ & 6 (1) & 136 & 1777 (1) & 8463 (10) & 17817 (36) & 48 \\
\bottomrule\end{tabular}

%% file: sections/08_conclusion.tex
\section{Conclusion}
\label{sec:conclusion}

Predicting with the last payoff, single-call scale-invariant regret matching$^+$ plays proportionally to clipped regrets and preserves regret length without step sizes. That regret norm saturates at a finite limit on every game with an unconditional certificate, making its inverse an effective step size. Near a unique strictly complementary equilibrium, where unused actions do strictly worse, play forms a smooth map on the frozen active face. Simultaneous play linearises it to a companion matrix with purely imaginary generator $\pm i\eta\sigma_j(\Umat)$, and alternation to one real cubic per singular value. Contraction holds below $\eta\sigma_{\max}=1/\sqrt3$ for simultaneous moves and $2/3$ for alternating turns, fastest at $1/\sqrt2$ in both cases, with higher limits unstable. Once that step size settles strictly below threshold, concurrent asymptotic theory makes this rate global, with a one-sided certificate below each threshold. Local results also cover strictly complementary limits under multiple equilibria. On \nInst{} random games, measured rates match the closed form and every rank-unique mixed game is certified, while restarts between thresholds separate schedules and alternating rounds run faster.

%% file: sections/16_imported.tex
\clearpage
\section{Imported Lemmas}
\label{app:imported}

The shift correction is $\delta^t:=\ip{m^t}{z^t}-\gamma^t$, with $\delta^t:=0$ when $q^t=0$.

\looseness=-1Here, we state four facts about the common update that we use repeatedly in our analysis. Fact (F1) pairs the energy identity of \citet{li2026last} with its lower bound, sharpening the intermediate inequality of \citet{zhang2025scale}. Fact (F2) combines the telescoping argument of \citet{li2026last}, which holds for any prediction, with the zero-sum identity in self-play, as in \citet{sullins2026irprmpreviousutilitypredictionconstant}. Facts (F3) and (F4) are elementary.
\begin{lemma}[Facts about the common update]\label{lem:facts}
The following facts hold, where (F1) concerns a player with $q^t\neq0$.
\begin{enumerate}
\item[(F1)] We have $\delta^t\norm{p^t}_1=\tfrac12\norm{p^t-q^t}_2^2$. Moreover,
  $\delta^t\ge \frac{N_t}{2n}\norm{z^t-\tilde z^t}_2^2$.
\item[(F2)] Suppose that $q^1_1=q^1_2=0$, under simultaneous self-play with arbitrary predictions or under the alternating schedule with previous-utility prediction. Then, for every $T$, we have
  $\sum_{t\le T}(\delta_1^t+\delta_2^t)\le N_{1,T+1}+N_{2,T+1}$.
\item[(F3)] Let $z\in\Delta_n$. Then, $\norm{e-\ip{e}{z}\one}_2\le K\norm{e}_2$.
\item[(F4)] Let $a,b\ge0$ with $a,b\neq0$. Then, $\norm{\N(a)-\N(b)}_2\le K\norm{a-b}_2/\norm{a}_1$.
\end{enumerate}
\end{lemma}
We now turn to (F2), where self-play enters the analysis. Combining \eqref{eq:shift} and \eqref{eq:accum} yields $q^{t+1}=\pos{q^t+\hat r^t+\delta^t\one}$, where we set $\hat r^t:=u^t-\ip{u^t}{z^t}\one$. Fix any comparator $p\in\Delta$. Note that $\ip{q^{t+1}}{p}\ge\ip{q^t}{p}+\ip{\hat r^t}{p}+\delta^t$, which implies that the external regret satisfies $\mathrm{Reg}(T):=\max_p\sum_{t\le T}\ip{u^t}{p-z^t}\le
N_{T+1}-\sum_{t\le T}\delta^t$ for any utility sequence. Under simultaneous self-play, the sum of the two regrets equals $T\,\Gap(\bar x_T,\bar y_T)\ge0$, the gap of the averaged profile, which establishes (F2). Under the alternating schedule, \cref{lem:altreg} shows that this sum is larger, so (F2) holds in that setting as well.

\begin{lemma}[Energy identity]\label{lem:energy}
Suppose that $q^t\neq0$. Then, $N_t\le N_{t+1}$ and $N_{t+1}^2\le N_t^2+\norm{g^t}_2^2$. Moreover,
$\norm{q^{t+1}-p^t}_2^2\le N_{t+1}^2-N_t^2$.
\end{lemma}
\looseness=-1The first two claims give the norm monotonicity of \citet{li2026last}, and we use the third in the analysis.

\begin{lemma}[Prediction error is a payoff action]\label{lem:gbound}
For \IRO{} with $t\ge2$, we have $\norm{g_1^t}\le K_1\sigma\norm{y^t-y^{t-1}}$ and $\norm{g_2^t}\le K_2\sigma\norm{x^t-x^{t-1}}$.
Under alternation, $g_2^t$ denotes the correction of the column accumulation made in round $t$, which books the standing play $y^{t-1}$ against $x^t$.
\end{lemma}

\begin{lemma}[Rank \citep{li2026last}]\label{lem:rank}
Suppose that \cref{ass:ne} holds. Let $\At:=A[S_1,S_2]$ and $\Umat:=\At-v\one\one^\top$.
Then, $\ker\Umat=\operatorname{span}(y^*|_{S_2})$ and $\ker\Umat^\top=\operatorname{span}(x^*|_{S_1})$.
In particular, $k_1=k_2=:k$ and $\operatorname{rank}\Umat=k-1$.
\end{lemma}

%% file: sections/07b_beyond.tex
\section{Beyond the Random Testbed}
\label{sec:beyond}

\suppressfloats[t]

\begin{figure}[t]
\centering
\begin{minipage}[t]{0.48\linewidth}\centering
\includegraphics[width=0.9\linewidth]{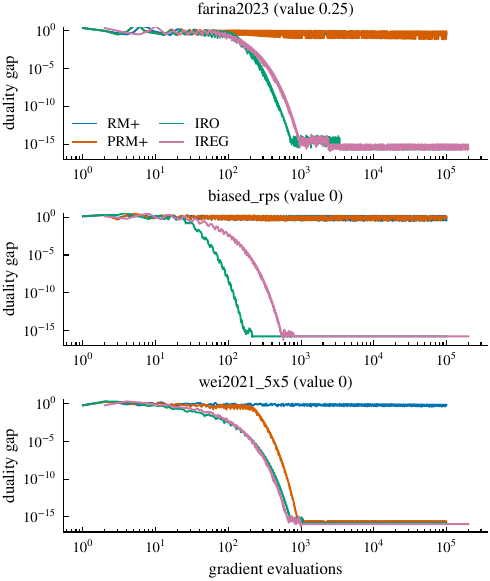}
\caption{Duality gap against payoff evaluations on the three games of
\cref{tab:structured}, all four learners from the uniform profile.}
\label{fig:structured}
\end{minipage}\hfill
\begin{minipage}[t]{0.48\linewidth}\centering
\includegraphics[width=0.9\linewidth]{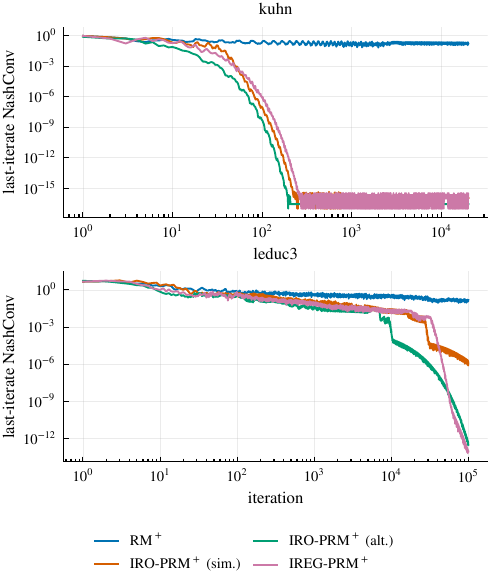}
\caption{Last-iterate NashConv along the iterations on Kuhn (top,
$2\cdot10^4$ iterations) and three-rank Leduc poker (bottom, $10^5$).}
\label{fig:efg}
\end{minipage}
\end{figure}
\paragraph{Games on which predictive regret matching fails.}
\begin{table}[t]
\caption{Median duality gap over the last decile of $10^5$ rounds and first round below $10^{-8}$ on three published games.}
\label{tab:structured}
\centering\small
\input{tables/tab_structured}
\end{table}
We test all four learners on three matrix games where the last iterate of clipped regret matching \RMp{} or its predictive version \PRMp{} is known to stall. These are the $3\times3$ game of \citet{farina2023regret}, a biased rock-paper-scissors, and the $5\times5$ game of \citet{wei2021linear}. Only the $5\times5$ game has multiple equilibria, and linear-programming reference equilibria for all three have best-response deviations below \sgBestDev{}. While \RMp{} stalls on all three games and \PRMp{} stalls on the first two (\cref{tab:structured,fig:structured}), both equal-norm shifted learners reach numerical precision everywhere. Single-call \IRO{} pushes the duality gap below $10^{-8}$ within a few hundred rounds.

At the landing, the measured long-run contraction rate of \IRO{} matches the companion-matrix prediction within a few percent on all three games. The ratio of measured to predicted rates is \sgBrpsIRORateRatio{} on biased rock-paper-scissors and \sgWeiIRORateRatio{} on the $5\times5$ game. Over the final two orders of magnitude in the Farina game (\cref{tab:farina-window}), this ratio reaches \sgFarinaWinLate{}. The $5\times5$ game violates \cref{ass:ne} and provides an example of \cref{thm:nonunique}. Its limit point satisfies strict complementarity, with strategy supports of sizes $\weiKone\times\weiKtwo$ and $\operatorname{rank}\Umat=\weiRank$. These values give $d_1=\weiDone$ and $d_2=\weiDtwo$. A finite-difference approximation of the Jacobian at this limit has $2+d_1+d_2=\weiUnit$ eigenvalues equal to one, matching the theoretical prediction.

Fitting a single rate over the broad window $[10^{-3},10^{-11}]$ on the Farina game yields a distorted ratio of \sgFarinaIRORateRatio{} because that interval averages over a two-mode transient. The effective step size settles before the window opens, so the companion spectrum is already fixed, with a second mode of modulus \sgFarinaRhoSecond{} alongside the leading mode of modulus \sgFarinaIRORhoCf{}. In the companion eigenvector basis, the initial regret deviation carries \sgFarinaModeRatioStart{} times more amplitude in this faster second mode than in the leading one. Across four two-decade windows, the mode ratio decays by the predicted factor $(\sgFarinaRhoSecond/\sgFarinaIRORhoCf)^{\Delta t}$, dropping to \sgFarinaModeRatioFourth{} by a gap of $10^{-11}$, after which the slower leading mode dominates. Extrapolating the linear part of \eqref{eq:perturbed} from round \frTstart{} with only these two conjugate pairs reproduces the measured gap trajectory to a median error of \frReconMed{} per round, and repeating the windowed fits on this two-mode reconstruction matches the empirical ratios within \frFitDevTwo{}.
\begin{table}[t]
\caption{Log-rate fits for simultaneous \IRO{} on the Farina $3\times3$ game across five gap windows against $\rho_{\mathrm{cf}}=\sgFarinaIRORhoCf{}$. Ratio columns give $\log\rho_{\mathrm{meas}}/\log\rho_{\mathrm{cf}}$ for the measured fit, the two-mode reconstruction, and the leading mode.}
\label{tab:farina-window}
\centering\small
\input{tables/tab_farina_window}
\end{table}

\paragraph{Extensive-form games.}
\begin{table}[t]
\caption{Last-iterate NashConv on Kuhn and three-rank Leduc poker up to $T=\efgT{}$, at $10^5$ iterations, and over the $2\cdot10^4$ last decile.}
\label{tab:efg}
\centering\small
\setlength{\tabcolsep}{3.5pt}
\input{tables/tab_efg}
\end{table}
An extensive-form panel shows the learner's behaviour outside the matrix setting. On Kuhn poker \citep{kuhn1950simplified} and three-rank Leduc poker \citep{southey2005bayes} (\efgLeducNinf{} information sets per player), we embed \IRO{} and extra-gradient \IREG{} into counterfactual regret minimisation (CFR), which runs an independent learner at each information set (a player's decision point given their observed history). Each local learner receives reach-weighted counterfactual action values (the expected payoffs at that decision point assuming the player plays to reach it) and solves the equal-norm shift over its available actions. We verify the game trees independently against the $64\times64$ normal form of Kuhn (minimax value $-1/18$) and the published Leduc value \efgLeducPublished{}, where standard CFR$^+$ reaches an averaged NashConv (the extensive-form duality gap) of \efgLeducCfrPlus{}.

On Kuhn poker, both shifted learning algorithms converge down to numerical round-off error (\cref{tab:efg}). Specifically, \IRO{} drives NashConv below $10^{-8}$ in \efgKuhnIROTconv{} iterations, whereas \RMp{} stops improving at \efgKuhnRMNcT{}. On Leduc poker, the final iterates oscillate during the first $2\cdot10^4$ iterations (\cref{fig:efg}). Because of these oscillations, simultaneous \IRO{} and \IREG{} trade the lead at early stopping times, while \RMp{} stays near \efgLeducRMNcT{}. We also test an alternating schedule where player one updates before player two in each round. Alternating updates leave the oscillatory phase sooner than simultaneous updates. It reaches numerical round-off on Kuhn poker in \efgKuhnAltTconv{} iterations. On Leduc poker at horizon $T$, its median NashConv over the final ten percent of iterations is \efgLeducAltMed{}, beating the \efgLeducIROMed{} achieved by simultaneous \IRO{}.

Running Leduc poker out to \efgLongT{} iterations ends the oscillations for all the shifted learning algorithms. Past that point, their final iterates shrink at a geometric rate of roughly one factor of ten every $10^4$ iterations. Alternating \IRO{} stays below $10^{-8}$ after iteration \efgLongAltStay{} and reaches \efgLongAltNc{}. Meanwhile, \IREG{} reaches \efgLongIregNc{}, and simultaneous \IRO{} reaches \efgLongIroNc{} while still dropping. By contrast, \RMp{} stalls at \efgLongRmNc{}. Extending our analysis of the equilibrium face to the sequence-form representation of extensive-form games \citep{von1996efficient} remains an open problem.

%% file: tables/tab_structured.tex
%% generated by scripts/make_tables.py; do not edit
\begin{tabular}{lrrrrrrrr}\toprule
game & \multicolumn{2}{c}{\RMp{}} & \multicolumn{2}{c}{\PRMp{}} & \multicolumn{2}{c}{\IRO{}} & \multicolumn{2}{c}{\IREG{}}\\
& gap & $t_{10^{-8}}$ & gap & $t_{10^{-8}}$ & gap & $t_{10^{-8}}$ & gap & $t_{10^{-8}}$\\\midrule
\citet{farina2023regret} $3\times3$ & $4.4\times10^{-1}$ & -- & $3.4\times10^{-1}$ & -- & $1.9\times10^{-16}$ & 393 & $5.0\times10^{-16}$ & 260 \\
biased RPS & $9.0\times10^{-1}$ & -- & $8.8\times10^{-1}$ & -- & $1.7\times10^{-16}$ & 99 & $1.7\times10^{-16}$ & 141 \\
\citet{wei2021linear} $5\times5$ & $6.3\times10^{-1}$ & -- & $2.2\times10^{-16}$ & 551 & $1.7\times10^{-16}$ & 354 & $1.1\times10^{-16}$ & 168 \\
\bottomrule\end{tabular}

%% file: tables/tab_farina_window.tex
%% generated by scripts/make_tables.py; do not edit
\begin{tabular}{lrrrrrrrr}\toprule
window & $t$ & $\eta_2\sigma_{\max}$ & local $\rho$ & $n_{\mathrm{pts}}$ & $R^2$ & $\log\rho_{\mathrm{meas}}/\log\rho_{\mathrm{cf}}$ & two-mode & leading only\\\midrule
$[10^{-3},10^{-5}]$ & 182--292 & 0.556/0.556 & 0.9514 & 75 & 0.846 & 1.126 & 1.126 & 0.987 \\
$[10^{-5},10^{-7}]$ & 269--367 & 0.556/0.556 & 0.9497 & 80 & 0.877 & 1.166 & 1.166 & 0.977 \\
$[10^{-7},10^{-9}]$ & 350--448 & 0.556/0.556 & 0.9499 & 77 & 0.907 & 1.161 & 1.161 & 0.974 \\
$[10^{-9},10^{-11}]$ & 437--550 & 0.556/0.556 & 0.9567 & 71 & 0.934 & 0.998 & 0.992 & 0.978 \\
$[10^{-11},10^{-13}]$ & 541--652 & 0.556/0.556 & 0.9574 & 64 & 0.982 & 0.982 & 0.978 & 0.993 \\
\bottomrule\end{tabular}

%% file: tables/tab_efg.tex
%% generated by scripts/make_tables.py; do not edit
\begin{tabular}{llrrrrrr}\toprule
game & learner & $t=2\cdot10^3$ & $5\cdot10^3$ & $10^4$ & $2\cdot10^4$ & $10^5$ & last decile\\\midrule
Kuhn & \RMp{} & $1.1\times10^{-1}$ & $2.1\times10^{-1}$ & $1.4\times10^{-1}$ & $1.6\times10^{-1}$ & $1.9\times10^{-1}$ & $1.7\times10^{-1}$ \\
 & simultaneous \IRO{} & $<10^{-15}$ & $<10^{-15}$ & $<10^{-15}$ & $<10^{-15}$ & $<10^{-15}$ & $<10^{-15}$ \\
 & alternating \IRO{} & $<10^{-15}$ & $<10^{-15}$ & $<10^{-15}$ & $<10^{-15}$ & $<10^{-15}$ & $<10^{-15}$ \\
 & \IREG{} & $<10^{-15}$ & $<10^{-15}$ & $<10^{-15}$ & $<10^{-15}$ & $<10^{-15}$ & $<10^{-15}$ \\
\midrule
Leduc (3 ranks) & \RMp{} & $3.3\times10^{-1}$ & $4.1\times10^{-1}$ & $2.9\times10^{-1}$ & $1.6\times10^{-1}$ & $1.5\times10^{-1}$ & $2.5\times10^{-1}$ \\
 & simultaneous \IRO{} & $7.9\times10^{-2}$ & $3.5\times10^{-2}$ & $1.9\times10^{-2}$ & $5.1\times10^{-3}$ & $1.3\times10^{-6}$ & $1.2\times10^{-2}$ \\
 & alternating \IRO{} & $2.8\times10^{-2}$ & $1.9\times10^{-2}$ & $6.5\times10^{-4}$ & $7.6\times10^{-6}$ & $3.2\times10^{-13}$ & $2.2\times10^{-5}$ \\
 & \IREG{} & $3.4\times10^{-2}$ & $4.0\times10^{-2}$ & $1.5\times10^{-2}$ & $1.2\times10^{-2}$ & $9.0\times10^{-14}$ & $1.8\times10^{-2}$ \\
\bottomrule\end{tabular}

%% file: sections/15_extra.tex
\section{Additional Statements}
\label{app:extra}

This appendix collects technical statements that are used in the main text without being stated there. Their proofs appear in \cref{app:proofs}.

\subsection{Activation and the rounds before $T_0$}
\label{app:activation}

\looseness=-1By \cref{lem:energy}, the norms are nondecreasing. Before $T_0$, one player is inactive and plays the uniform strategy, so the other observes a constant utility. In the following lemma, we characterize the behavior of the active player under a constant utility. We use this to bound $T_0$, while our certificates rely on the norm remaining unchanged.

\begin{lemma}[Constant utility]\label{lem:activation}
Suppose that a player runs \eqref{eq:shift} through \eqref{eq:accum} with $m^{t}=u^{t-1}$ on the same utility vector $u$ at every round, starting from $q^1=0$, and that $u$ is not constant across its actions. Let $t^*$ be the first round with $q^{t}\neq0$, and set $N^*:=N_{t^*}$. Then, $t^*=2$ and $q^{2}=\pos{u-\bar u\one}$, where $\bar u$ denotes the uniform average of $u$. Furthermore, for every $t\ge t^*$, we have $N_t=N^*$. At every such round, $\supp q^t$ is an upper level set $\{a:u_a\ge\theta_t\}$ of $u$, which is nonincreasing in $t$ and contained in $\{a:u_a>\bar u\}$. Moreover, among these rounds, consider any $t>t^*-1+N^*/\Delta_u$, where $\Delta_u:=\max_au_a-\max\{u_b:u_b<\max_au_a\}$, taking $N^*/\Delta_u:=0$ when $\supp q^{t^*}\subseteq\argmax u$ already. At every such round, we have $\supp q^t\subseteq\argmax u$, the state is fixed, and the played strategy is a best response to $u$. Finally, a player that never activates observes, at every round, a utility that is constant across its actions. Hence, each of its actions is a best response to the strategy played by its opponent at that round.
\end{lemma}

This lemma tracks the state prior to $T_0$. In particular, the active player's norm remains at $N^*$ while its strategy moves to a best response in finitely many rounds, which also bounds $T_0$. Under simultaneous updates, both players play the uniform strategy in round one. Consequently, the first player to activate does so at round two, with $N_i^*=\norm{\pos{u_i^1-\bar u_i^1\one}}$, $u_1^1=A\one/n_2$, and $u_2^1=-A^\top\one/n_1$. If both of these first accumulations are nonzero, then $T_0=2$ and both constants are closed-form in $A$. Note that this is the generic case for $n_1,n_2\ge2$, since the uniform profile leaves neither player indifferent.

Next, suppose that only player $i$ activates at round two. Then, as long as the other player stays inactive, the lemma fixes its strategy at a best response $\hat z_i$ to the uniform strategy starting at round $\tau_i:=\lfloor N_i^*/\Delta_{u_i}\rfloor+2\le2+N_i^*/\Delta_{u_i}$. If the other player activates before round $\tau_i$, then $T_0\le\tau_i$ already. Otherwise, at every subsequent round, the other player observes the utility of $\hat z_i$. That player either activates at some round $t_{-i}^*\le\tau_i+1$, giving $T_0\le\tau_i+1$ with $N_{-i}^*$ read off the run, or never activates. In the latter case, $(\hat z_i,\text{uniform})$ is an equilibrium with a constant orbit from round $\tau_i$ onward. If neither player activates, then the uniform profile is an equilibrium and the orbit is constant. Hence, we have either $T_0\le3+\max_iN_i^*/\Delta_{u_i}$, where here and below the maximum runs over the players whose first accumulation is nonzero, or $T_0=\infty$ where the orbit remains constant from some round onward with at most one active player.

We now consider row-first alternating updates, under which the column player accumulates before playing. Its round-one accumulation is $q_2^2=\pos{u_2^1-\bar u_2^1\one}$ with $u_2^1=-A^\top\one/n_1$. If this vector is nonzero, then $t_2^*=1$ and $N_2^*$ is its norm. Next, the row player observes $u_1^1=Ay^1$ at the column player's first play $y^1$ (shifted from $q_2^2$, or uniform when $q_2^2=0$). If $q_1^2=\pos{Ay^1-\ip{Ay^1}{\one/n_1}\one}$ is nonzero, then $t_1^*=2$ and $N_1^*$ is its norm. Note that these norms of each player's first accumulation are closed-form in $A$. Any player whose first accumulation vanishes activates later, if at all, with $N_i^*$ read off the run.

For example, for $A=\bigl(\begin{smallmatrix}2&-1\\-2&1\end{smallmatrix}\bigr)$, the column player's round-one accumulation is zero. The row player then activates at round two with $N_1^*=1/2$, and the column player activates within round two with $N_2^*=3/2$, giving $T_0=2$. Since the column player acts within the round, the same argument gives either $T_0\le2+\max_iN_i^*/\Delta_{u_i}$ or $T_0=\infty$. The certificates of \cref{sec:saturation} are explicit in $A$ and the activation norms for $T_0<\infty$. When $T_0=\infty$, at most one player activates and its norm stays constant by the lemma. Consequently, we obtain $M\le\max(N_1^*,N_2^*)$ with $G=0$, where $N_i^*:=0$ for a player that never activates, which bounds the norms on every $A$.

We next observe that prior to $T_0$, an active player's norm stays fixed under either schedule. Indeed, the inactive player plays the uniform strategy prior to $T_0$, so every accumulation of the active player observes a constant utility. In particular, under alternation, the row player's accumulations in rounds $t<t_2^*$ observe $Ay^t$ with $y^t$ uniform. Likewise, the column player's accumulations in rounds $t<t_1^*$ observe $-A^\top x^t$ with $x^t$ uniform. By the lemma, we have $N_{1,T_0}=N_1^*$ and $N_{2,T_0}=N_2^*$ for simultaneous updates. Under alternation, we similarly obtain $N_{1,T_0}=N_1^*$ as well as $N_{2,t+1}=N_2^*$ at every column accumulation from a nonzero vector in a round $t<T_0$.

In round $T_0$, the newly active player moves. Note that under alternation with $t_2^*=T_0$, the row player's accumulation in round $T_0$ already observes the column player's first active play $y^{T_0}$. In the example above, this gives $N_{1,3}^2=17/4$ compared to $N_1^{*2}=1/4$. Hence, the sums of \cref{sec:saturation} start at round $T_0$ for the row player. For the column player, they start at the first accumulation from a nonzero vector at or after $T_0$. From round $T_0$ onward, both plays are active, which we assume in what follows.

\subsection{Two lemmas for alternating saturation}
\label{app:altlemmas}

The alternating saturation theorem (\cref{thm:sat-alt}) relies on the following two lemmas.

\begin{lemma}[Utility improvement]\label{lem:improve}
Suppose that a player runs \eqref{eq:shift} through \eqref{eq:accum} with $m^{t+1}=u^t$ and $K=1+\sqrt n$. Then, for every $t\ge1$,
\[
  \ip{u^t}{z^{t+1}-z^t}\ \ge\ \frac{N_{t+1}}{4K^2}\norm{z^{t+1}-z^t}_2^2+\delta^{t+1}-\delta^t .
\]
\end{lemma}

This is Lemma 5.8 of \citet{nguyen2026convergencealternatingregretmatching} in the present notation.

\begin{lemma}[Alternating regret sum]\label{lem:altreg}
Suppose that row-first alternation is run from $q^1=0$, with $\mathrm{Reg}_1(T)$ taken over the row rounds $t=1,\dots,T$ and $\mathrm{Reg}_2(T)$ taken over the column rounds $k=0,\dots,T-1$.
Then,
\[
  \mathrm{Reg}_1(T)+\mathrm{Reg}_2(T)
  =T\,\Gap(\bar x_T,\bar y_T)+\sum_{k=0}^{T-1}\ip{u_2^k}{y^{k+1}-y^k}
  \ \ge\ \delta_2^T+\sum_{k=0}^{T-1}\frac{N_{2,k+1}}{4K_2^2}\norm{y^{k+1}-y^k}^2\ \ge\ \delta_2^T\ \ge0,
\]
In particular, $\sum_{t\le T}\delta_1^t+\sum_{k\le T}\delta_2^k\le N_{1,T+1}+N_{2,T+1}$, where $N_{2,T+1}=\norm{q_2^{T+1}}$ is the norm from which $y^T$ is played.
Note that this includes the column player's correction at its play in round $T$.
This gives the aligned form that (F2) takes under alternation, which is the form used in \cref{lem:disp}.
\end{lemma}

\subsection{Technical lemmas for the face and the thresholds}
\label{app:facelemmas}

The next three lemmas are used to prove \cref{thm:freeze,thm:dichotomy,thm:altcompanion}.

\begin{lemma}[Lipschitz shift]\label{lem:lip}
Let $\gamma(w,c)$ solve $\norm{\pos{w-\gamma\one}}_2=c>0$. Then,
$\abs{\gamma(w',c')-\gamma(w,c)}\le2\norm{w'-w}_2+\abs{c'-c}$. In particular, for
$q,q'\ne0$, we have $\abs{\gamma(q',m')-\gamma(q,m)}\le3\norm{q'-q}_2+2\norm{m'-m}_2$.
\end{lemma}

\begin{lemma}[Attracting manifold of fixed points]\label{lem:attract}
Let $F$ be a $C^1$ map on an open subset of $\R^N$. Let $M$ be an embedded $C^1$
submanifold of dimension $d$ consisting of fixed points of $F$, and fix $s^*\in M$.
Suppose that the eigenvalue $1$ of $DF(s^*)$ has algebraic multiplicity $d$, and that every
other eigenvalue has modulus at most $\rho_0<1$. Then, for every
$\rho\in(\rho_0,1)$, there exist a neighborhood $V$ of $s^*$ and a constant $C$
with the following property. Every orbit starting in $V$ converges to a point $s^\infty$ in
$M$, and moreover, $\norm{F^t(s)-s^\infty}\le C\rho^t$.
\end{lemma}

\begin{lemma}[Absorption under alternation]\label{lem:altfreeze}
Suppose that \cref{ass:ne} holds, and let
$R_{\rm alt}:=\{(c_1x^*,u_1^*,r_2^*(c_2),u_2^*):c_1,c_2>0\}$ with
$r_2^*(c_2):=c_2y^*+u_2^*+v\one$. Then, $r_2^*=c_2y^*$ on $S_2$, and
$r_2^*=u_2^*+v\one\le-\mu\one$ off $S_2$. Furthermore, every point of $R_{\rm alt}$ is fixed
by the alternating round. Moreover, for every $s^*\in R_{\rm alt}$, there exists
$r_{\rm alt}>0$ with the following properties. Consider any state within $r_{\rm alt}$ of $s^*$ in sup
norm. From this state, the three clipping operations of the round (row play, column
accumulation, and column play) and the two accumulations are supported in
$S_1\times S_2$, with every active coordinate positive. In particular, one round maps every state within $r_{\rm alt}$
of $s^*$ to a state of the face
$\mathcal F_{\rm alt}:=\{\supp q_1\subseteq S_1,\ \supp\pos{r_2}\subseteq S_2\}$,
with the off-support coordinates of $r_2'$ negative. In that round, the off-support
coordinates of the incoming state are read only through the
norm target $\norm{q_1}_2$ of the row player's shift. Finally, on $\mathcal F_{\rm alt}$
near $s^*$, the map $F_{\rm alt}$ coincides with the smooth map obtained as follows. We drop
$\pos{\cdot}$ on the support coordinates and set the off-support coordinates
of $q_1'$ and $q_2'$ to zero. We also carry the off-support coordinates of
$r_2'$ (which stay negative) as the functions $u_{2,b}-\gamma_2'$ of the
active coordinates. On this region, the off-support coordinates of $r_2$
are never read, and the orbit is determined by the $4k$ active coordinates
$(q_1|_{S_1},u_1^{\rm p}|_{S_1},r_2|_{S_2},m_2^{\rm p}|_{S_2})$.
\end{lemma}

We now examine the remaining coordinates on the full state, which fall into three groups. First, the off-support regrets vanish on the face and enter off it only through the norm target $\norm{q_1}_2$, whose partial derivatives in them vanish at zero. Second, the off-support predictions enter neither $\gamma$ nor the active corrections. Third, the off-support coordinates of $r_2$ are functions of the active coordinates that the round never reads. Consequently, each group contributes a zero column to the Jacobian of the full round map at $s^*$, giving a zero eigenvalue. Hence, the characteristic polynomial is \eqref{eq:altchar} times a power of $x$.

Note that the released code carries the column regret $q_2$ as a further unread block. In \cref{sec:experiments}, we take the finite-difference Jacobians on that $2n_1+3n_2$-dimensional state and match them against \eqref{eq:altchar} together with the predicted number of zero eigenvalues.

\subsection{Supporting statements for saturation and the face}
\label{app:support}

\begin{lemma}[Displacement control]\label{lem:disp}
Fix any $T\ge T_0$. Then, we have
$\sum_{t=T_0}^{T}\norm{z_i^t-\tilde z_i^t}^2\le a_i\,(N_{1,T+1}+N_{2,T+1})
\le a_i\,(N_1^*+N_2^*+\sqrt{G_1}+\sqrt{G_2})$.
The second inequality also holds with each $G_i$ replaced by its partial sum through round $T$.
\end{lemma}

\begin{theorem}[Alternating \IRO{} saturates, same certificate]\label{thm:sat-alt}
Under row-first alternation, fix any $A$ with $T_0<\infty$. Then, $G\le G^*(A)$ and $N_{i,\infty}^2\le N_i^{*2}+G^*$, where $G^*$ is as in \cref{thm:sat-iro}.
\end{theorem}

Alternation preserves or improves the self-play regret identity (\cref{lem:altreg} in \cref{app:extra}). Moreover, the surplus is bounded below by a weighted sum of squared movements, which is the mechanism behind the alternating regret-bounded-by-variation-in-utilities (RVU) property of \citet{nguyen2026convergencealternatingregretmatching}.

\paragraph{Nonzero initial regrets.}
Note that only the bookkeeping of the first active round depends on the zero initialization. Because the restarts in \cref{sec:experiments} initialize both players from nonzero vectors, we require statements for such starts.

\begin{proposition}[Nonzero initial regrets]\label{prop:init}
Suppose that both players start from nonzero $q_i^1\in\R^{n_i}_+$ with first prediction zero. Under alternation, the column player's standing strategy is $y^0=\N(q_2^1)$, taken as played from $q_2^1$ with prediction zero. Then, (F2) of \cref{lem:facts} holds. Moreover, \cref{thm:sat-iro,thm:sat-alt} hold with $T_0=1$ and $N_i^*=\norm{q_i^1}_2$, where we set $G^*:=0$ when $A=0$ (in which case $\Phi$ vanishes).
\end{proposition}

\begin{lemma}[Limit points are equilibria]\label{lem:limit}
Suppose that \IRO{} self-play is run under either schedule, starting from $q^1=0$ with $T_0<\infty$ or from initial regrets as in \cref{prop:init}. Then, $\norm{z_i^{t+1}-z_i^t}\to0$ for both players, and every limit point of the played profile $(x^t,y^t)$ is a Nash equilibrium. Moreover, under \cref{ass:ne}, the profile converges to $(x^*,y^*)$.
\end{lemma}

Convergence to the equilibrium set from zero initialization was shown by \citet{sullins2026irprmpreviousutilitypredictionconstant}, and under alternation by \citet{nguyen2026convergencealternatingregretmatching}. In \Cref{lem:limit}, we present the form we use below for either start.

\begin{lemma}[$R$ is a surface of fixed points]\label{lem:ray}
Let $s^*\in R$. Then, $F(s^*)=s^*$ holds with $\gamma_1=v$, $\gamma_2=-v$,
$p_i=q_i$, $z=(x^*,y^*)$, and $g=0$.
\end{lemma}

The equilibrium is reached on a two-parameter family of fixed points indexed by the two saturated regret norms. The rescaling $q_i\mapsto c\,q_i$ moves along this family while preserving the strategy. In contrast, the effective step size $\eta=1/\sqrt{c_1c_2}$ changes along $R$.

\begin{corollary}[Strict pure equilibria are reached exactly]\label{cor:pure}
Suppose that $k_1=k_2=1$, $x^*=e_a$, $y^*=e_b$, and round $t$ satisfies the hypothesis of
\cref{thm:freeze}. Then $s^{t+1}=(N_{1,t}e_a,N_{2,t}e_b,u_1^*,u_2^*)\in R$.
Consequently, from round $t+1$ on, the profile is $(e_a,e_b)$, the gap is zero, and the regrets
are constant.

Now consider alternating updates starting from a state within the radius of
\cref{lem:altfreeze}, stated in \cref{app:extra}, of a point of $R_{\rm alt}$, defined in \cref{sec:threshold}. In that round, the plays are
$(e_a,e_b)$, and the state after the round is the point of $R_{\rm alt}$ with
$c_1=\norm{q_1}_2$ and $c_2=\norm{\pos{r_2}}_2$. These values are the norms from which the two plays were made, read on the entering state. From then on, the profile is $(e_a,e_b)$,
the gap is zero, and the regrets are constant.
\end{corollary}
With \cref{cor:global} below, we conclude that every run covered by that corollary on a game with a unique strict pure equilibrium reaches it in finitely many rounds.

\begin{lemma}[Derivative of the shift]\label{lem:gamma}
On the face, the shift $\gamma$ is $C^\infty$ near $(q_1^*,u_1^*|_{S_1})$. Moreover, $d\gamma=\ip{x^*}{dm}$ holds at $(q_1^*,u_1^*|_{S_1})$.
\end{lemma}

\begin{lemma}[The prediction cancels in the accumulation]\label{lem:cancel}
At $s^*$, let $dp:=d(q+m-\gamma\one)$. Then,
\begin{gather*}
  dp_1=dq_1+P_1dm_1,\qquad dx=\Pi_1dp_1/c_1,\qquad dq_1'=dq_1+P_1du_1,\\
  du_1=\At\,dy,\qquad du_2=-\At^\top dx .
\end{gather*}
\end{lemma}

The prediction enters the played vector ($dp_1$) and leaves accumulated regret ($dq_1'$) untouched. This captures the first-order content of ``$\pos{v+g}$ with $g=(u-m)-\ip{u-m}{z}\one$''. Specifically, $+m$ in $v$ and $-m$ in $g$ cancel once we account for the shift derivative $\ip{x^*}{dm}$. Moreover, only this step differs between \IRO{} and \IREG{}, as $dm^{t+1}=du^{t}$ for \IRO{} while $dm^t=F'(d\tilde z^t)$ for \IREG{}.

\section{Certificates for the Landing}
\label{app:landing}

This appendix sharpens the runtime certificate of \cref{thm:certificate}. It explains how the dynamics select the landing point, bounds the remaining norm growth using data from a single round, and compares the resulting step size with the optimal one.

Throughout, a play of player $i$ is \emph{absorbed} when its unclipped vector $v_i$ is positive on $S_i$ and nonpositive off $S_i$. An accumulation is absorbed when $v_i+g_i$ is nonnegative on $S_i$ and nonpositive off $S_i$ and the play that it books is absorbed, so that $v_i=p_i$ on $S_i$ there. A round is absorbed when all its plays and accumulations are. By \cref{thm:freeze}, every round that starts within its radius is absorbed.

\Cref{cor:global} routes every run to $R$ and reads the rate at the reached point.
We now present two further facts describing how this point is selected. For the
simultaneous schedule and $t\ge T_0$, we define
$\eta_2(t):=\bigl(\norm{x^*}_2\norm{y^*}_2/(N_{1,t}N_{2,t})\bigr)^{1/2}$, which
equals $\eta=1/\sqrt{c_1c_2}$ at every point of $R$.

\begin{proposition}[One-way crossing and second-order freezing]\label{prop:landing}
Suppose that \cref{ass:ne} holds, and consider the orbit starting from $q^1=0$ with $T_0<\infty$, or from initial regrets as in \cref{prop:init}. Then, the following statements hold.
(i) The quantity $\eta_2(t)$ is nonincreasing and satisfies $\eta_2(t)\to\eta_\infty$. In particular, $\eta_2(t)\sigma_{\max}(\Umat)$ crosses $1/\sqrt3$ at most once and only downward.
(ii) Moreover, on every absorbed round, the energy identity is exact on the face and yields $N_{i,t+1}^2-N_{i,t}^2=\norm{g_i^t|_{S_i}}_2^2$, which for $t\ge2$ is at most $K_i^2\sigma^2\norm{z_{-i}^t-z_{-i}^{t-1}}_2^2$.
(iii) Suppose that every round from $t_0$ onward is absorbed and that $\norm{z_i^t-z_i^*}_2\le C\rho^t$ holds for both players and all $t\ge t_0-1$ with some $C<\infty$ and $0<\rho<1$, as in the proof of \cref{cor:global}. Then, for $t>t_0$,
\[
  0\le N_{i,\infty}^2-N_{i,t}^2\le 4K_i^2\sigma^2\sum_{s\ge t-1}\norm{z_{-i}^s-z_{-i}^*}_2^2
  \le \frac{4K_i^2\sigma^2C^2}{1-\rho^2}\,\rho^{2(t-1)},
  \qquad 0\le\frac{\eta_2(t)}{\eta_\infty}-1=O(\rho^{2t}).
\]
\end{proposition}
Note that the effective step size moves only while the strategies move, and by the squared increment. Once the orbit enters the linear regime, remaining growth is $O(\rho^{2t})$. Hence, in the linear regime the landing is fixed up to $O(\rho^{2t})$. \Cref{sec:experiments} measures the share of norm growth occurring between threshold crossing and absorption.

\paragraph{A sharp certificate below one half.}
The bound $\eps_{\rm lo}$ of \cref{thm:certificate} stems from the worst-case energy $G^*$ and sits far from the landing that it brackets. On the face, both the energy identity and the correction budget are exact. Together, they yield a bound on the remaining energy proportional to the current misalignment of the regrets with the equilibrium. Absorbed rounds are those defined at the start of this appendix. We write $x_1^*:=x^*$, $x_2^*:=y^*$, and $\Psi_t:=\ip{q_1^t}{x^*}+\ip{q_2^t}{y^*}$.

\begin{lemma}[Face identities]\label{lem:faceid}
Suppose that \cref{ass:ne} holds. Then, on an absorbed round $t$ of the simultaneous schedule, we have
(i) $N_{i,t+1}^2=N_{i,t}^2+\norm{g_i^t|_{S_i}}_2^2$,
(ii) $\Psi_{t+1}=\Psi_t+\delta_1^t+\delta_2^t$, and
(iii) $\delta_i^t\norm{p_i^t}_1=\tfrac12\norm{p_i^t-q_i^t}_2^2$.
On either schedule, (i) holds for each player at each of its absorbed accumulations, with the norm before and after that accumulation, and (iii) holds at every play from a nonzero vector. Moreover, on an absorbed round of the alternating schedule, $\Psi_{t+1}\ge\Psi_t+\delta_1^t+\delta_2^t$. Here, $\Psi_t$ is evaluated at the start of round $t$, and $q_2^t$ denotes the vector from which the column player played $y^{t-1}$.
\end{lemma}

\begin{theorem}[Tail energy on the face]\label{thm:tail}
Suppose that \cref{ass:ne} holds with $k\ge2$ and that the rounds $t_0-1,\dots,T$ of either schedule are absorbed. Let $\theta\ge\sqrt k\,\norm{z_i^t-z_i^*}_2$ hold for both players and all $t_0-1\le t\le T$. Define
\[
  \bar c_i:=\frac{N_{i,t_0}}{\norm{x_i^*}_2},\qquad
  \eps:=\frac{\sigma_{\max}(\Umat)}{\sqrt{\bar c_1\bar c_2}},\qquad
  m:=\sum_i\bigl(\norm{x_i^*}_2N_{i,t_0}-\ip{q_i^{t_0}}{x_i^*}\bigr),\qquad
  \bar e:=\sum_i\frac{\norm{g_i^{t_0-1}|_{S_i}}_2^2}{\bar c_i},
\]
$\Theta:=(1+\theta)^6$, and $H_T:=\sum_i\sum_{t=t_0}^{T}\norm{g_i^t|_{S_i}}_2^2/\bar c_i$. Suppose further that $4\Theta\eps^2<1$. Then,
\[
  H_T\ \le\ \bar H:=\frac{\Theta\eps^2\,(4m+2\bar e)}{1-4\Theta\eps^2},\qquad
  \sum_{t=t_0}^{T}(\delta_1^t+\delta_2^t)\le m+\tfrac12H_T,\qquad
  N_{i,T+1}^2\le N_{i,t_0}^2\Bigl(1+\frac{\bar H}{\bar c_i\norm{x_i^*}_2^2}\Bigr).
\]
\end{theorem}
Under the simultaneous schedule, $\eps=\eps_2(t_0)$. Under alternation, $\bar c_2$
is read from $N_{2,t_0}$, the norm before the column player's accumulation in
round $t_0$, which the budget includes. Consequently, $\eps$ is at least the
proxy $\eps_2(t_0)$ of \cref{thm:certificate}, which uses $N_{2,t_0+1}$. The
misalignment $m$ is nonnegative by the Cauchy-Schwarz inequality and vanishes
exactly when each $q_i^{t_0}$ is parallel to $x_i^*$. Both $m$ and $\bar e$ are
of second order, $m$ in the distance to $R$ and $\bar e$ in the preceding play increments whenever round $t_0-1$ predicts with a previous utility, so the bound $\bar H$ on the remaining energy is proportional to $m+\bar e/2$, with constant $4\eps^2/(1-4\eps^2)$ as
$\theta\to0$ at fixed $\eps<1/2$.

The hypotheses of \cref{thm:tail} concern future rounds. They can be discharged
at a single round because the resulting bound controls the assumed closeness.
\begin{theorem}[Sharp certificate]\label{thm:sharp}
Under \cref{ass:ne} with $k\ge2$ and the simultaneous schedule, suppose round $t_0-1$ is absorbed. Using the notation of \cref{thm:tail}, define $K:=1+\sqrt k$, $\bar c_{\min}:=\min_i\bar c_i$, and $A_{\max}:=\max_{a,b}\abs{A_{ab}}$. Let $\theta\in(0,1)$ satisfy
\begin{align*}
  &\sqrt k\,\norm{z_i^{t_0-1}-z_i^*}_2\le\theta\quad(i=1,2), &&\text{(C0)}\\
  &4\Theta'\eps^2<1,\qquad \Theta':=(1+\theta')^6,\quad
   \theta':=\max\Bigl\{\theta,\ \alpha+(1+\alpha)\,\frac{K(\sigma_{\max}(\Umat)+2\sqrt k\,\sigma)}{\bar c_{\min}}\,\theta\Bigr\},
   \quad\alpha:=\frac{\theta}{K}, &&\text{(C1)}\\
  &\sqrt k\,K^2\sqrt{2\bar m/\bar c_{\min}}\le\theta,\qquad
   \bar m:=m+\tfrac12\bar H,\quad
   \bar H:=\frac{\Theta'\eps^2(4m+2\bar e)}{1-4\Theta'\eps^2}, &&\text{(C2)}\\
  &\max\Bigl\{\frac{\tilde c\,\alpha}{\sqrt k},\ A_{\max}\theta\Bigr\}
   <r_0\Bigl(\frac{\bar c_{\min}}{1+\alpha}\Bigr),\qquad
   \tilde c:=\max_i\frac{\bar c_i\sqrt{1+\tau_i}}{1-\alpha},\quad
   \tau_i:=\frac{\bar H}{\bar c_i\norm{x_i^*}_2^2}, &&\text{(C3)}
\end{align*}
where $r_0(c):=\min\{\mu,\ c\min_{i,a\in S_i}x^*_{i,a}\}/(\max_i\kappa_i+2(\ell(c)+1))$, with $\ell(c):=A_{\max}\max_i2k\kappa_i/c$, is a lower bound on the radius of \cref{thm:freeze} at $c_{\min}=c$. Then every round $t\ge t_0$ is absorbed, $H_T\le\bar H$ holds for every $T$, and
\[
  N_{i,t_0}\le N_{i,\infty}\le N_{i,t_0}\sqrt{1+\tau_i},\qquad
  \frac{\eta_2(t_0)}{\bigl((1+\tau_1)(1+\tau_2)\bigr)^{1/4}}\le\eta_\infty\le\eta_2(t_0).
\]
Suppose also that the run starts as in \cref{cor:global}. Then the run converges linearly. Letting $\iota$ denote the interval of $\eta$ above, we also have $\rhoiro(s^*)\le\max_{\eta\in\partial\iota}\max\{r(\eta\sigma^+_{\min}(\Umat)),r(\eta\sigma_{\max}(\Umat))\}$.
\end{theorem}
The proof in \cref{sec:sharpproof} proceeds by induction on the round. The tail
bound through round $T$ places the state at round $T+1$ within the radius of
\cref{thm:freeze} and bounds its misalignment, extending the closeness
hypothesis of \cref{thm:tail} by one round. The certificate reads the run via
$(q^{t_0},z^{t_0-1},g^{t_0-1})$ and the game via $(x^*,y^*,\Umat,\mu)$, $\sigma$, $A_{\max}$ and the dimensions. It
requires $\eps_2(t_0)<1/2$, the level where the triangle inequality in the proof
 ceases to close, and therefore covers landings below the single-mode optimum.
\Cref{thm:lyap} covers the landings between $1/2$ and $1/\sqrt3$.

On the testbed, \cref{thm:tail} holds at all \sharpLemRounds{} simultaneous and
\sharpLemRoundsAlt{} alternating rounds whose future is absorbed with
$4\Theta\eps^2<1$, achieving $H_T/\bar H\le$ \sharpLemWorst{} when $\theta$ is
the realised closeness. Among the \nUniqueMixed{} rank-unique mixed games,
\sharpBelowHalf{} simultaneous runs end with $\eps_2<1/2$. Conditions (C0) to
(C3) fire on \sharpTrig{} of them (\cref{tab:sharp}), while the remaining
\sharpNoTrig{} at $n=50$ run out of rounds before coming close enough. Let
$w_\eta:=1-((1+\tau_1)(1+\tau_2))^{-1/4}$ denote the bracket's relative width.
The trigger fires at a median gap of \sharpTrigGapMed{}, giving a relative width
$w_\eta$ of \sharpShrinkMed{} at the median and \sharpShrinkMax{} at worst,
tightening the $10^{\sharpLoDecades}$ median factor separating $\eps_{\rm lo}$
from the landing in \cref{thm:certificate}. Every triggered run stays absorbed
and within $\theta$ after $t_0$, with realised energy below $\bar H$
(\cref{fig:sharp}). Narrow widths indicate diagnostic values at double precision.

\paragraph{A certificate up to the threshold.}
The energy route stops at $1/2$ because it bounds the two halves of a
displacement separately. Keeping the recursion whole removes the loss. In
suitable coordinates, the dynamics on the face is an exact companion recursion
with a second-order remainder. A Lyapunov norm of the companion matrix then
certifies any landing strictly inside the stability region, under either
schedule and at any state where its trigger holds.

For player $i$, let $\hat x_i:=x_i^*/\norm{x_i^*}_2$,
$\hat P_i:=I-\hat x_i\hat x_i^\top$, $d_i:=\hat P_i\one$ and
$\chi_i:=\norm{d_i}_2=(k-\norm{x_i^*}_2^{-2})^{1/2}$, which vanishes exactly when
$x_i^*$ is uniform. Set $U_1:=\Umat$, $U_2:=-\Umat^\top$, $v_1:=v$ and $v_2:=-v$.
For a strategy $z$ supported in $S_i$, write $\phi:=\hat P_iz$, with all vectors
restricted to the supports.

\begin{lemma}[Exact face coordinates]\label{lem:facecoord}
Suppose that \cref{ass:ne} holds. Then, the following statements hold.
(i) A strategy supported in $S_i$ is given by
$z=x_i^*(1-\ip{d_i}{\phi})+\phi$. Moreover, against an opponent strategy supported in
$S_{-i}$, the utility is $u_i|_{S_i}=v_i\one+U_i\phi_{-i}$.
(ii) Suppose that the plays of rounds $t-1$ and $t$ are supported in $S_1\times S_2$.
Suppose further that player $i$'s play and accumulation of round $t$ and its play of round $t+1$ are
absorbed, so that $v_i^t=p_i^t$ on $S_i$. Then, setting $c^t:=\norm{p_i^t}_1$, $\phi^t:=\phi_i^t$, and
\[
  \Delta:=U_i(2\phi_{-i}^t-\phi_{-i}^{t-1}),\qquad w:=U_i(\phi_{-i}^t-\phi_{-i}^{t-1}),
\]
we have $g_i^t|_{S_i}=w-\ip{w}{\phi^t}\one$ and
\begin{equation}\label{eq:facestep}
  c^{t+1}=h-k\varsigma,\qquad
  c^{t+1}\phi^{t+1}=c^t\phi^t+\Delta-\varsigma\,d_i,\qquad
  \varsigma=\frac{h-\sqrt{h^2-k\,\Xi}}{k},
\end{equation}
where $h:=c^t+\ip{d_i}{\Delta}$ and
$\Xi:=2c^t\ip{\phi^t}{\Delta}+\norm{\Delta}_2^2-\norm{g_i^t|_{S_i}}_2^2$. Moreover, under
alternation, the row player's step remains the same. The column player's step
from its play $y^t$ to $y^{t+1}$ is given by \eqref{eq:facestep} with
$\Delta=U_2(2\phi_1^{t+1}-\phi_1^t)$ and $w=U_2(\phi_1^{t+1}-\phi_1^t)$, provided that $x^t$ and $x^{t+1}$ are supported in $S_1$ and that the column player's accumulation and play producing $y^{t+1}$ are absorbed.
\end{lemma}

Fix a round $t_0$, and let $\bar N_i$ be the norm of the vector from which
$z_i^{t_0}$ is played. Define $\bar c_i:=\bar N_i/\norm{x_i^*}_2$,
$\bar\eta:=(\bar c_1\bar c_2)^{-1/2}$, $\eps_j:=\bar\eta\,\sigma_j(\Umat)$ for
$j\le k-1$, $\eps:=\eps_1$, and $\psi_i^t:=\sqrt{\bar c_i}\,\phi_i^t$. Dividing
\eqref{eq:facestep} by $\bar c_i$ gives
\begin{equation}\label{eq:perturbed}
  \psi^{t+1}=\psi^t+J(2\psi^t-\psi^{t-1})+\tilde e^t,\qquad
  J:=\bar\eta\begin{bmatrix}0&\Umat\\-\Umat^\top&0\end{bmatrix}
  \ \text{on}\ \mathcal E:=(x^*)^\perp\oplus(y^*)^\perp,
\end{equation}
with remainder $\tilde e_i^t:=\sqrt{\bar c_i}\,e_i^t$ and
$e_i^t:=\phi_i^{t+1}-\phi_i^t-\Delta/\bar c_i$. Because $J$ is skew-symmetric
with eigenvalues $\pm i\eps_j$ (\cref{lem:rank}), the companion matrix
$\bar C:=\bigl[\begin{smallmatrix}I+2J&-J\\I&0\end{smallmatrix}\bigr]$ of
$W^t:=(\psi^t,\psi^{t-1})\in\mathcal E^2$ is an orthogonal sum of $k-1$ real
$4\times4$ blocks, one per singular value, with spectral radius
$\bar\rho=\max_jr(\eps_j)$.

Under alternation, the column player reads the row player's new play via
$\psi_2^{t+1}=\psi_2^t-\bar\eta\,\Umat^\top(2\psi_1^{t+1}-\psi_1^t)+\tilde e_2^{t}$.
The row player reads $(\psi_2^t,\psi_2^{t-1})$ and nothing reads $\psi_1^{t-1}$,
making the state $W^t:=(\psi_1^t,\psi_2^t,\psi_2^{t-1})$. The $3\times3$ blocks
of its companion matrix $\bar C_{\rm alt}$ have characteristic polynomial
$\chi_{\eps_j}$, so $\bar\rho=\max_jr_{\rm alt}(\eps_j)$. In either case, for
$\rho\in(\bar\rho,1)$ let
\[
  \mathcal P:=\sum_{n\ge0}\rho^{-2n}(\bar C^\top)^n\bar C^n,\qquad
  \norm{W}_{\mathcal P}^2:=W^\top\mathcal PW,
\]
so that $\norm{W}_2\le\norm{W}_{\mathcal P}$ and
$\norm{\bar CW}_{\mathcal P}\le\rho\norm{W}_{\mathcal P}$. Let $M$ be the
$\mathcal P$-norm of the map injecting a one-round remainder into the state,
namely $\tilde e\mapsto(\tilde e,0)$ for simultaneous updates and
$(\tilde e_1,\tilde e_2)\mapsto(\tilde e_1,\tilde e_2-2\bar\eta\,\Umat^\top\tilde e_1,0)$
for alternating ones. Both $\mathcal P$ and $M$ are computed block by block from
a $4\times4$ or $3\times3$ Stein equation.

The remainder is controlled by four scalar functions. For $a,D,D_w,\tau\ge0$
and player $i$, with $\kappa_i:=\chi_i+\norm{x_i^*}_2^{-1}$, let
\begin{gather*}
  \varrho_i:=\frac{\tau+\kappa_ia}{1-\kappa_ia},\qquad
  \varsigma_i:=\frac{2(1+\varrho_i)aD+D^2+D_w^2(1+\sqrt k\,a)^2}{1-\varrho_i-\chi_iD},\qquad
  \varrho_i^+:=\varrho_i+\chi_iD+k\varsigma_i,\\
  e_i:=\frac{(\chi_iD+k\varsigma_i)\,a+\varrho_i^+D+\chi_i\varsigma_i}{1-\varrho_i^+},
\end{gather*}
defined when the three denominators are positive. Let $\mu_i$ be the
strict-complementarity margin of player $i$'s off-support actions,
$\alpha_1:=\max_{b\notin S_1}\norm{A[b,S_2]}_2$,
$\alpha_2:=\max_{b\notin S_2}\norm{A[S_1,b]}_2$, and
$x_{i,\min}:=\min_{a\in S_i}x^*_{i,a}$.

\begin{theorem}[Certificate up to the threshold]\label{thm:lyap}
Suppose that \cref{ass:ne} holds with $k\ge2$ under the simultaneous schedule, that rounds $t_0-1$ and $t_0$ are absorbed, and that $\bar\rho<1$, that is, $\eps<1/\sqrt3$. Fix $\bar\rho<\rho<\rho_1<1$ and set $\omega:=\norm{W^{t_0}}_{\mathcal P}>0$. Note that if $\omega=0$, then the plays are at equilibrium and the state after the round is a point of $R$. Let $a_0:=\omega/\sqrt{\min_i\bar c_i}$, $D:=\sqrt5\,\eps a_0$, $D_w:=\sqrt2\,\eps a_0$, and
\[
  \bar\tau_i:=\Bigl(1+\frac{D_w^2(1+\sqrt k\,a_0)^2}{\norm{x_i^*}_2^2\,(1-\rho_1^2)}\Bigr)^{1/2}-1 .
\]
We evaluate $\varrho_i,\varsigma_i,e_i$ at $(a_0,D,D_w,\bar\tau_i)$. Suppose that their denominators are positive and that
\begin{align*}
  &\rho+\frac M\omega\bigl(\bar c_1e_1^2+\bar c_2e_2^2\bigr)^{1/2}\le\rho_1, &&\text{(L1)}\\
  &\pi_i:=x_{i,\min}(1-\chi_ia_0)-a_0>0\quad\text{and}\quad(1-\varrho_i)\,\pi_i\ge D_w(1+a_0), &&\text{(L2)}\\
  &\mu_i>\alpha_i\sqrt k\,\norm{x_{-i}^*}_2\,a_0+\bar c_i\,(\varsigma_i+2D_wa_0) &&\text{(L3)}
\end{align*}
for $i=1,2$. Note that condition (L3) is void for any player without off-support actions. Then, for every $t\ge t_0$, round $t$ is absorbed,
\[
  \norm{W^t}_{\mathcal P}\le\rho_1^{\,t-t_0}\omega,\qquad
  \norm{z_i^t-x_i^*}_2\le\sqrt k\,\norm{x_i^*}_2\,\rho_1^{\,t-t_0}a_0,\qquad
  \bar N_i\le N_{i,t}\le\bar N_i(1+\bar\tau_i),
\]
so that the run converges linearly to a point of $R$ regardless of its history prior to $t_0$. Moreover, $\bar\eta\,\bigl((1+\bar\tau_1)(1+\bar\tau_2)\bigr)^{-1/2}\le\eta_\infty\le\bar\eta$.

The same conclusion holds under alternation when $\eps<2/3$ with the following adjustments. First, we use $\bar C_{\rm alt}$. Second, we read the column player's norm on its own clock by using $N_{2,t+1}$ (the norm from which $y^t$ is played) in place of $N_{2,t}$. In this case, the limit is a point of $R_{\rm alt}$. Third, we evaluate the column player's quantities $\bar\tau_2,\varrho_2,\varsigma_2,e_2$ at $D=3\eps a^+$ and $D_w=2\eps a^+$, and we use the same $D_w$ in its conditions (L2) and (L3). Finally, we substitute $a^+$ for $a_0$ in the first term of its condition (L3), where $a^+:=a_0(1+\sqrt5\,\eps)+\max\{1,(\bar c_1/\bar c_2)^{1/2}\}\,e_1$.
\end{theorem}
The proof appears in \cref{sec:lyapproof}. Relation \eqref{eq:perturbed} and the
bound $\norm{e_i^t}_2\le e_i$ give the contraction (L1), while (L2) and (L3)
keep the next round on the face. Along any run converging to a point $s^*$ of
$R$ strictly inside the stability region, the trigger fires from some round
onward. For fixed $\rho,\rho_1$ between the rate at $s^*$ and one, the reference
scales $\bar c_i$ converge, $\mathcal P$ and $M$ stay bounded, and $e_i$ and
$\bar\tau_i$ are $O(a_0^2)$, so (L1) to (L3) hold with slack once $\omega$ is
small. As the landing approaches the threshold, the constants degenerate through
$M$ and the margin $1-\bar\rho$. Because the conclusion bypasses global
convergence, it applies to any state where the trigger holds.

The same coordinates determine the tightness of this bracket. Along the linear
recursion, the energy of player $i$ is a quadratic form of the companion state.
With $\Gamma_i(W):=\norm{[J(\psi-\psi^-)]_i}_2^2$ for $W=(\psi,\psi^-)$, let
$\mathcal G_i:=\sum_{n\ge0}(\bar C^\top)^n\Gamma_i\bar C^n$.
\begin{proposition}[The remaining drift, to leading order]\label{prop:drift}
Suppose that the hypotheses of \cref{thm:lyap} hold for the simultaneous schedule. Then,
\[
  N_{i,\infty}^2-\bar N_i^2=\bar c_i\,\ip{W^{t_0}}{\mathcal G_iW^{t_0}}+O(\omega^3),
\]
with a constant that depends on the game, $\bar c_1,\bar c_2$, and $(\rho,\rho_1)$. Moreover, each $\mathcal G_i$ is positive definite on $\mathcal E^2$. Under alternation, the same conclusion holds with the energy forms of the alternating recursion on its state $(\psi_1^{t_0},\psi_2^{t_0},\psi_2^{t_0-1})$.
\end{proposition}
Let $\lambda_i>0$ denote the smallest eigenvalue of $\mathcal G_i$. Each
player's drift is at least $\bar c_i\lambda_i\omega^2/\norm{\mathcal P}$ up to
$O(\omega^3)$, while the bracket of \cref{thm:lyap} satisfies
$\bar\tau_i=O(\omega^2)$. Each player's remaining norm growth therefore has exact
order $\omega^2$, matching the bracket, and the quadratic form predicts it up to
a relative error proportional to the distance to $R$.

On the testbed, evaluating this certificate on the \nUniqueMixed{} rank-unique
mixed games at each new half-decade of the duality gap triggers \lyTrigSim{}
simultaneous and \lyTrigAlt{} alternating runs. The remaining \lyNoTrigSim{} and
\lyNoTrigAlt{} runs all have $n=50$ and reach the horizon first. Among runs
landing at $\eps_2\ge1/2$, \lyHiTrigSim{} of \lyHiSim{} simultaneous runs and
\lyHiTrigAlt{} of \lyHiAlt{} alternating runs fire, certifying landings up to
\lyEpsMaxSim{} and \lyEpsMaxAlt{} with median trigger gaps of \lyGapMedSim{} and
\lyGapMedAlt{}. After triggering, all \lyDecayRounds{} subsequent rounds stay
absorbed and satisfy $\norm{W^t}_{\mathcal P}\le\rho_1^{t-t_0}\omega$, with
realised tail energy at most \lyEnergyRatio{} of the bound. The quadratic form
of \cref{prop:drift} predicts each player's energy with median relative errors
of \lyPredTwo{}, \lyPredFour{}, and \lyPredSeven{} at gaps $10^{-2}$, $10^{-4}$,
and $10^{-7}$, scaling with the gap until double precision dominates.

The rate at a landing is set by the extreme singular values of $\Umat$, and the
best landing depends on their ratio.

\begin{proposition}[Oracle effective step size]\label{prop:oracle}
Let $\sigma_{\max}\ge\dots\ge\sigma_{\min}>0$ be the nonzero singular values of $\Umat$. Let $\kappa:=\sigma_{\max}/\sigma_{\min}$, and define $\rho(\eta):=\max_jr(\eta\sigma_j)$, where $r$ is the function from \cref{prop:scalar}. Then, the following properties hold:
(i) We have $\rho(\eta)=\max\{r(\eta\sigma_{\min}),r(\eta\sigma_{\max})\}$. Moreover, for $\eta\sigma_{\max}\le1/2$, we have $-\log\rho(\eta)=\eta^2\sigma_{\min}^2/2+O(\eta^4\sigma_{\min}^4)$.
(ii) The infimum $\rho^*:=\inf_\eta\rho(\eta)$ is attained at a unique $\eps^*:=\eta^*\sigma_{\max}\in[1/2,1/\sqrt3)$. Specifically, $\eps^*=1/2$ and $\rho^*=1/\sqrt2$ if and only if $\kappa=1$. Moreover, $\eps^*$ is nondecreasing in $\kappa$, and $\eps^*\to1/\sqrt3$ as $\kappa\to\infty$.
\end{proposition}
Near $1/2$, the rate has a square-root cusp
($\tfrac{d}{d\eps}r(\eps)^2\to-\infty$ as $\eps\uparrow1/2$). On a face with a
single singular value, a landing at $\eps_{\max}$ slightly below $1/2$ therefore
loses a share of order $(1/2-\eps_{\max})^{1/2}$ of the oracle log-rate. On an
ill-conditioned face, the oracle sits near the threshold, and a landing at
$\eps_{\max}<\eps^*$, where the smallest mode sets the rate, keeps the share
$(\eps_{\max}/\eps^*)^2\,(1+O((\eps^*/\kappa)^2))$ of the oracle log-rate. The
same proposition holds for the alternating schedule with $r_{\rm alt}$,
$2^{-3/4}$, $2/3$ and $\eta^2\sigma_{\min}^2$ in place of $r$, $1/2$, $1/\sqrt3$
and $\eta^2\sigma_{\min}^2/2$, since $r_{\rm alt}$ is unimodal
(\cref{prop:altscalar}). At the optimum, however, $r_{\rm alt}$ has a kink of
finite slope, so an alternating landing slightly below its optimum loses only
linearly.

\section{Further Related Work}
\label{app:relwork}

\paragraph{Optimistic and extra-gradient dynamics.}
Extra-gradient methods alternate a probe step with an update \citep{korpelevich1976extragradient}. \citet{popov1980modification} showed that the probe can reuse the previous update, matching the recursion derived from optimistic online learning with predictable sequences \citep{rakhlin2013online}. Both schemes share the regularity that accelerates game learning \citep{daskalakis2011near,syrgkanis2015fast} and admit a unifying proximal-point view \citep{pmlr-v108-mokhtari20a}.

Last-iterate theory for optimistic gradient methods centers on this two-step recursion \citep{daskalakis2018limit,daskalakis2019last,wei2021linear,gorbunov2022last,cai2022finite}. Linear rates hold under unique equilibria in extensive form \citep{lee2021last}, smoothed analysis \citep{anagnostides2024convergence}, and noisy feedback with mutation \citep{abe2023last}. Recent work extends these guarantees to multiplicative updates via optimistic exponential weights \citep{hadiji2026stability,orabona2026lastiterate,lazarsfeld2026optimism}. All these methods require an explicit step size. In \IRO{}, the inverse regret norm is an effective step size, growing dynamically and staying bounded by \cref{thm:sat-iro}. First-order methods with optimal condition-number dependence \citep{gilpin2008first} set the benchmark for hand-tuned step sizes, motivating the tuned baseline of \cref{sec:experiments}.

\paragraph{Alternating updates.}
Alternation has been standard in counterfactual regret minimisation plus (CFR$^+$) since its introduction \citep{tammelin2014solving,burch2019revisiting} and qualitatively alters the dynamics beyond constant factors. Alternating gradient descent-ascent maintains bounded regret and cycles on games where simultaneous updates diverge \citep{bailey2020finite}. Locally, it beats simultaneous rates by super-constant factors \citep{zhang2022near} and by an order in the condition number \citep{lee2024fundamental}, while achieving $O(1/T)$ average-iterate rates in constrained bilinear games \citep{wibisono2022alternating,katona2026symplectic,ICLR2026_7ff8961c}. In online learning, alternation weakens the adversary enough to break the $\Omega(\sqrt T)$ regret barrier \citep{cevher2023alternation,pmlr-v291-hait25a,tao2026optimalalternatingregretonline}.

For regret matching, \citet{li2026alternation} explain the speed-up of alternating CFR via the spectrum of the alternating linearization, and \citet{nguyen2026convergencealternatingregretmatching} establish the first convergence theory for alternating \RMp{}-type learners. In \cref{sec:threshold}, the cubic polynomial of the alternating companion matrix (the matrix advancing the two-step linear recursion by one step) provides the parallel spectral analysis for the shifted learner. \citet{li2026last} leave open whether a closed-form rate holds under alternation and beyond unique equilibria. \Cref{sec:threshold,sec:nonunique} answer both questions for the single-call learner. Alternation enlarges the stability region from $1/\sqrt3$ to $2/3$ and, at a common effective step size, doubles the local log-rate to leading order in $\eps$.

\section{Without Uniqueness: Strictly Complementary Limits}
\label{sec:nonunique}

Note that uniqueness enters our local theory in only one place, the rank lemma \cref{lem:rank} stated in \cref{app:imported}. The freezing and the linearisation rely on strict complementarity at the limit, meaning that every action outside the equilibrium support is strictly worse against the equilibrium. We now restate our results at such a limit. The equilibrium set then contributes directions along which nothing moves. Moreover, the spectrum of the companion matrix is carried by the nonzero singular values of the matrix restricted to the face of strategies using only the supported actions.
\begin{assumption}\label{ass:sc}
The pair $(x^*,y^*)$ is a strictly complementary equilibrium with value $v$.
Specifically, $S_1:=\supp x^*=\{a:(Ay^*)_a=v\}$ and $S_2:=\supp y^*=\{b:(A^\top x^*)_b=v\}$,
where the sizes $k_1,k_2$ and the margin $\mu>0$ are as in \cref{ass:ne}.
\end{assumption}
Let $X^*$ and $Y^*$ be the two equilibrium polytopes, and let $d_1,d_2$ denote their dimensions. The strictly complementary equilibria are the pairs of points in their relative interiors, and they share the supports $S_1,S_2$. By the strict complementarity of linear programming \citep{goldman1956theory}, such equilibria exist in every game. Note that \cref{ass:ne} corresponds to the case $d_1=d_2=0$. We keep $\At=A[S_1,S_2]$, $\Umat=\At-v\one\one^\top$, $T_x,T_y$, $\Pi_i,P_i$, and $L$ as in \cref{sec:face}, now with $T_x\subset\R^{k_1}$ and $T_y\subset\R^{k_2}$. We also define $Z_y:=\{\delta\in T_y:\At\delta=0\}$ and $Z_x:=\{\delta\in T_x:\At^\top\delta=0\}$.

\begin{lemma}[Rank without uniqueness]\label{lem:rankgen}
Suppose that \cref{ass:sc} holds. Then, the following statements hold:
(i) The space $Z_y$ is the tangent space of $Y^*$ at $y^*$, and $\ker\Umat=\operatorname{span}(y^*)\oplus Z_y$. The symmetric statements hold for $x$. In particular, we have $\operatorname{rank}\Umat=k_1-1-d_1=k_2-1-d_2=:r$.
(ii) On $T_x\oplus T_y$, we have $\ker L=Z_x\oplus Z_y$, and the eigenvalue $0$ of $L$ is semisimple. Moreover, the nonzero eigenvalues are $\pm i\,\eta\,\sigma_j(\Umat)$ for $j\le r$, ranging over the nonzero singular values.
\end{lemma}

\begin{theorem}[Local theory at a strictly complementary limit]\label{thm:nonunique}
Suppose that \cref{ass:sc} holds. Let
$R':=\{(c_1x,c_2y,Ay,-A^\top x):(x,y)\ \text{strictly complementary},\ c_1,c_2>0\}$,
and let $s^*\in R'$ be the point with $(x,y)=(x^*,y^*)$. Set $\eps_j:=\eta\sigma_j(\Umat)$
for each $j\le r$. For the alternating schedule, we read $R'$ with the column
component $c_2y+u_2+v\one$ of \cref{lem:altfreeze}. The proof of that lemma uses strict
complementarity of the pair alone. Then the following hold.
\begin{enumerate}
\item[(a)] The set $R'$ is a manifold of dimension $2+d_1+d_2$ consisting of fixed points of the simultaneous round map. Moreover, \cref{thm:freeze} holds at each of its points
  with the margin of that point.
\item[(b)] On the face state space $\R^{2(k_1+k_2)}$, we have
  $\operatorname{char}_J(x)=x^{2+d_1+d_2}(x-1)^{2+d_1+d_2}\prod_{j\le r}
  \bigl(x^2-(1+2i\eps_j)x+i\eps_j\bigr)\bigl(x^2-(1-2i\eps_j)x-i\eps_j\bigr)$.
  Here, the eigenspace of the eigenvalue $1$ coincides with the tangent space of the projection of $R'$ to the reduced coordinates.
  Moreover, for the alternating round map on the $2(k_1+k_2)$ active coordinates, we have
  $\operatorname{char}_{J_{\rm alt}}(x)=x^{2+r+d_1+d_2}(x-1)^{2+d_1+d_2}\prod_{j\le r}\chi_{\eps_j}(x)$.
\item[(c)] Suppose that $\eps_{\max}<1/\sqrt3$ for the simultaneous schedule or $\eps_{\max}<2/3$
  for the alternating schedule. Fix any $\rho$ above the closed-form rate of the
  schedule at $s^*$. Then, there exists a neighborhood of $s^*$ in the face from which
  every orbit converges to a point of $R'$ with $\Gap(z^t)=O(\rho^t)$. In particular, when
  $r=0$, this guarantee holds for every $\rho>0$. Moreover, if $\eps_{\max}$ is above the
  threshold, then part (b) of \cref{thm:dichotomy,thm:altdichotomy} holds at $s^*$.
\item[(d)] Suppose that a run of either schedule admits an accumulation point $s^*\in R'$
  with $\eps_{\max}(s^*)$ below the threshold of its schedule. Then, the run converges to
  $s^*$, and its gap decays at every rate above the closed-form rate at $s^*$.
\end{enumerate}
\end{theorem}
Note that uniqueness only restricts which limits exist, leaving the local dynamics unchanged once strict complementarity holds. At an equilibrium on the relative boundary of $X^*\times Y^*$, a best response lies outside its support, and hence the clipping remains active there. Whether a run reaches a strictly complementary limit in general remains open.

Existing remedies that guarantee linear last-iterate rates reintroduce a tuning parameter. Examples include a forgetting step \citep{cai2024fast}, smoothing and restarts \citep{cai2025lastiterate}, a perturbed-regularized objective \citep{meng2025lastiterate,meng2025efficient}, and regularization-free re-anchoring \citep{zhang2026teaching}. Separately, an adaptive regret domain recovers the average-iterate rate $O(1/T)$ without parameters \citep{meng2026a}, while a general reduction transforms average-iterate guarantees into last-iterate bounds through a modification of the dynamics \citep{cai2025from}.

\citet{zhang2025scale} remove the step size completely via an equal-norm shift of the regret vectors, subtracting a common scalar before clipping so the Euclidean length is preserved. This builds on parameter-free online learning \citep{chaudhuri2009parameter,orabona2016coin} and scale-free game dynamics \citep{grand2021conic,antonakopoulos2021adaptive,tsuchiya2026scaleinvariantfastconvergencegames}. \citet{dewasurendra2026selfboundingregretmatchingpotential} use a related one-step conservation law, paying for norm growth with forward utility gain, to analyze \RMp{} on products of simplices. \citet{li2026last} prove a linear last-iterate rate for the extra-gradient version of the shifted learner, which evaluates payoffs twice per round at a tentative profile. They show that the regret norm always saturates (meaning it stays bounded and converges) with an explicit bound. Their rate holds when the saturated norm meets a step-size condition.

Two concurrent papers study the single-call version, which reuses the previous utility with one payoff evaluation per round. \citet{sullins2026irprmpreviousutilitypredictionconstant} prove constant regret, bounded regret norms, and last-iterate convergence to the equilibrium set from zero initialization, covering both simultaneous updates (where players move at once, their Theorem 2 and Corollary 3) and alternating updates (where one moves and the other answers within the round, their Theorem 7). \citet{nguyen2026convergencealternatingregretmatching} introduce an alternating form of regret bounded by variation in utilities (RVU) under the schedule of \cref{sec:setup} (their Algorithm 2). They obtain the optimal average-iterate rate under alternation, prove convergence to a single equilibrium without assuming uniqueness (their Theorem 7.5), and exhibit a game where half-prediction \RMp{} converges under alternating updates while failing under simultaneous ones.

Both learners in \cref{sec:setup} share a Blackwell approachability step \citep{blackwell1956analog,abernethy2011blackwell,farina2021faster}, which also underlies recent solvers for optimization problems and tree-structured polytopes (treeplexes) \citep{grand2024solving,chakrabarti2024extensive}. An equivalence linking counterfactual regret minimization with online mirror descent \citep{liu2022equivalence} connects this step to optimistic-gradient tools. Our saturation proof in \cref{sec:saturation} relies on the same clipping identity and energy inequality that drive the constant-regret analysis of \citet{sullins2026irprmpreviousutilitypredictionconstant}. Finally, the Hoffman bound \citep{hoffman2003approximate} translates a small duality gap (the maximum payoff gain from unilateral deviation, which vanishes at equilibrium) into proximity to the equilibrium set, justifying the gap metric in \cref{sec:experiments}.

Looking forward, because the norm increment our monitor reads may become noise-dominated under sampled feedback like Monte Carlo counterfactual regret minimization \citep{lanctot2009monte}, applying variance reduction via correlated sampling \citep{li2026correlated} is a first step toward a monitor for that setting, while evaluating such a monitor would require variance-reduced estimators \citep{burch2018aivat,li2026av}. A companion direction to our rate certificate is certifying computed profiles as near-Nash in general-sum games \citep{li2026we} or strategies actually deployed in games \citep{li2026agents}. The same question about which object is deployed arises for large language model agents outputting the final answer of a bounded reasoning budget \citep{li2026pokerskill}.

%% file: sections/14_proofs.tex
\section{Deferred Proofs}
\label{app:proofs}

This appendix provides the proofs omitted from \cref{sec:setup,sec:saturation,sec:face,sec:threshold,sec:nonunique} and of the statements in \cref{app:extra,app:landing}. Each group follows the order the statements first appeared in the main development.

\subsection*{Proofs for \cref{sec:setup}}
\begin{proof}[Proof of \cref{lem:energy}]
Note that centering gives $\ip{p^t}{g^t}=\norm{p^t}_1\ip{z^t}{g^t}=0$. We first observe that coordinatewise, $\pos{a+b}^2\le\pos{a}^2+2\pos{a}b+b^2$. Summing this bound with $a=v^t_j$ and $b=g^t_j$, and using the first equality, we obtain $N_{t+1}^2\le\norm{p^t}^2+\norm{g^t}^2=N_t^2+\norm{g^t}^2$. We now turn to the lower bound. On coordinates with $p^t_j>0$, we have $q^{t+1}_j-p^t_j=\pos{p^t_j+g^t_j}-p^t_j\ge g^t_j$. Multiplying by $p^t_j\ge0$, summing, and again using the first equality gives $\ip{p^t}{q^{t+1}-p^t}\ge0$. Hence, we conclude that $N_{t+1}^2-N_t^2=\norm{q^{t+1}-p^t}^2+2\ip{p^t}{q^{t+1}-p^t}\ge\norm{q^{t+1}-p^t}^2\ge0$.
\end{proof}

\begin{proof}[Proof of \cref{lem:gbound}]
Note that $u_1^t-m_1^t=A(y^t-y^{t-1})$. Applying (F3) concludes the proof.
\end{proof}

\subsection*{Proofs for \cref{sec:saturation}}
\begin{proof}[Proof of \cref{lem:activation}]
We first consider round one. Since $q^1=0$, the player plays the uniform strategy and updates to $q^2=\pos{u-\bar u\one}$. If this vector vanishes, then $u$ is constant across actions, and every subsequent round repeats round one. Hence, we have $t^*=2$ whenever the player activates.
From round $t^*$ onward, we have $m^t=u^{t-1}=u^t$, which yields $g^t=0$ and $q^{t+1}=p^t=\pos{q^t+u-\gamma^t\one}$ with $\norm{q^{t+1}}=\norm{q^t}$. That is, the norm remains constant while the direction evolves.
Note that if $\gamma^t<\min_{a\in\supp q^t}u_a$, then $p^t_a>q^t_a$ holds on $\supp q^t$, and hence $\norm{p^t}>\norm{q^t}$. Since the norm is constant, we obtain $\gamma^t\ge\min_{a\in\supp q^t}u_a$.
We now show by induction on $t\ge t^*$ that $\supp q^t$ is an upper level set of $u$ on which $q^t_a-q^t_b$ has the sign of $u_a-u_b$. For the base case at $t^*$, this property holds since $q^2_a=u_a-\bar u$ on $S^0:=\{a:u_a>\bar u\}$. For the inductive step, any coordinate $b\notin\supp q^t$ satisfies $u_b<\min_{\supp q^t}u\le\gamma^t$, which gives $q^{t+1}_b=\pos{u_b-\gamma^t}=0$. Hence, deleted coordinates never return. On $\supp q^t$, the unclipped coordinate $q^t_a+u_a-\gamma^t$ is ordered like $u_a$, so its positive part is supported on an upper level set. Moreover, for $a,b$ both surviving, the difference $q^{t+1}_a-q^{t+1}_b=q^t_a-q^t_b+(u_a-u_b)$ preserves the sign of $u_a-u_b$. This completes the induction.
Fix $a^*\in\argmax u$. Since it has the largest coordinate and $p^t\neq0$, it never leaves the support. Meanwhile, any coordinate $b$ with $u_b<u_{a^*}$ that remains active at round $t$ satisfies $N^*\ge q^t_{a^*}-q^t_b=(t-t^*+1)(u_{a^*}-u_b)\ge(t-t^*+1)\Delta_u$. This inequality fails as soon as $t>t^*-1+N^*/\Delta_u$.
Once $\supp q^t\subseteq\argmax u$, the vector $u$ is constant on the support. Consequently, we have $\gamma^t=\max u$ and $p^t=q^t$, so the state remains fixed.
Finally, suppose that a player satisfies $q^t=0$ for all $t$. Then we have $\pos{u^t-\bar u^t\one}=0$ at every round, which implies that $u^t$ is constant across its actions. This concludes the proof.
\end{proof}

\begin{proof}[Proof of \cref{lem:disp}]
We first apply (F1) with $N_t\ge N_i^*$ and then apply (F2). Finally, iterating \cref{lem:energy} gives $N_{i,T+1}\le\sqrt{N_i^{*2}+G_i}\le N_i^*+\sqrt{G_i}$, concluding the proof.
\end{proof}

\begin{proof}[Proof of \cref{thm:sat-iro}]
Fix $T\ge T_0+1$. Throughout the proof, all sums run over $t\in[T_0,T]$, and we let $G,G_i,D_i$ denote the corresponding partial sums. Applying \cref{lem:gbound} together with the inequality $\norm{\alpha+\beta}^2\le2\norm{\alpha}^2+2\norm{\beta}^2$, we obtain
\[
  G_1\le K_1^2\sigma^2\sum\norm{y^t-y^{t-1}}^2\le 2K_1^2\sigma^2\,(D_2+C_2),
\]
where we define $C_2:=\sum\norm{\tilde y^t-y^{t-1}}^2$. A symmetric bound holds for $G_2$. We first bound the displacement part by combining \cref{lem:disp} with $\sqrt{G_1}+\sqrt{G_2}\le\sqrt{2G}$. We next bound the carried part. When $t\ge T_0+1$, note that $\tilde y^t=\N(q_2^t)$ and $y^{t-1}=\N(p_2^{t-1})$ with $\norm{q_2^t}_1\ge N_{2,t}$. Combining (F4) and \cref{lem:energy} yields
\[
  \norm{\tilde y^t-y^{t-1}}^2
  \le K_2^2\frac{\norm{q_2^t-p_2^{t-1}}^2}{N_{2,t}^2}
  \le K_2^2\frac{N_{2,t}^2-N_{2,t-1}^2}{N_{2,t}^2}
  \le K_2^2\!\int_{N_{2,t-1}^2}^{N_{2,t}^2}\!\frac{ds}{s},
\]
since $(\beta-\alpha)/\beta\le\int_\alpha^\beta ds/s$ for $0<\alpha\le\beta$. The sum telescopes to $K_2^2\log(N_{2,T}^2/N_2^{*2})\le K_2^2\log(1+G/N_2^{*2})$. Moreover, the $t=T_0$ term is at most $2$, which is the squared diameter of the simplex. Hence, we have $C_2\le2+K_2^2\log(1+G/N_2^{*2})$. Combining the four bounds gives $G\le\Phi(G)$ for every partial sum. Note that $\Phi$ is continuous, increasing, and concave, with $\Phi(0)=c_0>0$ and $\Phi(G)/G\to0$. Therefore, $\{G\ge0:G\le\Phi(G)\}=[0,G^*]$, which shows that every partial sum, and hence the full sum, is at most $G^*$. Finally, applying \Cref{lem:energy} iteratively establishes the norm bound. This concludes the proof.
\end{proof}

\begin{proof}[Proof of \cref{lem:improve}]
Note first that if $q^{t+1}=0$, then $q^t=0$. In that case, both strategies are uniform and both sides vanish.
Hence, we focus on the complementary case, and we write $p:=p^t$, with $p=0$ and $\delta^t=0$ if $q^t=0$. In that subcase, the next display reads $\ip{q^{t+1}}{p}=0=N_t^2$ directly.
Combining the inequality $q^{t+1}\ge q^t+\hat r^t+\delta^t\one$ with the orthogonality relation $\ip{\hat r^t}{p}=\norm p_1\ip{\hat r^t}{z^t}=0$ gives
$\ip{q^{t+1}}{p}\ge\ip{q^t+\delta^t\one}{p}=\ip{v^t-m^t+(\gamma^t+\delta^t)\one}{p}
=\norm p^2-\norm p_1\ip{m^t}{z^t}+\ip{m^t}{z^t}\norm p_1=N_t^2$,
where we used the property $\ip{v}{\pos v}=\norm{\pos v}^2$.
Hence, we obtain $\norm{q^{t+1}-p}^2\le N_{t+1}^2-N_t^2$.

We next express $q^{t+1}=\pos{q^t+w}$ with $w:=\hat r^t+\delta^t\one$, which gives the identity $\ip{w}{q^{t+1}}=\ip{q^{t+1}-q^t}{q^{t+1}}$.
Note also that $v^{t+1}-q^{t+1}=u^t-(\ip{u^t}{z^{t+1}}-\delta^{t+1})\one$.
Subtracting $\ip{v^{t+1}-q^{t+1}}{q^{t+1}}$ from the preceding identity yields
\[
  \norm{q^{t+1}}_1\bigl(\ip{u^t}{z^{t+1}-z^t}+\delta^t-\delta^{t+1}\bigr)
  =\ip{q^{t+1}-q^t}{q^{t+1}}+\ip{q^{t+1}-v^{t+1}}{q^{t+1}}
  \ge\tfrac12\norm{q^{t+1}-p^t}^2+\tfrac12\norm{q^{t+1}-p^{t+1}}^2,
\]
where we used the polarization identity $\ip{q^{t+1}-q^t}{q^{t+1}}=\frac12(N_{t+1}^2-N_t^2+\norm{q^{t+1}-q^t}^2)$.
Moreover, since $v^{t+1}\le p^{t+1}$, $q^{t+1}\ge0$, and $\norm{p^{t+1}}=N_{t+1}$, we have $\ip{q^{t+1}-v^{t+1}}{q^{t+1}}\ge\ip{q^{t+1}-p^{t+1}}{q^{t+1}}=\frac12\norm{q^{t+1}-p^{t+1}}^2$.

Finally, by (F4), we have $\norm{q^{t+1}-\norm{q^{t+1}}_1z^s}\le K\norm{q^{t+1}-p^s}$ for $s=t,t+1$.
Note that when $p^t=0$, this bound holds because $\norm{q-\norm q_1\one/n}\le2\norm q\le K\norm q$.
Combining the two bounds gives $\norm{q^{t+1}}_1^2\norm{z^{t+1}-z^t}^2\le2K^2(\norm{q^{t+1}-p^t}^2+\norm{q^{t+1}-p^{t+1}}^2)$.
Dividing by $\norm{q^{t+1}}_1\ge N_{t+1}$ concludes the proof.
\end{proof}

\begin{proof}[Proof of \cref{lem:altreg}]
By definition, we have $\mathrm{Reg}_1=T\max_pp^\top A\bar y_T-\sum_tx^{t\top}Ay^t$ and $\mathrm{Reg}_2=-T\min_{p'}\bar x_T^\top Ap'+\sum_kx^{k+1\top}Ay^k$. Summing these two equations combines the two sums into $\sum_{t=1}^Tx^{t\top}A(y^{t-1}-y^t)=\sum_t\ip{u_2^{t-1}}{y^t-y^{t-1}}$, which is the column player's improvement sum. By \cref{lem:improve}, and by the monotonicity of the norms, which lets us lower the weight $N_{2,k+2}$ of that lemma to $N_{2,k+1}$, we bound this improvement sum from below by the displayed sum plus $\delta_2^T-\delta_2^0$. Note that $\delta_2^0=0$ since $y^0$ is played from $q_2=0$. Next, we apply the per-player regret bound of \cref{sec:setup} to obtain $\mathrm{Reg}_1(T)\le N_{1,T+1}-\sum_{t\le T}\delta_1^t$. Counting the column player's own rounds, the same bound yields $\mathrm{Reg}_2(T)\le N_{2,T+1}-\sum_{k<T}\delta_2^k$. Combining the two inequalities and moving $\delta_2^T$ to the left establishes the last claim. This concludes the proof.
\end{proof}

\begin{proof}[Proof of \cref{thm:sat-alt}]
Fix $T\ge T_0'+1$, and we index both players by the round of \cref{alg:altround}.
Note that the row player plays $x^t$ from $q_1^t$ with correction $\delta_1^t$. Its accumulation in round $t$ has prediction error driven by $u_1^t-m_1^t=A(y^t-y^{t-1})$. Similarly, the column player plays $y^t$ from $q_2^{t+1}$ with correction $\delta_2^t$. Its accumulation in round $t$ transitions from $\pos{r_2^t}$ (the vector from which $y^{t-1}$ was played) to $q_2^{t+1}$. Its prediction error is driven by $u_2^t-m_2^{t-1}=-A^\top(x^t-x^{t-1})$, where $m_2^{t-1}$ is the prediction with which it played $y^{t-1}$. We let the row sums run over $t\in[T_0,T]$ and the column sums run over $t\in[T_0',T]$, where $T_0'=\max(T_0,t_2^*+1)$. Under this indexing, every counted accumulation starts from a nonzero vector, with starting norms $N_{1,T_0}=N_1^*$ and $N_{2,T_0'}=N_2^*$. Moreover, we count the row player's accumulation in round $T_0$, which observes the column player's first active play when $t_2^*=T_0$.
Hence, \cref{lem:gbound} holds with the same constants. Furthermore, we can iterate \cref{lem:energy} for each player along its counted accumulations starting from the norm $N_i^*$, and (F1) holds at every counted play.
Applying \Cref{lem:altreg} at horizon $T$ yields (F2), where both sums end at the plays of round $T$ and every term is nonnegative. Note also that both players play from a nonzero vector at every round $t\ge T_0$. In particular, the column player plays from $q_2^{t+1}$, which has norm at least $N_2^*$. Consequently, \cref{lem:disp} holds for both displacement sums $D_1,D_2$ over $[T_0,T]$. Note that only the energy sums $G_2$ and the carried terms of the column player start at $T_0'$.
We next bound the carried-to-previous terms of the row player's bound, which are $\norm{\tilde y^t-y^{t-1}}$ for $t\ge T_0+1$, with $\tilde y^t=\N(q_2^{t+1})$ and $y^{t-1}=\N(\pos{r_2^t})$. Since the column player's accumulation in round $t\ge T_0+1\ge t_2^*+1$ starts from a nonzero vector, we have $\norm{q_2^{t+1}-\pos{r_2^t}}^2\le N_{2,t+1}^2-N_{2,t}^2$. We bound these terms via this inequality and by telescoping from $N_{2,T_0+1}\ge N_2^*$, following the proof of \cref{thm:sat-iro}. Similarly, the carried-to-previous terms of the column player's bound are $\norm{\tilde x^t-x^{t-1}}$ for $t\ge T_0'+1$. We bound these using the row player's accumulation in round $t-1\ge T_0$ and by telescoping from $N_{1,T_0'}\ge N_1^*$.
For the initial terms, the row player's first term occurs at $t=T_0$ (where $y^{T_0-1}$ may be the uniform strategy). The column player's first term occurs at $t=T_0'$ (where $x^{T_0'-1}$ may be the uniform strategy), and each is at most $2$.
Combining the four bounds gives $G\le\Phi(G)$ for every partial sum with the same $\Phi$, and thus with the same $G^*$. Finally, iterating \cref{lem:energy} from $N_i^*$ establishes the norm bound. This concludes the proof.
\end{proof}

\begin{proof}[Proof of \cref{prop:init}]
We first note that the inequality $\ip{q^{t+1}}{p}\ge\ip{q^t}{p}+\ip{\hat r^t}{p}+\delta^t$ of \cref{sec:setup} holds for every prediction. Summing this inequality yields $\mathrm{Reg}(T)\le\max_p\ip{q^{T+1}-q^1}{p}-\sum_{t\le T}\delta^t\le N_{T+1}-\sum_{t\le T}\delta^t$, where the second step follows from $q^1\ge0$. Moreover, the self-play identity is unchanged. Under alternation, \cref{lem:improve} applies from the column player's play $y^0$ onward. Note that $\delta_2^0=0$, since a zero prediction gives a zero shift, and hence \cref{lem:altreg} and (F2) hold. Next, both players play from a nonzero vector in round one, and under alternation the standing play $y^0$ counts as the column player's active play $0$. Hence, we have $t_2^*=0$ and $T_0=T_0'=1$, while the norms $N_i^*$ are those of the initial vectors. A start with exactly one zero vector is not covered. That player is inactive in round one, leaving no positive initial norm for its constant $2n_i/N_i^*$. Finally, in the proofs of \cref{thm:sat-iro,thm:sat-alt}, only the first prediction error differs. That error equals the centred utility itself, with squared norm at most $K_i^2\sigma^2$. This bound is below the $2K_i^2\sigma^2\cdot2$ allotted to the first term in those proofs, and every remaining step holds as written. This concludes the proof.
\end{proof}

\begin{proof}[Proof of \cref{lem:limit}]\looseness=-1
By \cref{thm:sat-iro,thm:sat-alt,prop:init}, the norms are bounded, so we have $\sum_t\delta_i^t<\infty$ by (F2). Fix a player and $t\ge T_0$. Note that $\norm{z^{t+1}-z^t}\le\norm{z^{t+1}-\tilde z^{t+1}}+\norm{\tilde z^{t+1}-z^t}$. By (F1), the first term is at most $(2n\delta^{t+1}/N^*)^{1/2}$. Moreover, by (F4) and \cref{lem:energy}, the second term is $\norm{\N(q^{t+1})-\N(p^t)}\le K\norm{q^{t+1}-p^t}/N^*\le K(N_{t+1}^2-N_t^2)^{1/2}/N^*$. Hence, both terms tend to zero. Under alternation, the same argument runs along each player's own plays.

\looseness=-1We now consider a limit point $(\hat x,\hat y)$, and suppose that the row player has an action $a$ with $\beta:=(A\hat y)_a-\hat x^\top A\hat y>0$. Note that its regret increment $\hat r_{1,a}^t=(Ay^t)_a-x^{t\top}Ay^t$ is a continuous function of the profile. Hence, we have $\hat r_{1,a}^t\ge\beta/2$ whenever the profile lies within some radius $\bar r$ of $(\hat x,\hat y)$. Fix $L$. We choose a sufficiently late round $t_1$ so that the profile at that round lies within $\bar r/2$ of the limit point and moves by less than $\bar r/(2L)$ per round thereafter. Over the next $L$ rounds, the profile stays within that radius, which yields $q_{1,a}^{t+1}\ge q_{1,a}^t+\hat r_{1,a}^t+\delta_1^t\ge q_{1,a}^t+\beta/2$. Consequently, we obtain $N_{1,t_1+L}\ge L\beta/2$, contradicting boundedness once $L$ is large. Thus, no action of the row player improves on $\hat x$ against $\hat y$. The argument for the column player is symmetric. Under alternation, its increment for the play $y^t$ is a continuous function of $(x^{t+1},y^t)$, and we have $x^{t+1}-x^t\to0$. This concludes the proof.
\end{proof}

\subsection*{Proofs for \cref{sec:face}}
\begin{proof}[Proof of \cref{lem:rank}]
This is the kernel characterisation of \citet{li2026last} for a unique, strictly complementary equilibrium.
\end{proof}

\begin{proof}[Proof of \cref{lem:ray}]
We first analyze player 1. Note that $m_1=u_1^*$ equals $v$ on $S_1$ and is below $v-\mu$ off $S_1$.
Setting $\gamma=v$ gives $\pos{c_1x^*+u_1^*-v\one}=c_1x^*$, which has norm $N_1$.
Hence, uniqueness yields $\gamma_1=v$.
From this, we obtain $x=\N(c_1x^*)=x^*$, $u_1=Ay^*=m_1$, $g_1=0$, and $q_1'=\pos{v_1}=c_1x^*$.
The argument for player 2 follows by symmetry.
\end{proof}

\begin{proof}[Proof of \cref{lem:lip}]
Note that the function $h_w(\gamma):=\norm{\pos{w-\gamma\one}}_2$ is $1$-Lipschitz in $w$. On $(-\infty,\bar\gamma_w]$ with $\bar\gamma_w:=\max_aw_a$, it has slope $-\norm{\pos{w-\gamma\one}}_1/\norm{\pos{w-\gamma\one}}_2\le-1$ almost everywhere, and beyond $\bar\gamma_w$, it vanishes.

We first consider the case $\gamma'\le\bar\gamma_w$. Then, the $1$-expansion gives $\abs{\gamma'-\gamma}\le\abs{h_w(\gamma')-h_w(\gamma)}\le\norm{w'-w}_2+\abs{c'-c}$.

We now consider the case $\gamma'>\bar\gamma_w$. Then, $h_w(\gamma')=0$, so $c'\le\norm{w'-w}_2$. Moreover, we have $\bar\gamma_w-\gamma\le c\le\abs{c-c'}+\norm{w'-w}_2$ and $\gamma'<\bar\gamma_{w'}\le\bar\gamma_w+\norm{w'-w}_\infty$. Adding these last two inequalities gives $\gamma'-\gamma\le\abs{c-c'}+2\norm{w'-w}_2$. This concludes the proof.
\end{proof}

\begin{proof}[Proof of \cref{thm:freeze}]
Fix Player 1 and consider first a coordinate $a\notin S_1$. By \Cref{lem:lip} applied at $(q_1^*,u_1^*)$, where $\gamma=v$, we obtain $\abs{\gamma_1-v}\le5\sqrt{n_1}\,r$. Hence, $v_{1,a}=q_a+m_a-\gamma_1\le r+(v-\mu+r)-(v-5\sqrt{n_1}r)=-\mu+\kappa_1r<0$, and thus $p_a=0$. On $S_1$, we have $\abs{p_a-q^*_a}\le\kappa_1r$. This gives $\norm{\N(p_1)-x^*}_1\le2k_1\kappa_1r/c_1$, which in turn yields $\norm{u_2-u_2^*}_\infty\le A_{\max}\norm{x-x^*}_1\le\ell r$. Next, note that $q'_a=\pos{v_{1,a}+g_{1,a}}$, where $\abs{g_{1,a}}\le2\norm{u_1-m_1}_\infty\le2(\ell+1)r$. Returning to the first case, we find $v_{1,a}+g_{1,a}\le-\mu+(\kappa_1+2(\ell+1))r<0$. On $S_1$, we have $v_{1,a}\ge q^*_{1,a}-\kappa_1r>0$, and moreover $v_{1,a}+g_{1,a}\ge q^*_{1,a}-(\kappa_1+2(\ell+1))r>0$ by the second term in $r_0$. Combining these bounds shows that $\pos{\cdot}$ acts as the identity on both the played and the accumulated vector there, and hence $F=F_S$. Moreover, $F_S$ is smooth since $\gamma$ is smooth by \cref{lem:gamma}. Note that this second term is needed. With the radius $q^*_{i,a}/\kappa_i$ alone, a correction $g$ of size $2(\ell+1)r$ can clip an active coordinate inside the radius.
\end{proof}

\begin{proof}[Proof of \cref{cor:pure}]
By \cref{thm:freeze}, $p_1^t$ is supported on $\{a\}$ with norm $N_{1,t}$, giving $p_1^t=N_{1,t}e_a$ and $x^t=e_a$. Likewise, we have $y^t=e_b$, so $u_1^t=Ae_b=u_1^*$ holds exactly. Note that $g_{1,a}^t=(u^t-m^t)_a-\ip{u^t-m^t}{e_a}=0$, and hence we obtain $q^{t+1}_{1,a}=p^t_{1,a}=N_{1,t}$ while the other coordinates vanish. Consequently, the point lies in $R$ and is fixed by \cref{lem:ray}.

We now consider alternation. Since the entering $\pos{r_2}$ is supported on $\{b\}$, the standing play is $e_b$. Moreover, its correction vanishes on $b$, giving the accumulated vector $c_2e_b$. Next, the column play against $x=e_a$ has prediction $u_2^*$ and shift $-v$, which yields $r_2^*(c_2)$. Finally, the row accumulation against $e_b$ leaves $c_1e_a$ and stores $u_1^*$. This concludes the proof.
\end{proof}

\begin{proof}[Proof of \cref{lem:gamma}]
Note that on the face, $\gamma$ solves $f(\gamma;q,m):=\norm{q+m-\gamma\one}_2^2-\norm{q}_2^2=0$.
We observe that $\partial_\gamma f=-2\one^\top(q+m-\gamma\one)=-2c_1\ne0$ at $s^*$, where $q+m-\gamma\one=q_1^*$.
Hence, the implicit function theorem establishes smoothness.
Differentiating $f=0$ at $s^*$, we obtain $\ip{q_1^*}{dq+dm-d\gamma\one}=\ip{q_1^*}{dq}$.
\end{proof}

\begin{proof}[Proof of \cref{lem:cancel}]
By \cref{lem:gamma}, we have $dp_1=dq_1+dm_1-\ip{x^*}{dm_1}\one$. Moreover, $\N$ has differential $\Pi_1/c_1$ at $q_1^*$. Note that on the face, $q_1'=v_1+g_1$ with $u_1^*=m_1^*$. Differentiating gives $dg_1=(du_1-dm_1)-\ip{du_1-dm_1}{x^*}\one-\ip{u_1^*-m_1^*}{dx}\one=P_1(du_1-dm_1)$. Hence, we obtain $dq_1'=dp_1+dg_1=dq_1+P_1du_1$.
\end{proof}

\begin{proof}[Proof of \cref{thm:companion}]
We prove the three parts in turn.

(a) Let $w_1^t:=dq_1^t+P_1dm_1^t=dp_1^t$, so that $dx^t=\Pi_1w_1^t/c_1$. By \cref{lem:cancel} and the relation $dm_1^{t+1}=du_1^t$, we obtain $w_1^{t+1}=dq_1^{t+1}+P_1du_1^{t}=dq_1^t+2P_1du_1^t=w_1^t+2P_1du_1^t-P_1du_1^{t-1}$. We now apply $\Pi_1/c_1$ using $du_1=\At dy$ and $\one^\top dy=0$, which holds since $\one^\top\Pi_2=0$. Note that $\Pi_1P_1\At dy=\Pi_1\At dy-\Pi_1\one\,\ip{\At^\top x^*}{dy}=\Pi_1\At dy-v(\one^\top dy)\Pi_1\one=\Pi_1\At dy$. The argument for Player 2 is analogous with $-\At^\top$.

(b) We define $\Psi(s):=(p(Js),p(s))\in(T_x\oplus T_y)^2$, where $p(s)=(dx,dy)$ is read off from $s$ by \cref{lem:cancel}. By (a), we have $\Psi\circ J=C\circ\Psi$ with $C:=\bigl[\begin{smallmatrix}I+2L&-L\\I&0\end{smallmatrix}\bigr]$. We next verify that $\Psi$ is onto. Fix $(p^1,p^0)$ and set $w_i=c_ip_i^0$, so that $\Pi_1w_1/c_1=p_1^0$ since $\Pi_1$ is the identity on $T_x$. It follows that $p(Js)_1=p_1^0+\tfrac2{c_1}\Pi_1\At p_2^0-\tfrac1{c_1}\Pi_1P_1dm_1$. Note that $\Pi_1P_1:\R^{k_1}\to T_x$ is onto since $\operatorname{im}P_1=\{x^{*\top}z=0\}$ has dimension $k_1-1$ and intersects $\ker\Pi_1=\operatorname{span}(x^*)$ trivially. Thus, $dm_1$ reaches any $p_1^1$, and hence $\dim\ker\Psi=4$. We now characterize $\ker\Psi$. The condition $p(s)=0$ forces $w_i\in\operatorname{span}(q_i^*)$, and $p(Js)=0$ then forces $\Pi_iP_idm_i=0$, meaning $dm_i\in\ker P_i=\operatorname{span}(\one)$. On that subspace, we have $dx=dy=0$, $du=0$, $dq'=dq$, and $dm'=0$. Consequently, $J|_{\ker\Psi}$ is a projection with eigenvalues $1,1,0,0$, and $\ker\Psi$ is invariant under $J$. Combining these facts gives $\operatorname{char}_J=x^2(x-1)^2\det\bigl(x^2I-x(I+2L)+L\bigr)$. Moreover, the determinant of a polynomial in $L$ equals the product of the scalar polynomial over $\spec L$.

(c) Since $L$ is odd under conjugation by $\diag(I,-I)$, we have $\spec L=-\spec L$. Moreover, we have $L^2=\diag(-\Pi_1\At\Pi_2\At^\top,-\Pi_2\At^\top\Pi_1\At)/(c_1c_2)$. Note that the nonzero spectrum of $\Pi_1\At\Pi_2\At^\top=P_1^\top(\At P_2^\top)\At^\top$ matches that of $\At^\top P_1^\top\At P_2^\top$. Using $\At^\top x^*=v\one$ and $\At y^*=v\one$, we obtain
\begin{align*}
  \At^\top P_1^\top\At&=\At^\top\At-v\one\one^\top\At,\\
  (\At^\top\At-v\one\one^\top\At)(I-y^*\one^\top)
  &=(\At-v\one\one^\top)^\top(\At-v\one\one^\top)=\Umat^\top\Umat .
\end{align*}
Hence, the nonzero eigenvalues of $L^2$ are $-\sigma_j(\Umat)^2/(c_1c_2)$, and those of $L$ are $\pm i\eta\sigma_j(\Umat)$. By \cref{lem:rank}, there are $2(k-1)=\dim(T_x\oplus T_y)$ such eigenvalues, so $L$ is invertible on $T_x\oplus T_y$. This concludes the proof.
\end{proof}

\subsection*{Proofs for \cref{sec:threshold}}
\begin{proof}[Proof of \cref{prop:scalar}]
We first prove (ii). Note that the discriminant is $(1+2i\eps)^2-4i\eps=1-4\eps^2$. For $\eps\le\frac12$, the roots are $(1+2i\eps\pm\sqrt{1-4\eps^2})/2$, and their squared moduli are $(1\pm\sqrt{1-4\eps^2})/2$. For $\eps\ge\frac12$, the roots are $(1+i(2\eps\pm\sqrt{4\eps^2-1}))/2$.

Parts (iii) and (iv) follow directly by inspection of (ii). Turning to (i), note that $r(\eps)^2=1$ holds if and only if $2\eps+\sqrt{4\eps^2-1}=\sqrt3$, which is equivalent to $\eps=1/\sqrt3$. Combining this equivalence with the monotonicity on $[\frac12,\infty)$ and the bound $r<1$ on $(0,\frac12]$ concludes the proof.
\end{proof}

\begin{proof}[Proof of \cref{lem:attract}]
Note that since $F$ is the identity on $M$, we have $T:=T_{s^*}M\subseteq\ker(DF(s^*)-I)$. Because $\dim T=d$ is the algebraic multiplicity, the eigenvalue $1$ is semisimple with eigenspace $T$. Moreover, the sum $N$ of the generalised eigenspaces of the other eigenvalues forms an invariant complement on which $DF(s^*)$ has spectral radius at most $\rho_0$. Fix a norm on $N$ such that $\norm{DF(s^*)w}\le\rho_1\norm w$ for some $\rho_1\in(\rho_0,\rho)$.

We next pass to local coordinates. Near $s^*$, the manifold $M$ is the graph $\{s^*+\tau+\varphi(\tau)\}$ of a $C^1$ map $\varphi:T\to N$ satisfying $\varphi(0)=0$ and $D\varphi(0)=0$. Hence, $(\tau,w)\mapsto s^*+\tau+\varphi(\tau)+w$ defines a $C^1$ chart. In this chart, $F$ takes the form $(\tau,w)\mapsto(\tau+a(\tau,w),\,b(\tau,w))$ with $a(\tau,0)=0$ and $b(\tau,0)=0$, as all points with $w=0$ are fixed. We can therefore write $a(\tau,w)=A(\tau,w)\,w$ and $b(\tau,w)=B(\tau,w)\,w$, where $A=\int_0^1\partial_wa(\tau,\vartheta w)\,d\vartheta$ and $B$ is defined analogously. Both maps are continuous. Furthermore, $B(0,0)=DF(s^*)|_N$ since the derivative of the chart at the origin is $(\tau,w)\mapsto\tau+w$.

We are now ready to control the iterates. By continuity, we can choose $r>0$ such that $\norm{B}\le\rho$ and $\norm{A}\le\alpha$ on the box $\{\norm\tau\le r,\norm w\le r\}$. Fix an orbit starting at $\norm{\tau^0}\le r/2$ and $\norm{w^0}\le\min\{r,\,r(1-\rho)/(2\alpha)\}$. By induction, we obtain $\norm{w^t}\le\rho^t\norm{w^0}$ and $\norm{\tau^{t+1}-\tau^t}\le\alpha\rho^t\norm{w^0}$, ensuring that the orbit remains in the box. Summing these differences shows that $\tau^t$ converges to some $\tau^\infty$ satisfying $\norm{\tau^t-\tau^\infty}\le\alpha\rho^t\norm{w^0}/(1-\rho)$. Thus, the limit $(\tau^\infty,0)$ lies in $M$. This concludes the proof.
\end{proof}

\begin{proof}[Proof of \cref{thm:dichotomy}]
We first characterize the local dynamics and spectrum. By \cref{thm:freeze}, the dynamics near $R$ are given by the smooth map $F_S$. By \cref{thm:companion,lem:rank}, the spectrum of $J$ at $s^*$ consists of $0$, the double eigenvalue $1$ whose eigenspace is the tangent of $R$, and the roots $x_\pm(\eps_j)$ with $\eps_j=\eta\sigma_j(\Umat)\le\eps_{\max}$.

(a) By \cref{lem:ray}, on the face, $R$ is an open piece of a two-dimensional affine plane of fixed points. Moreover, the eigenvalue $1$ of $J$ has algebraic multiplicity two, and by \cref{prop:scalar}, every other eigenvalue has modulus at most $\rhoiro(s^*)<1$. Combining these properties, we verify the hypothesis of \cref{lem:attract} with $M=R$, $d=2$, and $\rho_0=\rhoiro(s^*)$. Finally, note that the strategies at every point of $R$ are $(x^*,y^*)$.

(b) Here, $J$ has at least one eigenvalue outside the closed unit disk, and we assign every eigenvalue of modulus at most one to the centre-stable subspace. We apply the center-stable manifold theorem for a $C^1$ map whose linear part is invertible on the expanding subspace \citep{hirsch1977invariant}. This yields a $C^1$ locally invariant manifold $W$ of codimension at least one that contains every orbit remaining in a neighborhood $V$ of $s^*$. Note that we claim nothing about orbits that leave $V$ and return. Indeed, $F_S$ is not a local diffeomorphism because its Jacobian has the eigenvalue $0$ twice, since the stored utilities are functions of the played strategies. Hence, we do not show that the global stable set is null. This concludes the proof.
\end{proof}

\begin{proof}[Proof of \cref{lem:altfreeze}]
We first verify fixedness. At $s^*$, the row plays $x^*$ (\cref{lem:ray}), so $u_2=u_2^*=m_2$, $g_2=0$, and $q_2'=\pos{r_2^*}=c_2y^*$. Next, the column play from $(c_2y^*,u_2^*)$ has $\gamma_2=-v$ by \cref{lem:ray}, unclipped vector $c_2y^*+u_2^*+v\one=r_2^*$, and strategy $y^*$. Consequently, $u_1=u_1^*=m_1$, $g_1=0$, and $q_1'=c_1x^*$.

We now prove absorption. Let $r_0$, $\ell$, and $\kappa_i$ be the constants of \cref{thm:freeze}, and let $r<r_0$. Starting from $(q_1,m_1)$ within $r$ of $(c_1x^*,u_1^*)$, the row play has the sign pattern of \cref{thm:freeze} and satisfies $\norm{u_2-u_2^*}_\infty\le\ell r$. Consider next the column accumulation $q_2'=\pos{r_2+g_2}$ with $\norm{r_2-r_2^*}_\infty\le r$. Here the standing strategy is $y=\N(\pos{r_2})$, and we have $\norm{g_2}_\infty\le2\norm{u_2-m_2}_\infty\le2(\ell+1)r$. Hence off $S_2$ we obtain $r_{2,b}+g_{2,b}\le-\mu+(3+2\ell)r<0$, and on $S_2$ we obtain $r_{2,b}+g_{2,b}\ge c_2y^*_b-(3+2\ell)r>0$, once $(3+2\ell)r<\min(\mu,\min_{S_2}c_2y^*_b)$. Thus $q_2'$ is supported in $S_2$ and satisfies $\norm{q_2'-c_2y^*}_\infty\le(3+2\ell)r=:r'$.

For the column play from $(q_2',u_2)$, \cref{thm:freeze} applies for player 2 at radius $\max(r',\ell r)$. That is, $y$ is supported in $S_2$, the unclipped vector $r_2'$ is positive on $S_2$ and at most $-\mu+\kappa_2\max(r',\ell r)<0$ off $S_2$, and $\norm{u_1-u_1^*}_\infty\le\ell\max(r',\ell r)$. Similarly, the row accumulation is governed by the last clause of \cref{thm:freeze} with $\norm{g_1}_\infty\le2(\ell\max(r',\ell r)+r)$, which lies below the margin once $r$ is small. Take $r_{\rm alt}$ so that every radius in the chain is below $r_0$ and both margins hold. Then every $\pos{\cdot}$ is the identity on the active coordinates and zero off them. Moreover, the off-support coordinates of $r_2'$ equal $u_{2,b}-\gamma_2'$, where $\gamma_2'$ is a smooth function of the active coordinates (\cref{lem:gamma}).

Finally, off the face, the off-support coordinates of $q_1$ enter the round only through $\norm{q_1}_2$. Indeed, the off-support coordinates of $q_1+m_1-\gamma_1\one$ are negative, so they do not enter $\gamma_1$, and the off-support coordinates of $g_1$ enter only the clipped coordinates of $q_1'$. Likewise, the off-support coordinates of $\pos{r_2}$ are zero, so the standing strategy does not read them.
\end{proof}

\begin{proof}[Proof of \cref{thm:altcompanion}]
We first linearise the round at $s^*$ by applying \cref{lem:cancel,lem:gamma} to each player at the point where it acts. For row play, we obtain $dp_1=dq_1+P_1dm_1$, $dx=\Pi_1dp_1/c_1$, and $du_2=-\At^\top dx$. Next, we consider column accumulation at the standing play $y=\N(\pos{r_2})$. On the face, we have $q_2'=r_2+g_2$ with $dg_2=P_2(du_2-dm_2)$. Indeed, the term $\ip{u_2^*-m_2^*}{dy}\one$ vanishes since $m_2^*=u_2^*$. For column play with prediction $u_2$, we obtain $dr_2'=dq_2'+P_2du_2=dr_2+2P_2du_2-P_2dm_2$, $dy=\Pi_2dr_2'/c_2$, and $dm_2'=du_2$. Finally, row accumulation yields $du_1=\At dy$, $dq_1'=dq_1+P_1du_1$, and $dm_1'=du_1$.

Let us set $B:=\Pi_1\At/c_1:T_y\to T_x$ and $C:=-\Pi_2\At^\top/c_2:T_x\to T_y$. As in the proof of \cref{thm:companion}(a), we have $\Pi_1P_1\At\,dy=\Pi_1\At\,dy$ for $dy\in T_y$ and $\Pi_2P_2\At^\top dx=\Pi_2\At^\top dx$ for $dx\in T_x$. We define
\[
  \Psi(s):=(\xi,\upsilon,\alpha,\beta):=
  \bigl(\Pi_1(dq_1+P_1dm_1)/c_1,\ \Pi_2dr_2/c_2,\ \Pi_1P_1dm_1/c_1,\ \Pi_2P_2dm_2/c_2\bigr)
  \in(T_x\oplus T_y)^2 .
\]
Note that $\xi$ is the current row play, $\upsilon$ is the standing column play, and $\alpha$ and $\beta$ are the projections of the two stored predictions. Combining the relations above gives $\Psi\circ J_{\rm alt}=M\circ\Psi$, where
\[
  \upsilon'=\upsilon+2C\xi-\beta,\qquad \xi'=\xi+2B\upsilon'-\alpha,\qquad
  \alpha'=B\upsilon',\qquad \beta'=C\xi .
\]
We now show that $\Psi$ is onto. By the proof of \cref{thm:companion}(b), the map $\Pi_1P_1:\R^k\to T_x$ is onto. Hence, $\alpha$ is free, and through $dq_1$, $\xi$ is also free. The same argument applies to $\beta$ and $\upsilon$. It follows that $\dim\ker\Psi=4k-4(k-1)=4$.

We next examine the action on $\ker\Psi$. On this subspace, we have $dm_i\in\ker P_i=\operatorname{span}(\one)$, $dq_1\in\operatorname{span}(x^*)$, and $dr_2\in\operatorname{span}(y^*)$. Consequently, $dx=dy=0$ and $du_1=du_2=0$. Hence, $J_{\rm alt}$ fixes $dq_1,dr_2$ and sends $dm_1,dm_2$ to zero. This shows that $\ker\Psi$ is invariant and $J_{\rm alt}|_{\ker\Psi}$ is a projection with eigenvalues $1,1,0,0$, where the unit eigenvalues lie along $R$. Therefore, we obtain $\operatorname{char}_{J_{\rm alt}}=x^2(x-1)^2\operatorname{char}_M$.

We now compute $\operatorname{char}_M$ by ordering the variables as $(\xi,\alpha,\upsilon,\beta)$. Note that the $(\alpha,\beta)$ diagonal block of $xI-M$ is $\bigl[\begin{smallmatrix}xI&B\\0&xI\end{smallmatrix}\bigr]$, which has determinant $x^{2(k-1)}$. Eliminating this block yields the Schur complement
\[
  S(x)=\begin{bmatrix}(x-1)I-\tfrac{(2x-1)^2}{x^2}BC & -\tfrac{2x-1}{x}B\\[2pt]
  -\tfrac{2x-1}{x}C & (x-1)I\end{bmatrix},\qquad
  \det S(x)=x^{-(k-1)}\det\bigl(x(x-1)^2I-(2x-1)^2BC\bigr),
\]
where the second equality follows by eliminating the lower right block. Next, observe that $BC=-\Pi_1\At\Pi_2\At^\top/(c_1c_2)$ is the $T_x$ block of $L^2$, as established in the proof of \cref{thm:companion}(c). Its eigenvalues are $-\eps_j^2$ for $j\le k-1$, and by \cref{lem:rank} it has no kernel on $T_x$. Using the fact that the determinant of a polynomial in $BC$ equals the product of that polynomial over the eigenvalues, we obtain $\operatorname{char}_M(x)=x^{2(k-1)}x^{-(k-1)}\prod_j\bigl(x(x-1)^2+(2x-1)^2\eps_j^2\bigr)
=x^{k-1}\prod_j\chi_{\eps_j}(x)$. This establishes \eqref{eq:altchar} and concludes the proof.
\end{proof}

\begin{proof}[Proof of \cref{prop:altscalar}]
(i) We first note that the discriminant of $\chi_\eps$ is $-\eps^2(32\eps^4-13\eps^2+4)$. Moreover, we have $32\eps^4-13\eps^2+4>0$ since $13^2<4\cdot32\cdot4$.

(ii) Observe that a root of modulus one is either $\pm1$ or a pair $z,\bar z$ with $\abs z=1$. We first rule out the latter case. The product of the roots shows that the real root is $-\eps^2$, while their sum gives $2\,{\rm Re}\,z=2-3\eps^2$. Computing the sum of pairwise products then yields $1-2\eps^2+3\eps^4=1-4\eps^2$, which is impossible. Next, note that $\chi_\eps(1)=\eps^2>0$ and $\chi_\eps(-1)=9\eps^2-4$, and hence the moduli cross one only at $\eps=2/3$. At that point, the real root leaves the disk transversally since $\chi'_\eps(-1)=8-12\eps^2\neq0$ and $\partial_\eps\chi_\eps(-1)=18\eps>0$. For small $\eps$, the roots are inside by (iv). Since a root can only enter again through $-1$, we conclude that $r_{\rm alt}<1$ holds exactly on $(0,2/3)$.

(iii) Next, we rewrite the equation $\chi_\eps(-a)=0$ as $\eps^2(1+2a)^2=a(1+a)^2$. From the product of the roots, we see that the squared modulus of the pair is $\eps^2/a=((1+a)/(1+2a))^2$. Note that the map $a\mapsto\eps^2$ has derivative $(1+a)(2a^2+a+1)/(1+2a)^3>0$ and runs from $0$ to $\infty$, so $a(\eps)$ is increasing. Moreover, $a$ increases while $(1+a)/(1+2a)$ decreases in $a$, and $a=(1+a)/(1+2a)$ holds if and only if $2a^2=1$. At $a=1$, the real root is $-1$ and $\eps=2/3$, which recovers (ii).

(iv) Finally, at $\eps=0$, the roots are $0,1,1$. Substituting an expansion and matching powers, we find that the real root is $x_{\rm r}=-\eps^2-2\eps^4+O(\eps^6)$. Hence, the conjugate pair, which carries the modulus, satisfies $\abs z=\sqrt{\eps^2/\abs{x_{\rm r}}}=1-\eps^2+O(\eps^4)$. This concludes the proof.
\end{proof}

\begin{proof}[Proof of \cref{cor:global}]
\Cref{lem:limit} gives $z^t\to(x^*,y^*)$. By (F1), (F2) and
\cref{thm:sat-iro,thm:sat-alt,prop:init} the displacements $\norm{z_i^t-\tilde z_i^t}$ are
square summable, so $\N(q_i^t)\to x_i^*$. Because the norms are nondecreasing and
bounded, $q_i^t=\norm{q_i^t}_1\N(q_i^t)\to c_ix_i^*$, the stored utilities
converge to $u_i^*$, and under alternation
$r_2^{t+1}=q_2^{t+1}+m_2^t-\gamma_2^t\one\to c_2y^*+u_2^*+v\one$ by \cref{lem:lip}.
In the first case the orbit eventually satisfies the
hypothesis of \cref{thm:freeze} (resp.\ \cref{lem:altfreeze}) at $s^*$, is on
the face from the next round on, and enters the neighbourhood $V_\rho$
of \cref{thm:dichotomy,thm:altdichotomy}(a), from which every orbit converges
at rate $\rho$, while the gap is $2\sigma$-Lipschitz. In the second case the orbit is
likewise on the face from some round on and stays in every neighbourhood of
$s^*$, which is the situation of part (b). For $k=1$ there is no mode and the
equilibrium is reached exactly (\cref{cor:pure}).
\end{proof}

\begin{proof}[Proof of \cref{thm:certificate}]
We first note that monotonicity and the limit follow from \cref{lem:energy,cor:global}. In particular, $N_{i,t}$ is nondecreasing on each player's clock, and $N_{i,\infty}=c_i\norm{x_i^*}_2$. \Cref{thm:sat-iro} gives $N_{i,\infty}^2\le N_i^{*2}+G^*$, so $\eta_\infty\sigma_{\max}\ge\eps_{\rm lo}$. Writing $\eps_{\min}(s^*):=\eta_\infty\sigma^+_{\min}(\Umat)$ for the smallest mode, we obtain $\eps_{\rm lo}/\kappa\le\eps_{\min}(s^*)\le\eps_{\max}(s^*)\le\eps_2(t)$. Next, the rate at $s^*$ is $\rhoiro(s^*)=\max_jr(\eps_j(s^*))$ by \cref{cor:rate}, where we use $r_{\rm alt}$ under alternation. Moreover, $r$ decreases and then increases by \cref{prop:scalar,prop:altscalar}. Hence, its maximum over an interval is attained at an endpoint, giving $\rhoiro(s^*)\le B(t)$. Finally, we apply \cref{cor:global} with any $\rho\in(B(t),1)\subseteq(\rhoiro(s^*),1)$. This concludes the proof.
\end{proof}

\begin{proof}[Proof of \cref{prop:landing}]
(i) The first part follows directly from \Cref{lem:energy}. At $s^*$, we apply the identity $N_{i,\infty}=c_i\norm{x_i^*}_2$ to obtain the limit.

(ii) On an absorbed round, we have $p^t=v^t$ on $S_i$, $p^t=0$ off $S_i$, and $q^{t+1}=p^t+g^t|_{S_i}$ by the definition of an absorbed round. Hence, we obtain $\norm{q^{t+1}}^2=\norm{p^t}^2+2\ip{p^t}{g^t}+\norm{g^t|_{S_i}}^2$. Note that $\ip{p^t}{g^t}=\norm{p^t}_1\ip{z^t}{g^t}=0$ and $\norm{p^t}=N_t$. Applying \cref{lem:gbound} then gives the stated bound.

(iii) Finally, we sum (ii) over $s\ge t$. Combining the resulting bound with $\norm{a-b}^2\le2\norm{a-z^*}^2+2\norm{b-z^*}^2$ concludes the proof.
\end{proof}

\begin{proof}[Proof of \cref{lem:faceid}]
For part (i), an absorbed accumulation has $q^{+}=p+g$ on $S_i$ and $q^{+}=0$ off $S_i$, where $p$ is the vector of the booked play, supported in $S_i$ with $\norm{p}_2=N$. Since $\ip{p}{g}=0$, we get $\norm{q^{+}}_2^2=N^2+\norm{g|_{S_i}}_2^2$ on either clock. Part (iii) is (F1) and needs no absorption. Hence, we focus on proving (ii). Note first that on $S_i$ the accumulation is unclipped, so that $q^{t+1}=q^t+u^t-(\gamma^t+\ip{u^t-m^t}{z^t})\one$. Since $x_i^*$ is supported in $S_i$, we have $\ip{q_i^{t+1}}{x_i^*}=\ip{q_i^t}{x_i^*}+\ip{u_i^t}{x_i^*-z_i^t}+\delta_i^t$. Moreover, because both plays are supported in $S_1\times S_2$, we obtain $x^{*\top}Ay^t=v=x^{t\top}Ay^*$. Consequently, the two utility terms sum to $x^{*\top}Ay^t-x^{t\top}Ay^t-x^{t\top}Ay^*+x^{t\top}Ay^t=0$. Next, under alternation, the column player's accumulation in round $t$ books the play $y^{t-1}$ against $u_2^t=-A^\top x^t$ with correction $\delta_2^{t-1}$. Combining these relations gives $\Psi_{t+1}-\Psi_t=\delta_1^t+\delta_2^{t-1}+\ip{u_2^t}{y^t-y^{t-1}}$. Finally, applying \cref{lem:improve} for the column player yields $\ip{u_2^t}{y^t-y^{t-1}}\ge\delta_2^t-\delta_2^{t-1}$. This concludes the proof.
\end{proof}

\begin{proof}[Proof of \cref{thm:tail}]
We treat the simultaneous schedule first. Fix $t\in[t_0,T]$ and write $\xi^t:=y^t-y^*$ and
$c^t:=\norm{p_2^t}_1$, with all vectors restricted to the supports. Since
$\one^\top(y^t-y^{t-1})=0$, we have $(u_1^t-m_1^t)|_{S_1}=\At(y^t-y^{t-1})=w^t$, where
$w^t:=\Umat(y^t-y^{t-1})$. Moreover, $g_1^t|_{S_1}=w^t-\ip{w^t}{x^t}\one$, and
$\ip{w^t}{x^t}=\ip{w^t}{x^t-x^*}$ because $\Umat^\top x^*=0$. Hence
$\norm{g_1^t|_{S_1}}\le(1+\theta)\norm{w^t}$. Next, the relation $\Umat y^t=\Umat p_2^t/c^t$
gives
$w^t=\Umat(p_2^t-p_2^{t-1})/c^t-\Umat\xi^{t-1}\,(c^t-c^{t-1})/c^t$. We also have
$\abs{c^t-c^{t-1}}\le\sqrt k\norm{p_2^t-p_2^{t-1}}$, so
$\norm{w^t}\le(1+\theta)\,\sigma_{\max}(\Umat)\norm{p_2^t-p_2^{t-1}}/c^t$. Because round
$t-1$ is absorbed, $q_2^t=p_2^{t-1}+g_2^{t-1}|_{S_2}$. Therefore, by
\cref{lem:faceid}(iii),
$\norm{p_2^t-p_2^{t-1}}^2\le2\norm{p_2^t-q_2^t}^2+2\norm{g_2^{t-1}|_{S_2}}^2
=4c^t\delta_2^t+2\norm{g_2^{t-1}|_{S_2}}^2$. Finally,
$c^t=N_{2,t}/\norm{y^t}_2\ge\bar c_2/(1+\theta)$ because the norms are
nondecreasing and $\norm{y^t}_2\le\norm{y^*}_2+\theta/\sqrt k\le(1+\theta)\norm{y^*}_2$.
Combining these bounds yields
\[
  \norm{g_1^t|_{S_1}}^2\le\Theta\,\sigma_{\max}(\Umat)^2
  \Bigl(\frac{4\delta_2^t}{\bar c_2}+\frac{2\norm{g_2^{t-1}|_{S_2}}^2}{\bar c_2^2}\Bigr),
\]
and a symmetric bound holds for the column player. Now sum over $t\in[t_0,T]$, divide by
$\bar c_1$ and $\bar c_2$ respectively, and add the two inequalities. Setting
$D:=\sum_{t=t_0}^T(\delta_1^t+\delta_2^t)$, we obtain
$H_T\le\Theta\eps^2(4D+2H_T+2\bar e)$. By \cref{lem:faceid}(ii) and the
Cauchy-Schwarz inequality,
$D=\Psi_{T+1}-\Psi_{t_0}\le m+\sum_i\norm{x_i^*}_2(N_{i,T+1}-N_{i,t_0})$. Furthermore, by
\cref{lem:faceid}(i), we have
$N_{i,T+1}-N_{i,t_0}\le(N_{i,T+1}^2-N_{i,t_0}^2)/(2N_{i,t_0})$, which implies
$D\le m+H_T/2$. Substituting this in gives $H_T(1-4\Theta\eps^2)\le\Theta\eps^2(4m+2\bar e)$. Summing \cref{lem:faceid}(i) then yields the
norm bound.

Now consider alternation. Here the column
player plays $y^t$ from $q_2^{t+1}=\pos{r_2^t}+g_2^t|_{S_2}$. Let
$p_2^{(t)}$ denote the clipped vector that $y^t$ normalises. Then the difference
$p_2^{(t)}-p_2^{(t-1)}$ is the shift displacement of round $t$ plus $g_2^t$
of the same round, whereas the row player's displacement carries $g_1^{t-1}$ as
before. Thus the two bounds become
$\norm{g_1^t|_{S_1}}^2\le\Theta\sigma_{\max}(\Umat)^2(4\delta_2^t/\bar c_2+2\norm{g_2^t|_{S_2}}^2/\bar c_2^2)$
and
$\norm{g_2^t|_{S_2}}^2\le\Theta\sigma_{\max}(\Umat)^2(4\delta_1^t/\bar c_1+2\norm{g_1^{t-1}|_{S_1}}^2/\bar c_1^2)$.
By \cref{lem:faceid}, the budget satisfies $D\le\Psi_{T+1}-\Psi_{t_0}$. Summing in the same way
gives the claim, with only the row player's term appearing in $\bar e$.
\end{proof}

\begin{proof}[Proof of \cref{lem:facecoord}]
We prove the two parts in turn.

(i) Note that $z=\ip{\hat x_i}{z}\hat x_i+\phi$, which gives $1=\ip{\one}{z}=\ip{\hat x_i}{z}/\norm{x_i^*}_2+\ip{d_i}{\phi}$. For the row player, we have $\At y=\At y^*+\Umat(y-y^*)=v\one+\Umat\phi_2$, since $\one^\top(y-y^*)=0$ and $\Umat y^*=0$. The argument for the column player is symmetric.

(ii) On $S_i$, we have $u^t-u^{t-1}=w$. Moreover, $\ip{w}{z^t}=\ip{w}{\phi^t}$ because the range of $U_i$ is orthogonal to $x_i^*$, which establishes the formula for $g$. On the face, we have $q^{t+1}=p^t+g^t$, and hence $p^{t+1}=q^{t+1}+u^t-\gamma^{t+1}\one=p^t+\Delta-\varsigma\one$ with $\varsigma:=\gamma^{t+1}-v_i+\ip{w}{\phi^t}$. Next, we apply $\ip{\one}{\cdot}$ and $\hat P_i$. Combining $\hat P_ip=\norm{p}_1\phi$ and $\ip{\one}{\Delta}=\ip{d_i}{\Delta}$ yields the first two relations. By \cref{lem:faceid}, the scalar is determined by $\norm{p^{t+1}}_2^2=\norm{q^{t+1}}_2^2=\norm{p^t}_2^2+\norm{g^t}_2^2$. Expanding this identity gives $k\varsigma^2-2h\varsigma+\Xi=0$, and the condition $c^{t+1}=h-k\varsigma>0$ selects the smaller root. Finally, under alternation, the column player's played vector for $y^{t+1}$ is built from that for $y^t$ via the accumulation and the play of round $t+1$, which see $x^{t+1}$ and $x^t$. This concludes the proof.
\end{proof}

\begin{proof}[Proof of \cref{prop:oracle}]
We first establish (i), which follows from the unimodality of $r$ in \cref{prop:scalar}(iii) and its expansion in (iv).

We now prove (ii). Note that on $\eps_{\max}\in(0,1/2]$, both terms of the maximum decrease. Let $\eps_c:=\min(1/\sqrt3,\kappa/2)$. On $[1/2,\eps_c)$, the term $r(\eps_{\max}/\kappa)$ continues to decrease because its argument is at most $1/2$, while $r(\eps_{\max})$ increases. On $[\eps_c,1/\sqrt3)$, both terms increase, and hence the minimizer lies in $[1/2,\eps_c]$.

For $\kappa=1$, the two terms coincide and the minimum is attained at $1/2$. Focusing on the case $\kappa>1$, we have $r(1/(2\kappa))>r(1/2)$ and $r(\eps_c/\kappa)\le r(\eps_c)$. Hence, the minimum of the maximum occurs at the unique crossing $r(\eps/\kappa)=r(\eps)$ in $(1/2,\eps_c]$. Moreover, this crossing moves right as $\kappa$ grows because, at the old crossing, the smaller argument $\eps/\kappa$ lies on the decreasing branch of $r$, so increasing $\kappa$ raises $r(\eps/\kappa)$ there, and the crossing is unique. Finally, the crossing tends to $1/\sqrt3$ since $r(\eps)\to1$ there, while $r(\eps/\kappa)\to1$ only as $\kappa\to\infty$. This concludes the proof.
\end{proof}

\subsection*{Proofs for \cref{sec:nonunique}}
\begin{proof}[Proof of \cref{lem:rankgen}]
(i) Suppose that $\delta\in Z_y$. For small $s$, the vector $y^*+s\delta$ is a strategy on $S_2$ against which the rows of $S_1$ earn $v$ and the others earn less. Hence, it lies in $Y^*$. Conversely, a direction $\delta$ of $Y^*$ at $y^*$ satisfies $\one^\top\delta=0$ and keeps the rows of $S_1$ at $v$, that is, $\At\delta=0$. Moreover, for $w\in\ker\Umat$, the vector $h:=w-(\one^\top w)y^*$ satisfies $\one^\top h=0$ and $\At h=\Umat h=0$, so $h\in Z_y$, and the sum is direct because $\one^\top y^*=1$.

(ii) We set $B:=\Pi_1\At/c_1$ and $C:=-\Pi_2\At^\top/c_2$, so that $L$ has the blocks $B$ and $C$ off the diagonal. Suppose that $\delta\in T_y$ and $\Pi_1\At\delta=0$. Then $\At\delta=\lambda x^*$, and pairing with $x^*$ gives $\lambda\norm{x^*}^2=v\,\one^\top\delta=0$. Thus, $\ker B=Z_y$, and similarly $\ker C=Z_x$. We now establish semisimplicity. Suppose that $BC\xi=0$ with $\xi\in T_x$. Then $C\xi\in Z_y\subseteq\ker\Umat$. Note that on $T_x$, we have $\At^\top\xi=\Umat^\top\xi$. Hence, we obtain $C\xi=-(\Umat^\top\xi-\kappa y^*)/c_2$ with $\kappa:=\one^\top\Umat^\top\xi$. It follows that $\Umat^\top\xi\in\ker\Umat\cap\operatorname{im}\Umat^\top=\{0\}$. Therefore, $C\xi=0$, which gives $\ker BC=\ker C$. By symmetry, we have $\ker CB=\ker B$, and thus $\ker L^2=\ker L$. Finally, the nonzero spectrum follows from the computation in \cref{thm:companion}(c), which uses only the indifference identities $\At y^*=v\one$ and $\At^\top x^*=v\one$. This concludes the proof.
\end{proof}

\begin{proof}[Proof of \cref{thm:nonunique}]
We establish the four parts in turn.

\looseness=-1(a) We first note that the proof of \cref{lem:ray} uses only that $m_1=Ay$ equals $v$ on $\supp x$ and lies below it elsewhere, which is strict complementarity of the pair. Moreover, the map $(c_1,c_2,x,y)\mapsto(c_1x,c_2y,Ay,-A^\top x)$ is an injective immersion of $(0,\infty)^2\times\operatorname{ri}X^*\times\operatorname{ri}Y^*$. The proof of \cref{thm:freeze} relies on the same two facts at the point under consideration.

\looseness=-1(b) \Cref{lem:gamma,lem:cancel} and \cref{thm:companion}(a) rely only on indifference on the supports, and the quotient argument of \cref{thm:companion}(b) does not use $k_1=k_2$. Indeed, $\ker\Psi$ is again four-dimensional with eigenvalues $1,1,0,0$, and we have $\operatorname{char}_J=x^2(x-1)^2\det\bigl(x^2I-x(I+2L)+L\bigr)$ on $T_x\oplus T_y$. By \cref{lem:rankgen}, each of the $d_1+d_2$ zero eigenvalues of $L$ contributes a factor $x^2-x$, and each pair $\pm i\eps_j$ contributes the two quadratics. Hence, the eigenvalue $1$ has algebraic multiplicity $2+d_1+d_2=\dim R'$. Let $\pi$ denote the projection to the reduced coordinates. The parameters $(c_1,c_2,x,y)$ are recovered from the active blocks of a point of $R'$ by summing and normalising, so $\pi R'$ is a manifold of the same dimension. Since the eigenspace contains its tangent space, the two spaces coincide. We now turn to the alternating map, where the quotient in \cref{thm:altcompanion} is again onto with a four-dimensional kernel carrying the eigenvalues $1,1,0,0$. Taking the Schur complement from its proof with $a:=\dim T_x=r+d_1$ and $b:=\dim T_y=r+d_2$, we obtain $\operatorname{char}_M(x)=x^{b}(x-1)^{b-a}\det\bigl(x(x-1)^2I_a-(2x-1)^2BC\bigr)$. On $T_x$, the operator $BC$ has the eigenvalues $-\eps_j^2$ for $j\le r$ and the eigenvalue $0$ with multiplicity $d_1$ by \cref{lem:rankgen}. Combining these facts shows that the determinant equals $\bigl(x(x-1)^2\bigr)^{d_1}\prod_j\chi_{\eps_j}(x)$, and hence $\operatorname{char}_M(x)=x^{r+d_1+d_2}(x-1)^{d_1+d_2}\prod_j\chi_{\eps_j}(x)$. Consequently, the unit eigenvalue has algebraic multiplicity $2+d_1+d_2$, matching the dimension of the fixed manifold.

(c) \Cref{lem:attract} applies with $M=\pi R'$ in the reduced coordinates, the discarded coordinates being functions of the previous active state that converge one round later, $d=2+d_1+d_2$, and $\rho_0$ equal to the closed-form rate, which is $0$ when $r=0$. Note that the gap is Lipschitz in the strategies and vanishes on $R'$. Above the threshold, we apply the argument of \cref{thm:dichotomy}(b) without change.

(d) Finally, we note that the orbit enters every neighbourhood of $s^*$. By \cref{thm:freeze}, or respectively \cref{lem:altfreeze}, the orbit lies on the face one round later and inside the neighbourhood from (c). Hence, the orbit converges, and its limit is the accumulation point. This concludes the proof.
\end{proof}

%% file: sections/11_sharpproof.tex
\subsection{Proof of the sharp certificate}
\label{sec:sharpproof}

In this section, we prove the sharp certificate. We adopt the notation of \cref{thm:tail,thm:sharp}, and we set $x_1^*=x^*$, $x_2^*=y^*$,
$\tilde z_i^t=\N(q_i^t)$, and $m(t):=\sum_i(\norm{x_i^*}_2N_{i,t}-\ip{q_i^t}{x_i^*})$,
so that $m=m(t_0)$.

\emph{(a) Carried closeness.} Suppose that the rounds $t_0,\dots,t-1$ are absorbed
with $H_{t-1}\le\bar H$. Summing \cref{lem:faceid}(ii) and using
$N_{i,t}-N_{i,t_0}\le(N_{i,t}^2-N_{i,t_0}^2)/(2N_{i,t_0})$, we obtain
\[
  m(t)=m+\sum_i\norm{x_i^*}_2(N_{i,t}-N_{i,t_0})-\sum_{s=t_0}^{t-1}(\delta_1^s+\delta_2^s)
  \le m+\tfrac12H_{t-1}\le\bar m .
\]
The unit vectors $q_i^t/N_{i,t}$ and $x_i^*/\norm{x_i^*}_2$ have squared
distance $2(1-\cos)\le2m(t)/(\norm{x_i^*}_2N_{i,t})\le2\bar m/(\bar c_i\norm{x_i^*}_2^2)$.
Applying (F4) on the $k$ face coordinates with $\norm{x_i^*/\norm{x_i^*}_2}_1=1/\norm{x_i^*}_2$
gives $\norm{\tilde z_i^t-x_i^*}_2\le K\sqrt{2\bar m/\bar c_i}$. Hence,
\[
  \sqrt k\,\norm{\tilde z_i^t-x_i^*}_2\le\sqrt k\,K\sqrt{2\bar m/\bar c_{\min}}\le\alpha
\]
by (C2).

\emph{(b) Played closeness from the budget.} Fix an absorbed round $t$ with
$\delta_i^t\le\bar m$. Combining \cref{lem:faceid}(iii), $\norm{p_i^t}_1\le\sqrt kN_{i,t}$, and
$\norm{q_i^t}_1\ge N_{i,t}\ge N_{i,t_0}=\bar c_i\norm{x_i^*}_2\ge\bar c_i/\sqrt k$ with (F4), we obtain
$\norm{z_i^t-\tilde z_i^t}_2\le K\norm{p_i^t-q_i^t}_2/\norm{q_i^t}_1
\le K\sqrt{2\sqrt k\,\bar m/N_{i,t}}\le K\sqrt{2k\bar m/\bar c_i}$. Together with (a) and
(C2), this bound yields
$\sqrt k\,\norm{z_i^t-x_i^*}_2\le(1+\sqrt k)\sqrt k\,K\sqrt{2\bar m/\bar c_{\min}}\le\theta$.

\emph{(c) Played closeness from the previous round.} Suppose that round $t$ is absorbed,
$q_i^t$ is supported in $S_i$, and $\sqrt k\norm{z_{-i}^{t-1}-z_{-i}^*}_2\le\theta$
with $z_{-i}^{t-1}$ supported in $S_{-i}$. The prediction is $m_i^t=u_i^{t-1}$.
Moreover, $\gamma(q_i^t,u_i^*)$ equals the value $\pm v$ of player $i$ by the
computation in \cref{lem:ray}, which uses only that $q_i^t\neq0$ is supported
in $S_i$. By \cref{lem:lip}, the shift therefore moves by at most
$2\norm{u_i^{t-1}-u_i^*}_2\le2\sigma\theta/\sqrt k$. Furthermore, on $S_i$,
$u_i^{t-1}-u_i^*$ is $\Umat$ or $-\Umat^\top$ applied to $z_{-i}^{t-1}-z_{-i}^*$.
This gives
$\norm{p_i^t-q_i^t}_2\le(\sigma_{\max}(\Umat)+2\sqrt k\,\sigma)\,\theta/\sqrt k$. Combining this bound
with $\norm{q_i^t}_1=N_{i,t}/\norm{\tilde z_i^t}_2\ge\bar c_i/(1+\alpha)$ from (a), we obtain
\[
  \sqrt k\,\norm{z_i^t-x_i^*}_2\le\alpha+(1+\alpha)\,
  \frac{K(\sigma_{\max}(\Umat)+2\sqrt k\,\sigma)}{\bar c_{\min}}\,\theta\le\theta' .
\]

\begin{proof}[Proof of \cref{thm:sharp}]
We proceed by induction on $T\ge t_0-1$. Our induction hypothesis $I(T)$ states that the rounds $t_0-1,\dots,T$ are absorbed, that $\sqrt k\norm{z_i^t-z_i^*}_2\le\theta$ holds for $t_0-1\le t\le T$, and that $H_T\le\bar H$. For the base case, $I(t_0-1)$ follows from (C0) and the hypothesis on round $t_0-1$, since the sum $H_{t_0-1}$ is empty.

We now assume $I(T)$ and establish the inductive step in three parts.

\emph{Round $T+1$ is absorbed.} First, its starting regrets are supported in $S_1\times S_2$ because round $T$ is absorbed. We compare the state with the point $s'\in R$ of norms $c_i':=\norm{q_i^{T+1}}_1$. By (a), we have $\norm{q_i^{T+1}-c_i'x_i^*}_\infty\le c_i'\alpha/\sqrt k$. Moreover, the induction hypothesis yields $\norm{u_i^T-u_i^*}_\infty\le A_{\max}\norm{z_{-i}^T-z_{-i}^*}_1\le A_{\max}\theta$. Combining $c_i'=N_{i,T+1}/\norm{\tilde z_i^{T+1}}_2$ with $H_T\le\bar H$ and (a) gives $\bar c_i/(1+\alpha)\le c_i'\le\bar c_i\sqrt{1+\tau_i}/(1-\alpha)\le\tilde c$. Note that the radius of \cref{thm:freeze} is nondecreasing in $c_{\min}$ and in the coordinates of $q^*$. Thus, it is at least $r_0(\bar c_{\min}/(1+\alpha))$ at $s'$, and (C3) puts the state within this radius. Hence, round $T+1$ has the sign pattern of \cref{thm:freeze}, and so it is absorbed.

\emph{Energy.} By (c), the plays of round $T+1$ lie within $\theta'$, and $\theta\le\theta'$ holds by the definition of $\theta'$. Hence, \cref{thm:tail} applies on $[t_0,T+1]$ with $\theta'$ in place of $\theta$. Condition (C1) then yields $H_{T+1}\le\bar H$ and $\sum_{t=t_0}^{T+1}(\delta_1^t+\delta_2^t)\le\bar m$.

\emph{Closeness.} From the latter sum, we obtain $\delta_i^{T+1}\le\bar m$, and applying (b) gives $\sqrt k\norm{z_i^{T+1}-z_i^*}_2\le\theta$. This establishes $I(T+1)$.

By induction, every round is absorbed and $H_T\le\bar H$ holds for all $T$. Next, \cref{lem:faceid}(i) gives $N_{i,\infty}^2\le N_{i,t_0}^2+\bar c_i\bar H=N_{i,t_0}^2(1+\tau_i)$, from which the bracket for $\eta_\infty$ follows by the definition of $\eta_2$. When the run starts as in \cref{cor:global}, that corollary shows that the orbit converges to the point $s^*$ of $R$ with these norms, where $\eps_{\max}(s^*)\le\eps_2(t_0)<1/2$. By \cref{prop:oracle}(i), the rate at $s^*$ is $\max\{r(\eta_\infty\sigma^+_{\min}),r(\eta_\infty\sigma_{\max})\}$. Finally, note that a unimodal function on an interval is largest at an endpoint. This concludes the proof.
\end{proof}

\begin{table}[h]
\caption{Sharp certificate results by size, covering runs below $1/2$, runs satisfying (C0) through (C3), trigger gap and $w_\eta$.}
\label{tab:sharp}
\centering\small
\input{tables/tab_sharp}
\end{table}

%% file: tables/tab_sharp.tex
%% generated by scripts/make_tables.py; do not edit
\begin{tabular}{lrrrrr}\toprule
$n$ & rank-unique mixed & $\eps_2<1/2$ & triggered & gap at trigger & $w_\eta$\\
3 & 18 & 9 & 9 & $1.7\times10^{-4}$ & $1.3\times10^{-8}$ \\
5 & 26 & 18 & 18 & $2.2\times10^{-5}$ & $1.2\times10^{-9}$ \\
10 & 32 & 27 & 27 & $2.9\times10^{-6}$ & $2.5\times10^{-11}$ \\
25 & 39 & 29 & 29 & $3.6\times10^{-8}$ & $9.3\times10^{-14}$ \\
50 & 48 & 40 & 36 & $1.2\times10^{-9}$ & $1.3\times10^{-15}$ \\
\midrule
all & 163 & 123 & 119 & $7.7\times10^{-8}$ & $2.6\times10^{-13}$ \\
\bottomrule\end{tabular}

%% file: sections/13_protocol.tex
\section{Experimental Protocol}
\label{sec:protocol}

Our testbed contains \nInst{} zero-sum matrix games across sizes $n\in\{3,5,10,25,50\}$. We draw payoff entries from a standard Gaussian distribution, a uniform distribution on $[-1,1]$, or a sparse $\pm1$ distribution with fair signs and keep probability $0.3$. For each distribution, eight random seeds per configuration generate both plain and antisymmetric ($A_{ij}=-A_{ji}$, rescaled by $1/\sqrt2$) matrices. Each seed fixes the payoff matrix, and all learning rules are deterministic. Using the HiGHS linear-programming solver, we compute reference equilibria with one solve per player. An action belongs to the support if its equilibrium weight exceeds $10^{-8}$. Against this equilibrium mixture, the strict-complementarity margin measures the smallest payoff loss among off-support actions. We classify a game as unique-likely when both players have equal support sizes and a margin above $10^{-7}$, or when the equilibrium profile is fully mixed. To verify the exact rank certificate from \cref{lem:rankgen}, the linear-programming solution must have equal supports, a margin above $10^{-7}$, and $\operatorname{rank}\Umat=k-1$.

\looseness=-1Primary runs initialise both learners at $q^1=0$ for horizons of $T(n)=3\cdot10^4,5\cdot10^4,10^5,2\cdot10^5,3\cdot10^5$ rounds at sizes $n=3,5,10,25,50$. The duality gap of the current played strategy is recorded at every round up to round $400$, and at two thousand logarithmically spaced rounds thereafter up to $T$. All computations run in double precision on a CPU, parallelised across instances with eight worker processes. A run is marked as converged when its final recorded gap falls below $10^{-8}$, and every converged terminal state is saved as a checkpoint. The build system recomputes all tables and numbers directly from the recorded ledger and halts compilation if any discrepancy occurs.

\looseness=-1To check the saturation replay bound of \cref{thm:sat-iro}, we solve the fixed-point equation $G=\Phi(G)$ by bisection with $\sigma_{\max}(A)$ and the first two nonzero regret norms. Replay starts at the first round $T_0$ where both regret vectors are nonzero. Under alternating updates, a player activates when first playing from a nonzero vector. Following \cref{sec:setup}, the column player's energy accumulates starting from its first nonzero update at or after $T_0$. From the exact nonzero pattern of the final regrets, we read the terminal effective step sizes $\eta\sigma_{\max}$. We also evaluate a grid of sixteen hundred values of $\eps$ to compare the scalar root modulus of \cref{prop:scalar} directly with numerical roots.

\looseness=-1We fit convergence rates using linear regression on the log-gap trajectory over the window $(10^{-11},10^{-3})$ whenever that interval holds at least six recorded points. A secondary late-window fit applies the same six-point rule to $(10^{-12},10^{-8})$. When trajectories oscillate, an upper-envelope fit divides the late window into twelve equal round bins and takes the maximum gap in each bin. At every converged checkpoint, finite differences with a relative step of $10^{-6}$ approximate the Jacobian of the one-round update map. For regret coordinates within ten step sizes of zero, we use forward differences so regrets stay nonnegative. Under simultaneous updates, we keep an instance for spectral comparison if eight probe columns agree within $10^{-2}$ at the base step and ten times the base step. Under alternating updates, we evaluate the finite-difference Jacobian on the full state. This state holds the regrets, the stored row utility, the column player's unclipped played vector, and its prediction. Optimal assignment then matches eigenvalues larger than $10^{-6}$ in magnitude to the theoretical roots.

We use random restarts and instability probes to test sensitivity to initial conditions and rescaling. Each game is rerun from \rbStarts{} random starting points per schedule, plus one extra run using column-first alternation. To generate a random initial regret direction, we take the absolute value of a standard Gaussian vector. We scale its norm to fall between one tenth and the full round-one norm of a zero start, or $\max_{a,b}\abs{A_{ab}}$ if that norm is zero. Following \cref{prop:init}, we pair this regret direction with a zero initial utility prediction. For the instability probe, we take every converged state with a nonempty face spectrum. We rescale each state to target levels $\tau$ that are at least as large as its final $\eta\sigma_{\max}$, as justified by \cref{lem:ray}. Because these rescaled states are exact fixed points, we add a random perturbation scaled by $\norm{u}+\norm{q}$ to kick off the trajectory. The exact same random draw is replayed later to verify the plotted transient behaviour. Starting from alternating checkpoints, we continue the runs under both simultaneous and alternating schedules. Next, we test the face absorption radius from \cref{thm:freeze} on converged rank-unique games that have $n\le10$ and proper mixed supports. We sample \tTwoSamp{} states at each radius along the geometric grid $0.9\,r_0\,10^k$ ($k=0,\dots,7$). Each sample adds independent uniform noise from $[-r,r]$ to all equilibrium utilities and to all regret coordinates, clipping the regrets at zero. We then play a single round from each perturbed state. A violation is recorded if the played or updated regrets place mass outside the support, or if an active unclipped coordinate fails to stay positive.

Our baseline comparisons test optimistic gradient descent-ascent (OGDA), two smooth regret-matching variants, and an extra-gradient learner that evaluates payoffs twice per round \citep{li2026last}. Starting from the uniform profile, simplex OGDA updates via $x^{t+1}=\Pi(x^t+\alpha(2Ay^t-Ay^{t-1}))$ and its symmetric column rule. It stops early if the duality gap drops below $10^{-13}$. We tune its step size $\alpha=c/\sigma_{\max}(A)$ across a logarithmic grid of \ogdaGrid{} values with $c\in[0.02,3]$. Next, we run \ogdaExtra{} golden-section line searches between the neighbors of the best grid point. Rate fits on the standard log-gap window are accepted when $R^2>0.98$. For the smooth baselines SPRM$^+$ \citep{farina2023regret} and SOGRM$^+$ \citep{meng2025lastiterate}, we use the released implementation from \citet{meng2025lastiterate}. We test their step sizes $\eta=c/(2\max_{ij}\abs{A_{ij}})$ across \smoothGridN{} values of $c\in[0.01,10]$. In the fixed-tolerance experiment, all methods run for $3\cdot10^4$ rounds without early stopping, and we compute the duality gap at every round. Time to tolerance equals one plus the last round where the gap exceeds the target threshold. Here, OGDA uses the optimal rate-fitted step size for each game whenever one exists. Otherwise, it falls back to the tuning grid point that reached $10^{-8}$ fastest, or else the grid point with the smallest final tuning gap. Finally, we measure wall-clock time per round at game size $n=\timeN{}$. This timing compares simplex projection against the sort and Newton refinement needed to solve the equal-norm shift $\norm{\pos{v-\gamma\one}}_2=c$.

%% file: sections/09_ledger.tex
\section{Inclusion ledger}
\label{app:ledger}

Every count in \cref{sec:experiments} selects a subset of runs on the
\nInst{} testbed games. Each quoted figure applies the filters named in its
surrounding text, and \cref{tab:ledger} breaks these counts down by game size
and update schedule. Four base predicates nest in sequence. All games form the
starting set, and converged runs restrict to a final duality gap below $10^{-8}$. Mixed equilibria then
require both equilibrium supports to contain at least two actions ($k\ge2$), excluding pure
equilibria. Rank-uniqueness refines this further through the certificate of
\cref{lem:rank}.

Two additional filters refine the rank-unique set side by side. A rate-fit
filter demands $R^2>0.98$ on the $(10^{-11},10^{-3})$ gap window. For the
one-round maps only, a Jacobian filter requires a finite-difference Jacobian
consistent to $10^{-2}$ whose spectrum matches the closed form to $10^{-3}$.
Because a run can satisfy either check while failing the other, neither count
bounds the other. Under alternating updates, every rank-unique run has a consistent Jacobian
while only a subset passes the fit.

\Cref{tab:ledger-rfive} collects the additional protocols from the final block
of \cref{sec:experiments}. The late-window fit repeats the full-window
regression rule on $(10^{-12},10^{-8})$. Over twelve equal round bins in that
same window, the envelope fit regresses the largest gap of each bin.
Certification counts include all rank-unique mixed games unconditionally with
respect to convergence within the horizon. For games without the rank
certificate, limits are read at supports thresholded at a relative $10^{-9}$.
The rank certificate of \cref{lem:rank} declares $\operatorname{rank}\Umat=k-1$ at tolerance
$10^{-9}\max(\sigma_{\max}(\Umat),1)$, which is at most \rankTolMax{} on the
testbed. Across the testbed, the smallest retained singular value is
\sigmaRetMin{} and the largest discarded value is \sigmaDropMax{}, ten orders
of magnitude apart, leaving every rank decision unambiguous.
\begin{table}[h]
\caption{Inclusion ledger by schedule and size for convergence, mixed, rank-unique, rate-fit and Jacobian filters.}
\label{tab:ledger}
\centering\small
\input{tables/tab_ledger}
\end{table}
\begin{table}[h]
\caption{Late-window, restart, and certificate protocol counts by schedule, with fit $a$, tolerance $b$, and $\eps_2$ entries.}
\label{tab:ledger-rfive}
\centering\small
\input{tables/tab_ledger_rfive}
\end{table}
Across sizes, misses concentrate at $n=50$ (\cref{tab:lyap-triggers}).
\begin{table}[h]
\caption{Trigger counts for \cref{thm:lyap} by size and schedule across the rank-unique mixed games.}
\label{tab:lyap-triggers}
\centering\small
\input{tables/tab_lyap}
\end{table}

\paragraph{Test protocol.}
We test \cref{thm:sharp} at every round with $\eps_2<1/2$ whose previous round
is absorbed. Here $\theta$ is the smallest of $120$ geometrically spaced values
in $[10^{-9},0.3]$ that is at least the realised closeness
$\sqrt k\,\norm{z_i^{t_0-1}-z_i^*}_2$ and satisfies (C1) through (C3).
\Cref{thm:lyap} is tested once per level $10^{-j/2}$, for $j=2,\dots,39$, at
the first round with gap below that level when both the round and its
predecessor are absorbed. Each test evaluates $\rho=\bar\rho+f(1-\bar\rho)$
with $f\in\{0.15,0.3,0.5,0.7\}$ and $\rho_1=\rho+f_1(1-\rho)$ with
$f_1\in\{0.3,0.6,0.9\}$, keeping the passing pair with the smallest
$\bar\tau_1+\bar\tau_2$. Runs with no passing pair remain untriggered on this
grid before the horizon. Retrospective energies sum the per-round support
corrections $\norm{g_i^t|_{S_i}}_2^2$ directly up to the last recorded round,
avoiding differences of squared norms. Misalignment is computed as
$m_i=\tfrac12\norm{x_i^*}_2N_i\norm{q_i/N_i-x_i^*/\norm{x_i^*}_2}_2^2$, and
width as
$w_\eta=-\operatorname{expm1}(-\tfrac14\operatorname{log1p}(\tau_1+\tau_2+\tau_1\tau_2))$.

The ledger released with the code has \nLedgerRows{} rows and \nLedgerCols{}
columns, storing one row per game and update schedule. Six
identifier columns give the instance, size $n$, distribution, generator, seed,
and schedule. Equilibrium data follow, recording the support sizes $k_1,k_2$,
support shape, margin, $\operatorname{rank}\Umat$, singular values
$\sigma_{\min},\sigma_{\max}$, rank tolerance, retained and dropped singular
values, and the rank and heuristic uniqueness flags. Next come the run
summaries $T$, final gap, converged and mixed flags, number of modes, final
norms, and $\eta\sigma_{\min},\eta\sigma_{\max}$.

Fit columns record $R^2$, point count, window, $\rho_{\rm meas}$,
$\rho_{\rm cf}$, the extra-gradient prediction, log-rate ratios, and the first
recorded round with gap below $10^{-8}$. Alongside the Jacobian consistency and
closed-form deviation, certificate columns report the terminal $\eps_2$,
crossing round, monotonicity, and certification status. The remaining entries
store the tuned step size, rate, and fit flag for optimistic gradient
descent-ascent.

\looseness=-1Every paired comparison in \cref{sec:experiments} joins ledger rows on the
instance identifier. Comparing update order pairs alternating runs with
simultaneous runs under single-call prediction.
Per-evaluation comparisons pair simultaneous single-call runs with simultaneous
extra-gradient runs. Two companion tables released with the code supply \cref{tab:ttt}, one logging the five learners' crossing rounds per game and the other the two smooth baselines.

\Cref{sec:beyond} examines three published matrix games plus Kuhn
poker and Leduc hold'em. Listed in \cref{tab:matrices} and
\cref{tab:ledger-beyond}, those benchmarks are disjoint
from the \nInst{} testbed games counted in \cref{sec:experiments}.

%% file: tables/tab_ledger.tex
%% generated by scripts/make_tables.py; do not edit
\begin{tabular}{lrrrrrr}\toprule
filter & $n=3$ & $n=5$ & $n=10$ & $n=25$ & $n=50$ & total\\
\midrule
\multicolumn{7}{l}{\emph{simultaneous \IRO{}}}\\
all games & 48 & 48 & 48 & 48 & 48 & 240 \\
converged & 48 & 48 & 48 & 48 & 44 & 236 \\
mixed ($k\ge2$) & 18 & 29 & 40 & 47 & 44 & 178 \\
rank-unique equilibrium & 18 & 26 & 32 & 39 & 44 & 159 \\
rate fit ($R^2>0.98$) & 17 & 22 & 27 & 37 & 32 & 135 \\
consistent Jacobian & 18 & 17 & 15 & 14 & 19 & 83 \\
FD agrees to $10^{-3}$ & 18 & 17 & 15 & 14 & 19 & 83 \\
\midrule
\multicolumn{7}{l}{\emph{simultaneous \IREG{}}}\\
all games & 48 & 48 & 48 & 48 & 48 & 240 \\
converged & 48 & 48 & 48 & 48 & 48 & 240 \\
mixed ($k\ge2$) & 18 & 29 & 40 & 47 & 48 & 182 \\
rank-unique equilibrium & 18 & 26 & 32 & 39 & 48 & 163 \\
rate fit ($R^2>0.98$) & 16 & 22 & 27 & 36 & 35 & 136 \\
\midrule
\multicolumn{7}{l}{\emph{alternating \IRO{}}}\\
all games & 48 & 48 & 48 & 48 & 48 & 240 \\
converged & 48 & 48 & 48 & 48 & 46 & 238 \\
mixed ($k\ge2$) & 18 & 29 & 40 & 47 & 46 & 180 \\
rank-unique equilibrium & 18 & 26 & 32 & 39 & 46 & 161 \\
rate fit ($R^2>0.98$) & 18 & 24 & 28 & 38 & 34 & 142 \\
consistent Jacobian & 18 & 26 & 32 & 39 & 46 & 161 \\
FD agrees to $10^{-3}$ & 18 & 26 & 32 & 39 & 46 & 161 \\
\bottomrule\end{tabular}

%% file: tables/tab_ledger_rfive.tex
%% generated by scripts/make_tables.py; do not edit
\begin{tabular}{lrr}\toprule
count & simultaneous & alternating\\
rank-unique mixed games & 163 & 163 \\
certified below the threshold of the schedule & 163 & 163 \\
of which converge within the horizon & 159 & 161 \\
rate fit, full window (within 5\%) & 135 (122) & 142 (134) \\
rate fit, late window (within 5\%) & 131 (131) & 143 (143) \\
late-window envelope (within 10\%) & 158 (158) & 157 (157) \\
random restarts: converged / above / crossed threshold & 815: 795 / 0 / 815 & 815: 803 / 0 / 815 \\
column-first alternation: converged / above / crossed threshold & -- & 163: 162 / 0 / 163 \\
sharp certificate: $\eps_2<1/2$ / triggered & 123 / 119 & -- \\
games without the rank certificate & 54 & 54 \\
of which strictly complementary limits & 35 & 39 \\
with $2+d_1+d_2$ unit eigenvalues & 35 & 39 \\
\bottomrule\end{tabular}

%% file: tables/tab_lyap.tex
%% generated by scripts/make_tables.py; do not edit
\begin{tabular}{lrrrr}\toprule
$n$ & \multicolumn{2}{c}{simultaneous} & \multicolumn{2}{c}{alternating}\\
 & triggered & games & triggered & games\\\midrule
3 & 18 & 18 & 18 & 18 \\
5 & 26 & 26 & 26 & 26 \\
10 & 32 & 32 & 32 & 32 \\
25 & 39 & 39 & 39 & 39 \\
50 & 42 & 48 & 43 & 48 \\
\midrule
all & 157 & 163 & 158 & 163 \\
\bottomrule\end{tabular}

%% file: sections/10_instances.tex
\section{Instances outside the random testbed}
\label{app:instances}

\Cref{tab:matrices} gives the three matrices from \cref{sec:beyond} entry by entry, transcribed from Appendix E of \citet{farina2023regret}. The third matrix is the $5\times5$ game of \citet{wei2021linear}. \Cref{tab:ledger-beyond} reports the defined ledger columns.

The first two games pass \cref{lem:rank}'s certificate. At equilibrium, every action receives positive probability, giving full-game support and infinite margin. In both cases $\operatorname{rank}\Umat=k-1=2$, and the face matrix is well conditioned with $\kappa=$ \sgFarinaKappa{} and \sgBrpsKappa{}. The $5\times5$ game lies outside \cref{ass:ne}. Its equilibrium set contains multiple points, the linear-programming solution has margin zero, and $\operatorname{rank}\Umat=2$. The run converges to a strictly complementary limit with margin \weiMargin{}. Thus \cref{sec:nonunique}'s analysis applies.

\begin{table}[h]
\caption{Matrices of \cref{sec:beyond} and their game values $v$, transcribed from \citet{farina2023regret} and \citet{wei2021linear}.}
\label{tab:matrices}
\centering\small
\input{tables/tab_matrices}
\end{table}

\begin{table}[h]
\caption{Defined columns of \cref{tab:ledger}, rank and extreme nonzero singular values of $\Umat=A[S_1,S_2]-v\one\one^\top$, and \cref{lem:rank} status for the three matrix games.}
\label{tab:ledger-beyond}
\centering\small
\input{tables/tab_ledger_beyond}
\end{table}

Among the game trees in \cref{sec:beyond}, only Kuhn poker has a normal form of practical size. Its $\efgKuhnNpure$ pure strategies per player yield a $\efgKuhnNpure\times\efgKuhnNpure$ matrix with linear-programming minimax value $-1/18$. Our solver's best response matches this value within \efgKuhnBrDev{}. Three-rank Leduc poker has \efgLeducNinf{} information sets per player. Counterfactual regret minimisation (CFR$^+$) reaches \efgLeducValue{} against the published value \efgLeducPublished{}, a difference of \efgLeducValueDev{}. All instances in this appendix are separate from the \nInst{} random games in \cref{sec:experiments}.

%% file: tables/tab_matrices.tex
%% generated by scripts/make_tables.py; do not edit
\begin{tabular}{ccc}\toprule
\citet{farina2023regret} & biased RPS & \citet{wei2021linear}\\\midrule
$\begin{pmatrix}3 & 0 & -3 \\ 0 & 3 & -4 \\ 0 & 0 & 1\end{pmatrix}$ & $\begin{pmatrix}0 & 1 & -3 \\ -1 & 0 & 1 \\ 3 & -1 & 0\end{pmatrix}$ & $\begin{pmatrix}0 & -1 & 1 & 0 & 0 \\ 1 & 0 & -1 & 0 & 0 \\ -1 & 1 & 0 & 0 & 0 \\ -1 & 1 & 0 & 2 & -1 \\ -1 & 1 & 0 & -1 & 2\end{pmatrix}$\\
$v=0.25$ & $v=0$ & $v=0$\\
\bottomrule\end{tabular}

%% file: tables/tab_ledger_beyond.tex
%% generated by scripts/make_tables.py; do not edit
\begin{tabular}{lrrrrrrrll}\toprule
instance & $n$ & supp. & LP margin & $\operatorname{rank}\Umat$ & $\sigma_{\max}$ & $\sigma_{\min}^{>0}$ & $\kappa$ & eq.\ set & certified\\\midrule
\citet{farina2023regret} & 3 & $3\times3$ & $\infty$ & 2 & 5.97 & 2.99 & 1.99 & single & yes \\
biased RPS & 3 & $3\times3$ & $\infty$ & 2 & 3.32 & 3.32 & 1.00 & single & yes \\
\citet{wei2021linear} & 5 & $3\times3$ & 0 & 2 & 1.73 & 1.73 & 1.00 & multiple & no \\
\bottomrule\end{tabular}

%% file: sections/12_lyapproof.tex
\section{Proof of the Certificate up to the Threshold}
\label{sec:lyapproof}

In this section, we prove the certificate up to the threshold, adopting the notation of \cref{lem:facecoord,thm:lyap}. We restrict all vectors to the supports and drop the player index $i$ where no confusion arises. Moreover, we set $a_t:=\norm{W^t}_2/\sqrt{\min_i\bar c_i}$, so that $\norm{\phi_i^t}_2,\norm{\phi_i^{t-1}}_2\le a_t$ holds for both players (under alternation for the three strategies $\phi_1^t,\phi_2^t,\phi_2^{t-1}$ of the state).

\paragraph{One step.}
We first analyze a single step. Let a player's strategy of round $t\ge t_0$ be supported in its support, with coordinates $(c^t,\phi^t)$, and let $\Delta,w$ be orthogonal to $x^*$. We define $g:=w-\ip{w}{\phi^t}\one$ and the \emph{candidate} $(c^{t+1},\phi^{t+1})$ by \eqref{eq:facestep}, and we do not assume absorption of the next play. Finally, let $a\ge\norm{\phi^t}_2$, $\norm{\Delta}_2\le\bar cD$, $\norm{w}_2\le\bar cD_w$, and $0\le N_t/\bar N-1\le\tau$, and let the denominators of $\varrho,\varsigma_i,e_i$ at $(a,D,D_w,\tau)$ be positive.
\begin{enumerate}
\item[(S1)] $z-x^*=\phi-x^*\ip{d}{\phi}$ is an orthogonal sum, so
  $\norm{z-x^*}_2^2=\norm{\phi}^2+\norm{x^*}^2\ip{d}{\phi}^2\le k\norm{x^*}_2^2\norm{\phi}_2^2$,
  using $1+\norm{x^*}^2\chi^2=k\norm{x^*}^2$, and
  $1-\chi\norm{\phi}\le\norm{z}_2/\norm{x^*}_2\le1+\kappa\norm{\phi}$.
\item[(S2)] $c^t=N_t/\norm{z^t}_2$ and $\bar c=\bar N/\norm{x^*}_2$, so by (S1)
  $c^t/\bar c\in[(1+\kappa a)^{-1},(1+\tau)(1-\chi a)^{-1}]$ and
  $\abs{c^t/\bar c-1}\le\varrho$.
\item[(S3)] $\norm{g}_2\le\norm{w}+\sqrt k\,\abs{\ip{w}{\phi^t}}\le\bar cD_w(1+\sqrt k\,a)$,
  and coordinatewise $\abs{g_a}\le\norm{w}_2(1+a)$.
\item[(S4)] $\abs{\Xi}\le\bar c^2\bigl[2(1+\varrho)aD+D^2+D_w^2(1+\sqrt k\,a)^2\bigr]$ and
  $h\ge\bar c(1-\varrho-\chi D)>0$. Positivity of $1-\varrho^+$ gives
  $k\abs{\Xi}<h^2$, so the root of \eqref{eq:facestep} is real,
  $\varsigma=\Xi/(h+\sqrt{h^2-k\Xi})$, and
  $\abs{\varsigma}\le\abs{\Xi}/h\le\bar c\,\varsigma_i$.
\item[(S5)] $c^{t+1}/\bar c-1=(c^t/\bar c-1)+(\ip{d}{\Delta}-k\varsigma)/\bar c$, hence
  $\abs{c^{t+1}/\bar c-1}\le\varrho^+$.
\item[(S6)] From \eqref{eq:facestep},
  $e^t=(c^t/c^{t+1}-1)\phi^t+(1/c^{t+1}-1/\bar c)\Delta-(\varsigma/c^{t+1})d$ with
  $c^t-c^{t+1}=k\varsigma-\ip{d}{\Delta}$, so by (S4), (S5)
  $\norm{e^t}_2\le\bigl[(\chi D+k\varsigma_i)a+\varrho^+D+\chi\varsigma_i\bigr]/(1-\varrho^+)=e_i$.
\item[(S7)] If the play of round $t$ is strictly positive on the support and
  zero off it, from a vector $q$ supported in the support with prediction $v_i\one+V$, $V$ orthogonal to $x^*$ and
  $\norm{V}_2\le\bar cD$, then $\abs{\gamma^t-v_i}\le\bar c\,\varsigma_i$. Indeed
  $q=p^t-V+\beta\one$ with $\beta:=\gamma^t-v_i$ and $\norm{q}_2=\norm{p^t}_2$, so
  $k\beta^2+2h'\beta+\Xi'=0$ with $h':=c^t-\ip{d}{V}\ge\bar c(1-\varrho-\chi D)$ and
  $\Xi':=\norm{V}^2-2c^t\ip{\phi^t}{V}$, $\abs{\Xi'}\le\bar c^2[2(1+\varrho)aD+D^2]$. The
  mass $\ip{\one}{q}=h'+k\beta$ is positive, which selects the root with
  $\abs{\beta}\le\abs{\Xi'}/h'$.
\end{enumerate}
We note that the functions $\varrho,\varsigma_i/a,\varrho^+$ and $e_i/a$, evaluated at $(a,\lambda a,\lambda_wa,\tau)$ for fixed $\lambda,\lambda_w,\tau$, are nondecreasing in $a$. Indeed, each is built from sums, products, and quotients $1/(1-\cdot)$ of nondecreasing nonnegative terms.

\begin{proof}[Proof of \cref{thm:lyap}, simultaneous schedule]
We use induction on $t\ge t_0$ with the hypothesis $I(t)$, namely that the rounds $t_0-1,\dots,t$ are absorbed and $\norm{W^s}_{\mathcal P}\le\rho_1^{\,s-t_0}\omega$ for $t_0\le s\le t$. The base case $I(t_0)$ holds by assumption. Now assume $I(t)$. Then $a_s\le\rho_1^{\,s-t_0}a_0$ for $s\le t$.

\emph{Norms.} For the opponent data of round $s$, we have $\norm{\Delta}\le\sigma_{\max}(\Umat)\norm{2\phi_{-i}^s-\phi_{-i}^{s-1}}\le\sqrt5\,\sigma_{\max}(\Umat)\norm{W^s}_2/\sqrt{\bar c_{-i}}
=\sqrt5\,\eps\,\bar c_i\norm{W^s}_2/\sqrt{\bar c_i}\le\bar c_i\sqrt5\,\eps a_s$. Likewise, $\norm{w}\le\bar c_i\sqrt2\,\eps a_s$. By \cref{lem:faceid}(i) and (S3), the energies on the support satisfy $N_{i,t+1}^2=\bar N_i^2+\sum_{s=t_0}^t\norm{g_i^s}^2
\le\bar N_i^2+\bar c_i^2D_w^2(1+\sqrt k\,a_0)^2/(1-\rho_1^2)$. That is, $N_{i,s}\le\bar N_i(1+\bar\tau_i)$ for $s\le t+1$.

\emph{Contraction.} Define $(c',\phi')$ for each player by \eqref{eq:facestep} from the data of round $t$, and set $W':=(\psi',\psi^t)$. By (S6) at $(a_t,\sqrt5\,\eps a_t,\sqrt2\,\eps a_t,\bar\tau_i)$ and monotonicity, we have $\norm{e_i^t}\le(a_t/a_0)e_i$. Hence $\norm{\tilde e^t}_2\le(a_t/a_0)(\bar c_1e_1^2+\bar c_2e_2^2)^{1/2}
\le\norm{W^t}_{\mathcal P}(\bar c_1e_1^2+\bar c_2e_2^2)^{1/2}/\omega$. Moreover, $W'=\bar CW^t+(\tilde e^t,0)$ by \eqref{eq:perturbed}. Applying (L1) gives $\norm{W'}_{\mathcal P}\le\rho\norm{W^t}_{\mathcal P}+M\norm{\tilde e^t}_2\le\rho_1\norm{W^t}_{\mathcal P}$.

\emph{The candidate is the play of round $t+1$.} Since $\norm{\phi'}\le\norm{W'}_2/\sqrt{\bar c_i}\le a_0$, every coordinate of $z':=x^*(1-\ip{d}{\phi'})+\phi'$ is at least $\pi_i>0$. Also, $c'>0$ by (S5). Thus $p':=c'z'$ is positive on $S_i$. By the choice of $\varsigma$, this vector has the Euclidean norm of $q^{t+1}$, and it equals $q^{t+1}+u^t-\gamma'\one$ on $S_i$ with $\gamma':=v_i+\varsigma-\ip{w}{\phi^t}$. Off the support, $q^{t+1}$ vanishes because round $t$ is absorbed. Furthermore, for $b\notin S_i$, by (S1) and (S4) we have $u^t_b-\gamma'\le-\mu_i+\alpha_i\norm{z_{-i}^t-x_{-i}^*}_2+\abs{\varsigma}+\abs{\ip{w}{\phi^t}}
\le-\mu_i+\alpha_i\sqrt k\,\norm{x_{-i}^*}_2a_0+\bar c_i(\varsigma_i+D_wa_0)<0$, where the strict inequality holds by (L3). Hence $\pos{q^{t+1}+u^t-\gamma'\one}$ equals $p'$ on $S_i$, vanishes off it, and has the norm of $q^{t+1}$. By uniqueness of the shift, we get $\gamma^{t+1}=\gamma'$. Therefore, the play of round $t+1$ is supported in the supports, and $W^{t+1}=W'$.

\emph{The accumulation of round $t+1$ is absorbed.} On $S_i$, we apply (S2) and the coordinate bound of (S3) at round $t+1$ to obtain $p^{t+1}_a+g^{t+1}_a\ge c^{t+1}\pi_i-\norm{w^{t+1}}(1+a_0)
\ge\bar c_i\bigl[(1-\varrho_i)\pi_i-D_w(1+a_0)\bigr]\ge0$, where the final inequality follows from (L2). Off $S_i$, the unclipped accumulated coordinate satisfies $u_b^{t+1}-\gamma'-\ip{w^{t+1}}{\phi^{t+1}}\le-\mu_i+\alpha_i\sqrt k\,\norm{x_{-i}^*}_2a_0+\bar c_i(\varsigma_i+2D_wa_0)<0$ by (L3). This establishes $I(t+1)$.

Finally, the bound on $z_i^t-x_i^*$ is (S1). Because the profile converges to $(x^*,y^*)$ and the norms converge, $p_i^t=c_i^tz_i^t$ and $q_i^{t+1}=p_i^t+g_i^t$ converge to $c_ix_i^*$, and the stored utilities converge to $u_i^*$. Thus the limit is a point of $R$, and the bracket for $\eta_\infty$ is the definition of $\eta_2$.
\end{proof}

\begin{proof}[Proof of \cref{thm:lyap}, alternating schedule]
The state after round $t$ is $W^t=(\psi_1^t,\psi_2^t,\psi_2^{t-1})$. The
column player's quantities are read on its own clock, meaning that $c_2^t$ and $N_{2,t+1}$
belong to the vector from which $y^t$ is played. The statement $I(t)$ is as before. Given the
masses and the norm budgets, every strategy coordinate used below belongs to
$W^t$ or to the new row candidate. Specifically, the row step
reads $(\phi_2^t,\phi_2^{t-1})$, the column step and its accumulation read
$\phi_1^t$, $\phi_2^t$, and the row candidate, and (S7) reads $U_2\phi_1^t$. Assume
$I(t)$. Round $t+1$ consists of four operations in order. First is the row player's play, second is the column player's
accumulation of its standing strategy $y^t$ against $x^{t+1}$, third is the column
player's play, and fourth is the row player's accumulation.

\emph{Candidates.} As in the simultaneous case, the row candidate $(c_1',\phi_1')$ is defined by
\eqref{eq:facestep} from the data $(\phi_2^t,\phi_2^{t-1})$ of $W^t$. Then (S6) gives $\norm{e_1^t}\le(a_t/a_0)e_1$. Hence
$\norm{\phi_1'}\le\norm{\phi_1^t}+\norm{\Delta}/\bar c_1+\norm{e_1^t}$. Weighting each term
before bounding it also yields
$\sigma_{\max}(\Umat)\norm{\phi_1'}\le\eps\,\bar c_2\bigl[a_t(1+\sqrt5\,\eps)+(\bar c_1/\bar c_2)^{1/2}\norm{e_1^t}\bigr]$.
Consequently, both $\norm{\phi_1'}$ and $\sigma_{\max}(\Umat)\norm{\phi_1'}/(\eps\bar c_2)$ are at most
$(a_t/a_0)a^+$. Next, the column candidate $(c_2',\phi_2')$ is defined by
\eqref{eq:facestep} from $(c_2^t,\phi_2^t)$ using
$\Delta=U_2(2\phi_1'-\phi_1^t)$, $w=U_2(\phi_1'-\phi_1^t)$, and the formal support
accumulation $g=w-\ip{w}{\phi_2^t}\one$. Its data satisfy
$\norm{\Delta}\le\bar c_2\,3\eps a^+(a_t/a_0)$ and $\norm{w}\le\bar c_2\,2\eps a^+(a_t/a_0)$.

\emph{Norms.} The row player's norm is bounded as before. By (S3) and
$I(t)$, the column player's completed accumulations of rounds $t_0+1,\dots,t$ have energies at most $(\bar c_2\,2\eps a^+(1+\sqrt k\,a_0))^2\rho_1^{\,2(s-1-t_0)}$.
Therefore $N_{2,t+1}\le\bar N_2(1+\bar\tau_2)$, where $\bar\tau_2$ is computed from the column
player's $D_w$.

\emph{Contraction.} Applying (S6) to the column candidate at
$(a_t,3\eps a^+a_t/a_0,2\eps a^+a_t/a_0,\bar\tau_2)$ gives
$\norm{e_2^t}\le(a_t/a_0)e_2$. Because the column player's linear part reads the row candidate including its remainder, the candidate state satisfies
$W'=\bar C_{\rm alt}W^t+(\tilde e_1,\tilde e_2-2\bar\eta\,\Umat^\top\tilde e_1,0)$.
Then (L1) gives $\norm{W'}_{\mathcal P}\le\rho_1\norm{W^t}_{\mathcal P}$.

\emph{The four operations of round $t+1$.} By (L2), (L3), and the uniqueness of the shift, the row candidate is the row
player's play, just as in the
simultaneous case. Now consider the column player's accumulation of $y^t$. On $S_2$, we have
$p_{2,a}^t+g_a\ge\bar c_2[(1-\varrho_2)\pi_2-D_w(1+a_0)]\ge0$ by (S2), (S3), and (L2).
Off $S_2$, the unclipped accumulated coordinate is
$u_{2,b}^{t+1}-\gamma_2^t-\ip{w}{\phi_2^t}$, where $\gamma_2^t$ is the shift of the play
$y^t$ made with prediction $v_2\one+U_2\phi_1^t$. Applying (S7) with
$\norm{U_2\phi_1^t}\le\bar c_2\eps a_0$ gives $\abs{\gamma_2^t-v_2}\le\bar c_2\varsigma_2$ both at the base
round $t=t_0$ and at every later round. By (L3), this coordinate is therefore at most
$-\mu_2+\alpha_2\sqrt k\,\norm{x^*}_2a^++\bar c_2(\varsigma_2+D_wa_0)<0$. Thus the
accumulation is absorbed, and the formal norm target of the column candidate
equals the actual one. By the arguments of the simultaneous case, using the column player's constants for its play and the
row player's constants for its accumulation, the column candidate is the play $y^{t+1}$ and the
row player's accumulation of round $t+1$ is absorbed. This establishes $I(t+1)$. Finally,
the limit is a point of $R_{\rm alt}$.
\end{proof}

\begin{proof}[Proof of \cref{prop:drift}]
We first establish the expansion. By \cref{lem:faceid}(i) and \cref{thm:lyap}, we have $N_{i,\infty}^2-\bar N_i^2=\sum_{t\ge t_0}\norm{g_i^t}^2$, which sums the energies on the support. Next, by \cref{lem:facecoord}, we obtain $\norm{g_i^t}^2=\norm{w}^2-2\ip{w}{\phi^t}\ip{w}{d_i}+k\ip{w}{\phi^t}^2
=\norm{w}^2(1+O(a_t))$. Note that $\norm{w}^2=\norm{U_i(\phi_{-i}^t-\phi_{-i}^{t-1})}^2=\bar c_i\,\Gamma_i(W^t)$ holds exactly. Unrolling \eqref{eq:perturbed} yields $W^{t_0+n}=\bar C^nW^{t_0}+\sum_{j<n}\bar C^{\,n-1-j}(\tilde e^{t_0+j},0)$. By (S6), evaluating $e_i$ at $(a,\sqrt5\,\eps a,\sqrt2\,\eps a,\bar\tau_i)$ gives $O(a^2+\bar\tau_ia)$, and we also have $\bar\tau_i=O(a_0^2)$. Hence, we find $\norm{\tilde e^{t_0+j}}=O(\omega^2\rho_1^{\,j})$, which implies $\norm{W^{t_0+n}-\bar C^nW^{t_0}}_{\mathcal P}=O(\omega^2n\rho_1^{\,n-1})$. Since $\Gamma_i$ is a bounded quadratic form and both states are $O(\omega\rho_1^{\,n})$, we obtain $\Gamma_i(W^{t_0+n})-\Gamma_i(\bar C^nW^{t_0})=O(\omega^3n\rho_1^{\,2n-1})$, which is summable. Summing over $n$ yields the expansion. Under alternation, we establish the expansion in the same way for $\bar C_{\rm alt}$. In that setting, the forms are $\norm{\bar\eta\,\Umat(\psi_2^n-\psi_2^{n-1})}^2$ for the row player and $\norm{\bar\eta\,\Umat^\top(\psi_1^{n+1}-\psi_1^n)}^2$ for the column player, where the new row strategy is taken along the linear orbit. Note that the column player's sum starts at its accumulation of round $t_0+1$, which is the first round after the vector of norm $\bar N_2$.

We now prove positivity under either schedule for the row player. For the column player, we exchange the roles of the two recursions. Specifically, under alternation, its form implies that $\psi_1^n$ is constant for $n\ge0$. Moreover, the row recursion makes $\psi_2$ halve from $\psi_2^{-1}$ onward, while the column recursion shifts it by $-\bar\eta\,\Umat^\top\psi_1^0$. By \cref{lem:rank}, $\Umat$ is injective on $(y^*)^\perp$ and $\Umat^\top$ is injective on $(x^*)^\perp$. Suppose that $\ip{W}{\mathcal G_1W}=0$. Then, $\psi_2^n=\psi_2^{-1}$ holds for all $n\ge0$ along the linear orbit from $W$. The column recursion then yields $\Umat^\top(2\psi_1^{m}-\psi_1^{m-1})=0$, with $m=n$ for simultaneous updates and $m=n+1$ for alternating updates. Hence, the row strategy halves at every step starting from the first step read by the column player. Meanwhile, the row recursion gives $\psi_1^{n+1}-\psi_1^n=\bar\eta\,\Umat\psi_2^0=:b$ for all $n\ge0$. Any sequence that both halves and shifts by the constant $b$ must be constant, and hence zero. Therefore, we have $b=0$, which gives $\psi_2^0=\psi_2^{-1}=0$ and $\psi_1^n=0$ for $n\ge0$. For simultaneous updates, we further have $\psi_1^{-1}=2\psi_1^0=0$. Combining these identities gives $W=0$. This concludes the proof.
\end{proof}